%% file: master.tex
\documentclass[11pt,a4paper]{article}

\input{Prolog}

\subheader{\hfill DESY 13-243, NIKHEF-2013-039}

\title{Correlations in double parton distributions: effects of evolution}

\author[a]{Markus Diehl,}
\author[b]{Tomas Kasemets}
\author[a]{and Shane Keane}

\affiliation[a]{Deutsches Elektronen-Synchroton DESY, 22603 Hamburg,
  Germany} 
\affiliation[b]{Nikhef and Department of Physics and Astronomy, VU
  University Amsterdam, De Boelelaan 1081, 1081 HV Amsterdam, The
  Netherlands}

\emailAdd{markus.diehl@desy.de}
\emailAdd{kasemets@nikhef.nl}
\emailAdd{shane.keane13@imperial.ac.uk}

\abstract{We numerically investigate the impact of scale evolution on
  double parton distributions, which are needed to compute multiple hard
  scattering processes.  Assuming correlations between longitudinal and
  transverse variables or between the parton spins to be present at a low
  scale, we study how they are affected by evolution to higher scales,
  i.e.\ by repeated parton emission.  We find that generically evolution
  tends to wash out correlations, but with a speed that may be slow or
  fast depending on kinematics and on the type of correlation.  Nontrivial
  parton correlations may hence persist in double parton distributions at
  the high scales relevant for hard scattering processes.}

\begin{document}

\maketitle

\input{Introduction}
\input{DPDs}
\input{EvoGen}
\input{EvoTrans}
\input{EvoPol}
\input{Factorize}
\input{Conclusions}


\section*{Acknowledgments}

We are greatly indebted to J.~Gaunt for the permission to use his DPD
evolution code and for his help with all matters arising during its
modification.  We thank D.~Boer for correspondence regarding references
\cite{Boer:2009nc,Boer:2010zf,Qiu:2011ai} and W. Vogelsang for calling to
our attention the results in \cite{Vogelsang:1998yd}.  T.K.\ acknowledges
financial support from the European Community under the ``Ideas'' program
QWORK (contract 320389).


\input{Appendices}
\input{Bibliography}

\end{document}

%% file: Prolog.tex
\usepackage{jheppub}

\usepackage[english]{babel}

\usepackage{subfig}

\usepackage{afterpage}

\newcommand{\sect}[1]{section~\protect\ref{#1}}
\newcommand{\sects}[1]{sections~\protect\ref{#1}}
\newcommand{\app}[1]{appendix~\protect\ref{#1}}

\newcommand{\fm}{\operatorname{fm}}
\newcommand{\gev}{\operatorname{GeV}}

\newcommand{\jpsi}{J\mskip -2mu/\mskip -0.5mu\Psi}

\newcommand{\vek}[1] {\boldsymbol{#1}}

\newcommand{\y}{\vek{y}}
\newcommand{\ytilde}{\tilde{\vek{y}}}

\newcommand{\dd}{d}

\newcommand{\half}{{\textstyle\frac{1}{2}}}

\newcommand{\ms}{\mskip 1.5mu}

\allowdisplaybreaks[3]

%% file: Introduction.tex
\section{Introduction}

An intriguing aspect of proton-proton collisions at high energies is
double parton scattering (DPS), where two partons from each proton
interact in two separate hard subprocesses.  While the description of
single hard scattering has become an area of precision calculations, our
understanding of double hard scattering (and of its extension to three or
more hard subprocesses) remains fragmentary, both at the conceptual and at
the quantitative level.  The initial state of double parton scattering is
quantified by double parton distributions (DPDs), which quantify the joint
distribution of two partons in a proton, depending on their quantum
numbers, their longitudinal momentum fractions and their relative
transverse distance from each other.  A better knowledge of these
distributions is important because DPS processes can contribute to many
final states of interest at the LHC
\cite{Bartalini:2011jp,Platzer:2012gla,Abramowicz:2013iva} and,
furthermore, because they quantify characteristic aspects of proton
structure beyond the information contained in the familiar parton
distribution functions (PDFs) for a single parton.

DPDs depend on a scale, which in a physical process is given by the typical
scale of the hard scattering, just as for PDFs.  The scale dependence of
DPDs is described by a generalization of the familiar DGLAP evolution
equations.  Two versions of this have been discussed in the literature: a
homogeneous equation describing the separate evolution of each of the two
partons and an inhomogeneous equation, which has an additional term
describing the perturbative splitting of one parent parton into the two
partons that will undergo a hard scattering \cite{Kirschner:1979im,%
  Shelest:1982dg,Snigirev:2003cq,Gaunt:2009re,Ceccopieri:2010kg}.  Which
version is adequate for the description of double hard scattering
processes remains controversial in the literature
\cite{Gaunt:2011xd,Gaunt:2012dd,Diehl:2011tt,Diehl:2011yj,%
  Ryskin:2011kk, Manohar:2012pe,Blok:2011bu,Blok:2013bpa}.  In this work
we use the homogeneous equation, having in mind that a systematic theory
of double hard scattering will treat the physics associated with the
inhomogeneous splitting term separately.

The joint distribution of two partons in the proton can be subject to
various correlations.  A number of arguments suggest a nontrivial
interplay between the dependence of DPDs on the longitudinal momentum
fractions $x_1,$ $x_2$ of the partons, as well as between their momentum
fractions and their relative transverse distance $\vek{y}$
\cite{Diehl:2013mma}.  Moreover, the polarizations of two partons can be
correlated even in an unpolarized proton
\cite{Mekhfi:1983az,Mekhfi:1985dv,Diehl:2011yj}.  A recent study in the
MIT bag model \cite{Chang:2012nw} finds indeed large quark spin
correlations in its range of validity, i.e.\ at $x_i$ above, say $0.1$,
and at a low scale.  Two-parton correlations in color, quark flavor or
fermion number have also been discussed in the literature
\cite{Mekhfi:1985dv,Diehl:2011yj,Manohar:2012jr}, but will not be
considered in the present work.

Given the large range of energy scales relevant in LHC processes, it is
important to understand how correlations in the distribution of two
partons evolve with the scale.  One might for instance expect that spin
correlations that exist at a low scale become diluted by the subsequent
parton radiation that is described by DGLAP evolution to higher scales.
The purpose of the present paper is to study the evolution behavior of
different correlations in DPDs in a quantitative manner.  First results of
our study have been presented in \cite{Kasemets:2013nma}.

This paper is organized as follows.  In \sect{sec:DPDs} we recall some
basics of DPDs and in \sect{sec:evo} some details of their scale
evolution.  The behavior of correlations between $x_1, x_2$ and $\vek{y}$
under evolution is examined in \sect{sec:trans}.  A large part of our
investigation, presented in \sect{sec:pol}, is the study of how spin
correlations evolve.  In \sect{sec:factorize} we investigate the stability
under evolution of the ansatz that two unpolarized partons are distributed
independently of each other.  Our conclusions are given in
\sect{sec:concl}.  In \app{ap:pdfs} we motivate the choice of PDFs used in
our studies, and in \app{ap:split} we collect some analytical expressions
needed in \sect{sec:pol}.

\clearpage

%% file: DPDs.tex
\section{Double parton distributions}
\label{sec:DPDs}

If one assumes factorization, then the cross section for a double parton
scattering process can be written as
\begin{align}
  \label{eq:cross}
\frac{d\sigma}{dx_1\, dx_2\, dx_3\, dx_4}
&= \frac{1}{C}\! \sum_{p_1, p_2, p_3, p_4} \hspace{-0.8em}
   \hat{\sigma}_{p_1 p_3}(x_1\ms x_3)\, \hat{\sigma}_{p_2 p_4}(x_2\ms x_4)
   \int d^2\vek{y}\;  F_{p_1 p_2}(x_1,x_2,\vek{y})\,
                      F_{p_3\ms p_4}(x_3,x_4,\vek{y})
\nonumber \\[0.2em]
&\quad + \text{ \{color, flavor and fermion number interference terms\} }
\,,
\end{align}
where $C$ is a combinatorial factor, $\hat{\sigma}_{p_i p_j}$ is the
tree-level cross section for the hard scattering initiated by partons
$p_i$ and $p_j$ and $F_{p_i p_j}$ is a DPD for partons $p_i$ and $p_j$ in
the proton.  The formula can be extended to include radiative corrections
for $\hat{\sigma}_{p_i p_j}$ and then involves convolution integrals over
parton momentum fractions, just as for single hard scattering.  There is
no complete proof that factorization as in \eqref{eq:cross} actually
holds, but many important ingredients to such a proof can be found in
\cite{Diehl:2011yj,Manohar:2012jr}.  We shall not be concerned with the
interference terms alluded to in \eqref{eq:cross}, but note that the
DPDs describing color or fermion number interference are accompanied by
Sudakov double logarithms and have a different scale evolution than the
one we are studying in this work.  In the parlance of \cite{Diehl:2011yj},
the distributions $F_{p_i p_j}$ in \eqref{eq:cross} are color singlet
DPDs.

It is understood that the DPS cross section in \eqref{eq:cross} needs to
be added to the one for single hard scattering.  (One also needs to add
the interference between single and double hard scattering, which has
received only little attention in the literature so far and will not be
discussed here). Double parton scattering can compete with the single
scattering mechanism in parts of phase space and even in inclusive cross
sections when the single parton scattering contribution is suppressed by
multiple small coupling constants.  In particular, DPS is enhanced for
small momentum fractions $x_i$, because the joint distribution of two
small-$x$ partons increases faster with $1/x$ than the distribution of a
single one.  As an immediate consequence, DPS tends to be more important
at the LHC than at previous hadron colliders.

We can include the effects of parton spin correlations in \eqref{eq:cross}
by denoting with $p_i$ the type of a parton and its polarization at the
same time.  $F_{p_i p_j}$ ($\hat{\sigma}_{p_i p_j}$) are then sums or
differences of DPDs (subprocess cross sections) for different polarization
states of the partons. Following \cite{Diehl:2011yj}, we write $q,
\bar{q}, g$ for unpolarized partons, $\Delta q, \Delta\bar{q}, \Delta g$
for longitudinally polarized ones, $\delta q$ or $\delta\bar{q}$ for
transversely polarized quarks or antiquarks, and $\delta g$ for linearly
polarized gluons.  For each label $\delta q$ or $\delta\bar{q}$ the
corresponding DPDs and hard-scattering cross sections carry one Lorentz
index in the transverse plane (corresponding to the transverse
polarization vector), whereas for each index $\delta g$ we need two
transverse indices.
In DPDs for two quarks, the polarization combinations allowed by parity
and time reversal invariance are~\cite{Diehl:2011yj}
\begin{align}
\label{eq:def-qq}
  F_{qq}(x_1,x_2,\y) & = f_{qq}(x_1,x_2,y) \,,
  \nonumber\\
  F_{\Delta q \Delta q}(x_1,x_2,\y) & =
    f_{\Delta q \Delta q}(x_1,x_2,y) \,,
  \nonumber\\
  F_{q \ms \delta q}^j(x_1,x_2,\y) & =
    \ytilde^j M f_{q \ms \delta q}(x_1,x_2,y) \,,
  \nonumber\\
  F_{\delta q \ms q}^j(x_1,x_2,\y) & =
    \ytilde^j M f_{\delta q \ms q}(x_1,x_2,y) \,,
  \nonumber\\
  F_{\delta q \delta q}^{jj'}(x_1,x_2,\y) & =
    \delta^{jj'} f_{\delta q \delta q}(x_1,x_2,y)
    +  2\tau^{jj'\!,kk'} \y^k \y^{k'} 
       M^2 f_{\delta q \delta q}^t(x_1,x_2,y) \,,
\end{align}
where we write $y=\sqrt{\y^2}$ and introduce the proton mass $M$ so that
all functions $f$ have the same mass dimension. Furthermore, we use
$\ytilde^j = \epsilon^{jj'} \y^{j'}$ with the antisymmetric symbol
$\epsilon^{jj'}$ in two dimensions, and
\begin{align}
  \tau^{jj'\!,kk'} = \half \ms \bigl( \delta^{jk}\delta^{j'k'} 
     + \delta^{jk'}\delta^{j'k} - \delta^{jj'}\delta^{kk'} \bigr) \,.
\end{align}
As shown in \cite{Diehl:2013mla}, one can write
\begin{align}
\label{eq:def-qg}
  F_{qg}(x_1,x_2,\y) & = f_{qg}(x_1,x_2,y) \,,
  \nonumber\\
  F_{\Delta q \Delta g}(x_1,x_2,\y) & =
     f_{\Delta q \Delta g}(x_1,x_2,y) \,,
  \nonumber\\
  F_{q \ms \delta g}^{jj'}(x_1,x_2,\y) & =
     \tau^{jj'\!,kk'}  \y^k \y^{k'} M^2 f_{q \ms \delta g}(x_1,x_2,y) \,,
  \nonumber\\
  F_{\delta q \ms g}^j(x_1,x_2,\y) & =
     \ytilde^j M f_{\delta q \ms g}(x_1,x_2,y) \,,
  \nonumber\\
  F_{\delta q \delta g}^{j,kk'}(x_1,x_2,\y) & =
     \!\! {}- \tau^{jj', kk'} \ytilde^{j'} M f_{\delta q \delta g}(x_1,x_2,y)
  \nonumber\\
  & \quad - \tau^{kk'\!,l\ms l'}\bigl(  \ytilde^j  \y^l
                 +  \y^j \ytilde^{l} \bigr)\,
               \y^{\ms l'} M^3 f_{\delta q \delta g}^t(x_1,x_2,y)
\end{align}
\vspace{-0.1cm}
for DPDs of one quark and one gluon, and
\begin{align}
\label{eq:def-gg}
  F_{gg}(x_1,x_2,\y) & = f_{gg}(x_1,x_2,y) \,,
  \nonumber\\
  F_{\Delta g \Delta g}(x_1,x_2,\y) & =
    f_{\Delta g \Delta g}(x_1,x_2,y) \,,
  \nonumber\\
  F_{g \ms \delta g}^{jj'}(x_1,x_2,\y) & =
    \tau^{jj'\!,kk'} \y^k \y^{k'} M^2 f_{g \ms \delta g}(x_1,x_2,y) \,,
  \nonumber\\
  F_{\delta g \ms g}^{jj'}(x_1,x_2,\y) & =
    \tau^{jj'\!,kk'} \y^k \y^{k'} M^2 f_{\delta g \ms g}(x_1,x_2,y) \,,
  \nonumber\\
  F_{\delta g \delta g}^{jj',kk'}(x_1,x_2,\y) & =
    \half\ms \tau^{jj'\!,\,kk'} f_{\delta g \delta g}(x_1,x_2,y) \,,
  \nonumber\\                           
  & \hspace{-4em} + \tau^{jj'\!,l\ms l'}
                   \tau^{kk'\!,mm'} \bigl(  \ytilde^{\ms l} \ytilde^m
                 -  \y^l \y^m \bigr) \y^{l'} \y^{m'}\,
      M^4 f_{\delta g \delta g}^t(x_1,x_2,y)
\end{align}
\vspace{-0.1cm}\noindent
for two gluons.  Expressions analogous to \eqref{eq:def-qq} and
\eqref{eq:def-qg} hold if one or two quarks are replaced by antiquarks.
\vspace{-0.1cm}
\subsubsection*{Polarization effects in double parton scattering}

DPDs for polarized partons contribute to the cross section
\eqref{eq:cross} if the cross section differences $\hat{\sigma}_{p_i p_j}$
for the relevant hard subprocesses are nonzero.  A systematic discussion
would go beyond the scope of this work, but let us mention a few important
examples.  A detailed discussion of the impact of parton spin correlations
on the production of two electroweak gauge bosons ($\gamma^*$, $Z$ or $W$)
has been given in \cite{Manohar:2012jr,Kasemets:2012pr}.  One finds
nonzero cross section differences $\hat{\sigma}_{\Delta q \Delta\bar{q}}$,
and for $Z$ and $W$ production also $\hat{\sigma}_{\Delta q\ms \bar{q}}$
and $\hat{\sigma}_{q \Delta\bar{q}}$ thanks to their parity violating
couplings.  If the corresponding DPDs for longitudinal quark and antiquark
polarization are nonzero, this influences both the overall rate of DPS and
the distribution in transverse momentum and rapidity of the leptons into
which the gauge bosons decay.  For $\gamma^*$ and $Z$ production there is
a nonzero cross section difference $\hat{\sigma}_{\delta q\ms
  \delta\bar{q}}$, which leads to an azimuthal correlation between the
decay planes of the two bosons, provided that the transverse polarizations
of two quarks or antiquarks in the proton are correlated as well.

The cross section differences $\hat{\sigma}_{\Delta a \Delta b}$ for
longitudinal polarization in jet production are nonzero for most
combinations $a, b$ of quarks, antiquarks and gluons, and the same holds
for prompt photon production (see e.g.\ table 4.1 in
\cite{Bourrely:1987gp}).  This will impact the overall rate of DPS as well
as the transverse-momentum and rapidity distributions of the jets if there
are longitudinal spin correlations between two partons in the proton.  For
transverse (anti)quark polarization, there are nonzero cross section
differences $\hat{\sigma}_{\delta q\ms \delta\bar{q}}$ and
$\hat{\sigma}_{\delta q\ms \delta q}$ for jet production and
$\hat{\sigma}_{\delta q\ms \delta\bar{q}}$ for the prompt photon channel
$q\bar{q}\to g\gamma$ \cite{Jaffe:1996ik}, which together with transverse
polarization correlations in the proton induce azimuthal correlations
between the relevant jet planes.  Azimuthal correlations in the final
state can also be induced by linearly polarized gluons, with a nonzero
cross section difference $\hat{\sigma}_{\delta g\ms \delta g}$ for jet
production \cite{Boer:2009nc}.  The cross section difference
$\hat{\sigma}_{g \delta g}$ is zero in that case, but it is nonzero for
the production $gg\to Q\bar{Q}$ of heavy quarks \cite{Boer:2010zf} and the
production $gg\to \gamma\gamma$ of a photon pair \cite{Qiu:2011ai}.  We
thus see that a number of important DPD channels will be impacted by spin
correlations of partons in the proton.

%% file: EvoGen.tex
\section{Evolution of double parton distributions}
\label{sec:evo}

As discussed in the introduction, we consider the homogeneous evolution
equation of DPDs.  For two unpolarized quarks we then have
\begin{align}
  \label{eq:evol-parton-1}
  \frac{\dd f_{qq}(x_1,x_2,y; Q)}{\dd\ln Q^2}
    & = \frac{\alpha_s(Q)}{2\pi} \Bigl[
        P_{qq} \otimes^{}_1 f_{qq}
      + P_{qg} \otimes^{}_1 f_{gq} 
      + P_{qq} \otimes^{}_2 f_{qq}
      + P_{qg} \otimes^{}_2 f_{qg} \Bigr] \,,
\end{align}
where
\begin{align}
\label{eq:otimes}
    P_{ab}( \, .\, ) \otimes^{}_1
       f_{bc}( \, .\, , x_2, y;Q) 
    & = \int_{x_1}^{1-x_2} \frac{d z}{z}
        P_{ab}\left( \frac{x_1}{z} \right)
          f_{bc}(z, x_2, y;Q) \,,
\nonumber \\[0.2em]
    P_{ab}( \, .\, ) \otimes^{}_2
       f_{bc}(x_1, \, .\, , y;Q) 
    & = \int_{x_2}^{1-x_1} \frac{d z}{z}
        P_{ab}\left( \frac{x_2}{z} \right)
          f_{bc}(x_1, z, y;Q)
\end{align}
is a convolution in the first or second argument of the DPDs with the
splitting functions $P_{ab}$ known from the DGLAP evolution of
single parton distributions.  We use the leading-order (LO) approximation
of the splitting functions throughout this work.  The explicit evolution
equations for all unpolarized and polarized DPDs, as well as a list of the
associated splitting functions, are collected in appendix A of
\cite{Diehl:2013mla}.  We note that the splitting functions for antiquarks
are identical with those for quarks.

Let us briefly recapitulate the pattern of evolution for small $x_i$,
starting with gluon distributions.  For small argument $x$, we have
\begin{align}
  \label{eq:asy-splitting}
P_{gg}(x) & \to 2N_c /x \,, &
P_{gq}(x) & \to 2C_F /x \,
\nonumber \\
P_{\Delta g \Delta g}(x) & \to 4N_c \,, &
P_{\Delta g \Delta q}(x) & \to 2C_F \,,
\end{align}
where $N_c$ is the number of colors and $C_F = 2N_c^{} \,/ (N_c^2-1)$.  In
each case the correction to the asymptotic behavior is one power higher in
$x$.  For small $x_1$ (and $x_2$ not too large) the first convolution
integral in \eqref{eq:otimes} can receive a substantial contribution from
the region $x_1 \ll z \ll 1-x_2$ where the splitting functions take their
asymptotic forms \eqref{eq:asy-splitting} while the DPDs are far away from
the kinematic limit $z = 1-x_2$, where they become small.  An analogous
statement holds of course for the second integral in \eqref{eq:otimes} at
small $x_2$ (and not too large $x_1$).  This explains the steep rise of
the unpolarized gluon distribution with $Q^2$.  For longitudinal gluon
polarization, this rise is weaker since $P_{\Delta g \Delta g}$ and
$P_{\Delta g \Delta q}$ lack the $1/x$ singularity of their unpolarized
counterparts.

For linearly polarized gluons (which do not mix with quarks under
evolution) the situation is special.  The small-$x$ limit of the splitting
function reads
\begin{align} 
\label{eq:lin-split}
P_{\delta g\delta g}(x) \to 2N_c\ms x +
  \frac{\alpha_s}{2\pi}\, \frac{N_c^2 + (N_c^{} - 2C_F)\, n_F}{6x} \,,
\end{align}
where $n_F$ is the number of active quark flavors.  Here we have included
the small-$x$ limit of the NLO contribution computed in
\cite{Vogelsang:1998yd} because it has a
$1/x$ enhancement while the LO term vanishes like $x$ for $x\to 0$.
Unless this NLO effect is very large (i.e.\ unless one considers scales
where $\alpha_s$ is large) one hence expects that distributions for
linearly polarized gluons have at most a moderate growth with $Q^2$ at
small momentum fractions.

For quark distributions the splitting functions $P_{qq}(x)$, $P_{qg}(x)$,
$P_{\Delta q \Delta q}(x)$ and $P_{\Delta q \Delta g}(x)$ all tend towards
constant values at small $x$, whereas $P_{\delta q \delta q}(x) \to 2
C_F\ms x$.  Compared with unpolarized gluons, one thus expects a much
milder growth with $Q^2$ for quarks at small $x_i$, irrespective of their
polarization.  The strongest increase is to be expected for unpolarized
quarks since they mix with the large unpolarized gluon distribution.


\subsection{Numerical implementation}
\label{sec:code}

To solve the evolution equations numerically, we use a modified version of
the code described in \cite{Gaunt:2009re,Gaunt:thesis}.  The original code
was written to solve the inhomogeneous evolution equations of
Refs.~\cite{Kirschner:1979im, Shelest:1982dg} for unpolarized DPDs.  We
have modified the code by removing the inhomogeneous term and by adding
the splitting functions for polarized partons.

The code solves the double DGLAP equations in a variable flavor number
scheme.  It works in $x$-space, on a grid in $x_1$, $x_2$ and $t = \ln
Q^2$, performing the evolution stepwise in $t$ by a fourth-order
Runge-Kutta method.  The $x_i$ grid points are evenly spaced in
$\log\bigl[ {x_i}/{(1-x_i)} \bigr]$, with an equal number of points in
both directions.  They are bounded from above by the kinematic limit
$x_1+x_2 \leq 1$ and from below by the choice of $x_{\text{min}}=10^{-6}$.
The grid points in $t$ are evenly spaced, ranging from $t_0$ to $t_{max}$
for which we chose different values in different parts of our
investigation.  We used 240 grid points in each of the $x_i$ directions
and 60 grid points in $t$.  The code is supplemented with a routine that
interpolates between different grid points.  We made some small changes to
this routine, making in particular sure that it can handle polarized
distributions, which may have zero crossings and negative values.

The accuracy of the original code was investigated in \cite{Gaunt:2009re},
with estimated errors below $1\%$ for $x_i\leq 0.3$ and evolution from
$Q^2=1\gev^2$ to $Q^2=10^4\gev^2$.  We have updated these estimates after
our modification of the code and with our grid settings.  We find again an
accuracy better than $1\%$ for the evolution of unpolarized DPDs, whereas
for polarized distributions the relative error increases up to $4\%$ in
some regions at moderate $x_i$.  In the vicinity of zero crossings, the
relative error diverges and is of course no longer a useful measure for
numerical uncertainties.

For the solution of the evolution equations, the code transforms the DPDs
to a straightforward generalization of the single parton ``evolution
basis''.  This basis is defined by \cite{Ellis:1991qj}
\begin{align}
\label{eq:evobase}
  \Sigma & = \sum_i q_i^+ \,, & V_i   & = q_i^-\,, 
  \nonumber\\
  T_3   & = u^+ - d^+ \,,     & T_8   & = u^+ + d^+ - 2s^+
\end{align}
and similar combinations including heavier quarks, where $q_i^\pm = q_i^{}
\pm \bar{q}_i^{}$.  Analogous linear combinations are formed for polarized
partons.  In this basis single parton evolution is particularly simple,
since mixing only occurs between the singlet ($\Sigma$) and the gluon
while the other combinations evolve separately.  For the up and down
quarks, $V_i$ corresponds to the valence contributions $u_v$ and $d_v$.
The evolution code makes use of the basis \eqref{eq:evobase} for both
partons.

%% file: EvoTrans.tex
\section{Correlations between $x_1, x_2$ and $y$}
\label{sec:trans}

Various studies of generalized parton distributions suggest a nontrivial
interplay between the distribution of partons in longitudinal momentum and
in transverse space \cite{Diehl:2013mma}.  Specifically, the impact
parameter dependent single parton distribution $f_a(x,\vek{b})$, i.e.\ the
probability density to find parton $a$ with momentum fraction $x$ at a
transverse distance $\vek{b}$ from the proton center, is not simply the
product between a function of $x$ and a function of $\vek{b}$.  It is
therefore natural to assume that there is also a correlation between the
longitudinal variables $x_1$, $x_2$ and the transverse distance $\vek{y}$
in DPDs.  In this section we investigate how such a correlation behaves
under scale evolution.  We consider only unpolarized partons and focus on
the region of small momentum fractions $x_i$, which is relevant for a
large range of DPS processes at the LHC.

\subsection{Initial conditions}
\label{sec:init-cond}

As a model for the DPD at the starting scale of evolution, we take the
simple ansatz that follows if one assumes the two partons to be
independent.  The DPD can then be written as a convolution
\begin{align}
  \label{eq:dpd-gpd}
  F_{ab}(x_1,x_2,\y) = \int d^2\vek{b}\;
                       f_a(x_1,\vek{b}+\y) \, f_b(x_2,\vek{b})
\end{align}
of impact parameter dependent single parton distributions
$f_a(x,\vek{b})$, as shown for instance in \cite{Diehl:2011yj}.  For these
distributions, we assume a Gaussian $\vek{b}$ dependence with an $x$
dependent width, namely
\begin{align}
  \label{eq:tran-gau}
f_a(x,\vek{b}) = f_a(x)\, \frac{1}{4\pi h_{a}(x)}\,
    \exp\biggl[ -\frac{\vek{b}^2}{4\ms h_{a}(x)} \biggr]
\end{align}
with
\begin{align}
  \label{eq:tran-log}
  h_{a}(x) = \alpha_{a}' \ln \frac{1}{x} + B_{a} \,.
\end{align}
Here $f_a(x)$ denotes the usual parton densities, for which we take the LO
set of the MSTW 2008 analysis \cite{Martin:2009iq}.  We return to the
choice of this PDF set below.  For the starting scale where the ansatz
\eqref{eq:tran-gau} is assumed we take $Q_0^2 = 2 \gev^{2}$.
We should note that the form \eqref{eq:tran-log} is tailored for the
region of $x$ up to $10^{-1}$ and not meant to be realistic for larger
$x$, see e.g.\ the discussion in section 7.3 of \cite{Diehl:2004cx}.  We
take different parameters in \eqref{eq:tran-log} for gluons and for the
sum $q^+ = q+\bar{q}$ and difference $q^- = q-\bar{q}$ of quark and
antiquark distributions,
\begin{align}
  \label{eq:trans-param}
  \alpha_{q^-}' & = 0.9 \gev^{-2}\,, &
  B_{q^-}       & = 0.59 \gev^{-2}\,,
\nonumber \\
  \alpha_{q^+}' & = 0.164 \gev^{-2}\,, &
  B_{q^+}       & = 2.4 \gev^{-2}\,,
\nonumber \\
  \alpha_{g}'   & = 0.164 \gev^{-2}\,, & 
  B_{g}         & = 1.2 \gev^{-2} \,.
\end{align}
The parameter values for $q^-$ were obtained in a model dependent
determination of generalized parton distributions from electromagnetic
form factor data \cite{Diehl:2004cx}.  For the remaining parameters we use
input from hard exclusive scattering processes.  At leading order in
$\alpha_s$, the scattering amplitudes for exclusive $\jpsi$
photoproduction and for deeply virtual Compton scattering (DVCS) are
described by generalized parton distributions for gluons and for the sum
$q^+$ of quarks and antiquarks, respectively.  Up to an uncertainty from
the so-called skewness effect, one can thus connect the measured $t$
dependence in those processes with the Fourier transform of the
distribution \eqref{eq:tran-gau} to transverse-momentum space.  The values
of $\alpha_{\smash{g}}'$ and $B_{g}$ given in \eqref{eq:trans-param} have
been determined in \cite{Diehl:2007zu} to match the measurement of elastic
$\jpsi$ photoproduction in \cite{Aktas:2005xu}.  Experimental
uncertainties do not allow us to extract a value for
$\alpha_{\smash{q^+}}'$ from DVCS, and we take the same value as for
gluons in this case.  The value of $B_{q^+}$ in \eqref{eq:trans-param} has
been chosen to correspond to a DVCS cross section $d\sigma/dt \propto
e^{bt}$ with $b \approx 7 \gev^{-2}$ at $x \approx 10^{-3}$ and $Q^2
\approx 2 \gev^2$, guided by a fit to the $t$ dependence in
\cite{Aaron:2007ab}.  

Let us emphasize that the functional form and numerical values in
\eqref{eq:tran-gau} to \eqref{eq:trans-param} are not meant to be a
precision extraction of impact parameter dependent parton densities, but
as a simple ansatz in rough agreement with phenomenology.  The focus of
our interest is how correlations of this type are affected by scale
evolution.

Inserting \eqref{eq:tran-gau} into \eqref{eq:dpd-gpd} we obtain our ansatz
for the unpolarized DPDs,
\begin{align}
  \label{eq:g-ansatz}
F_{ab}(x_1,x_2,\y) =f_a(x_1) \ms f_b(x_2)\,
   \frac{1}{4\pi h_{ab}(x_1,x_2)}\,
      \exp\biggl[ - \frac{\y^2}{4\ms h_{ab}(x_1,x_2)} \biggr]
\end{align}
at the starting scale $Q_0^2 = 2 \gev^2$, with
\begin{align}
  \label{eq:dpd-trans}
h_{ab}(x_1,x_2) = h_a(x_1) + h_b(x_2)
 = \alpha_{a}'\ln\frac{1}{x_1} + \alpha_{b}'\ln\frac{1}{x_2}
   + B_{a} + B_{b} \,.
\end{align}
Note that the $\y$ dependence in \eqref{eq:g-ansatz} does not factorize
into separate contributions from each of the two partons.

To specify the mixing between gluon and quark singlet distributions, we
take the parameters $\alpha_{\smash{q^+}}'$ and $B_{q^+}$ in
\eqref{eq:trans-param} for all light quark flavors, $u, d, s$ at the
starting scale $Q_0$.  The charm distribution is negligibly small there,
since for the MSTW 2008 distribution we have $m_c \approx Q_0$.  In the
following we will consider the combinations $u^-$ and $u^+$ as
representatives of the quark sector for definiteness.  We have checked for
a few example cases that no qualitatively new features arise for
distributions where one or two $u$ quarks are replaced by $d$ quarks.
This is not surprising since the PDFs for these quarks are similar in
shape and the $y$ dependence in our ansatz has a trivial flavor structure.


\subsection{Change under evolution}

According to \eqref{eq:evol-parton-1} DPDs evolve independently at each
value of the interparton distance $y$.  However, the nontrivial interplay
between $y$ and the momentum fractions $x_1$ and $x_2$ in the starting
conditions \eqref{eq:g-ansatz} has consequences for the scale evolution at
different values of $y$. The exponential factor in \eqref{eq:g-ansatz}
leads to a suppression of the large $x_i$ region, which becomes more
important as $y$ increases.  Furthermore, the relative size of gluon and
$q^+$ distributions, which mix under evolution, changes with $y$ because
our ansatz implies $h_g(x) < h_{q^+}(x)$ and thus has a broader $y$
profile for quarks than for gluons.

\begin{figure}[p]
  \centering
  \subfloat[]{\includegraphics[width=0.478\textwidth]{%
      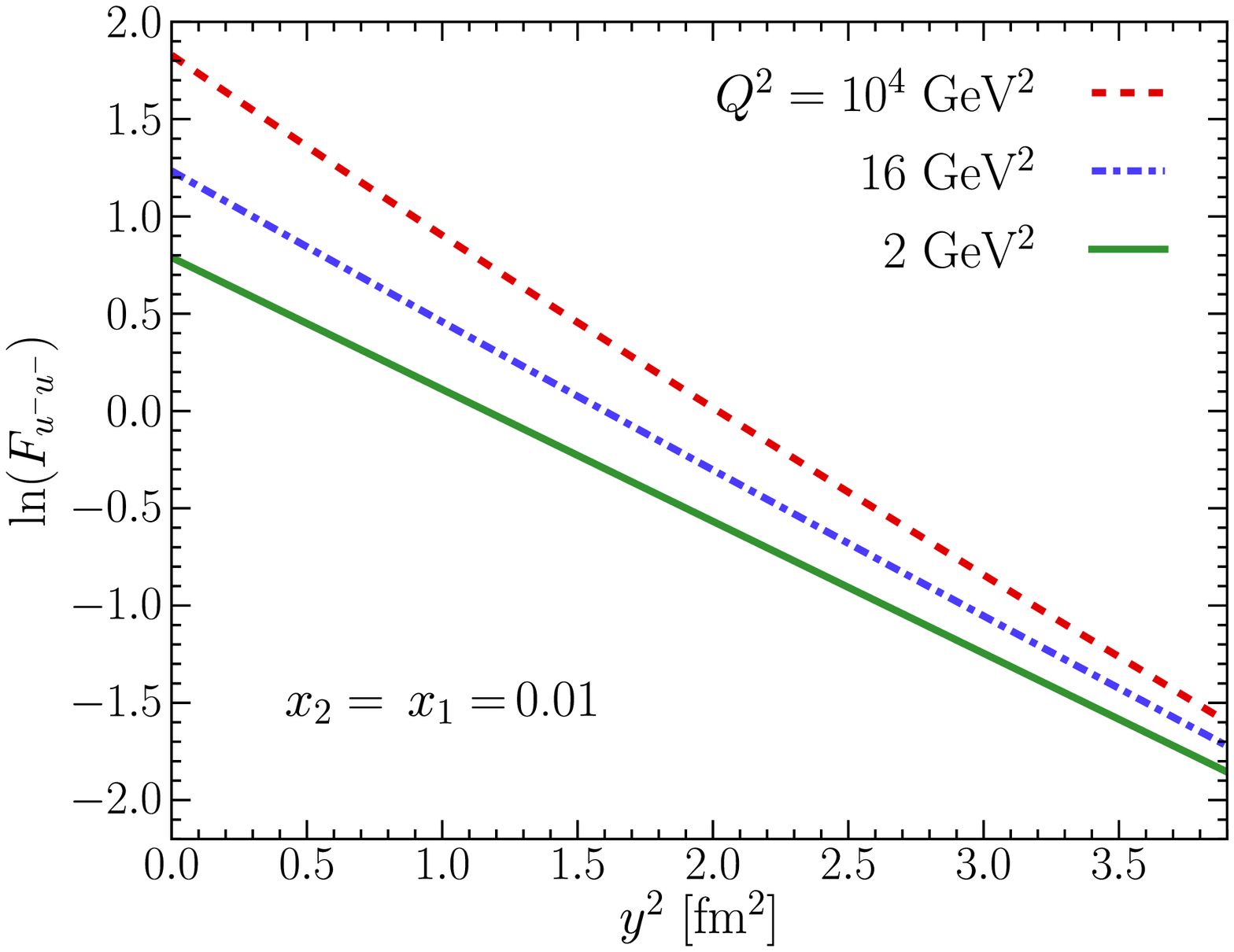}}
  \subfloat[]{\includegraphics[width=0.501\textwidth]{%
      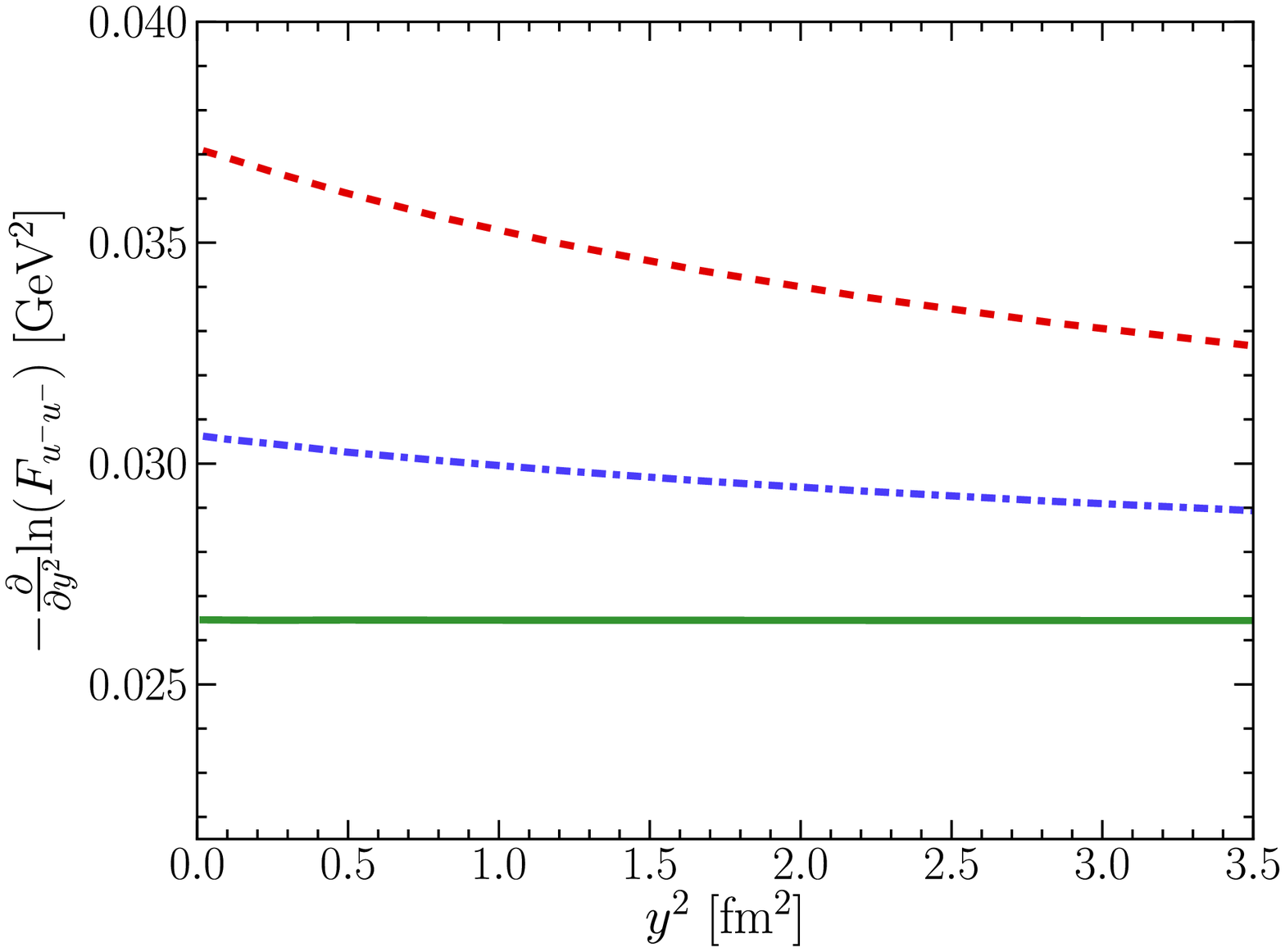}}

  \vspace{0.5em}

  \subfloat[]{\includegraphics[width=0.465\textwidth]{%
      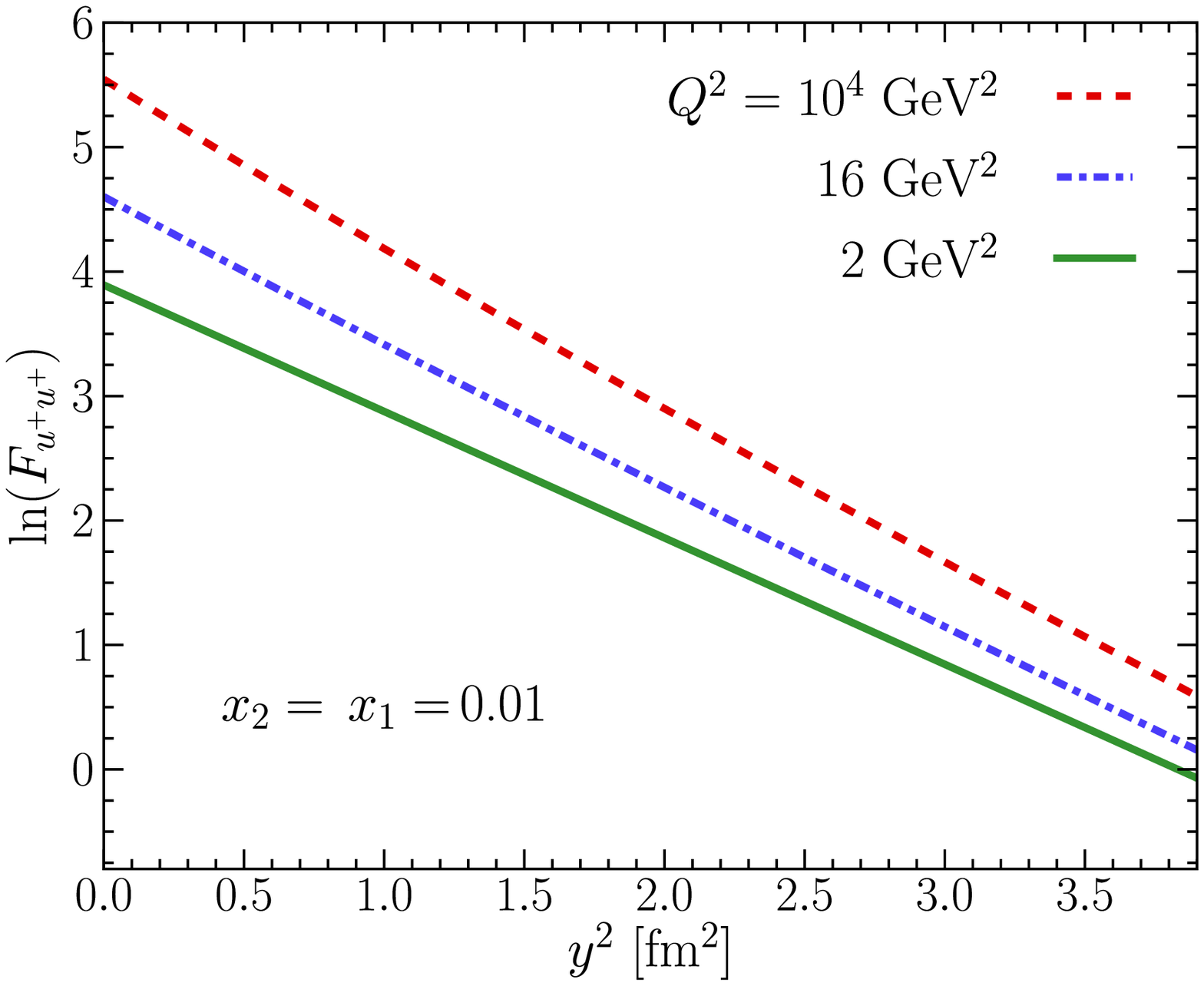}}
  \subfloat[]{\includegraphics[width=0.515\textwidth]{%
      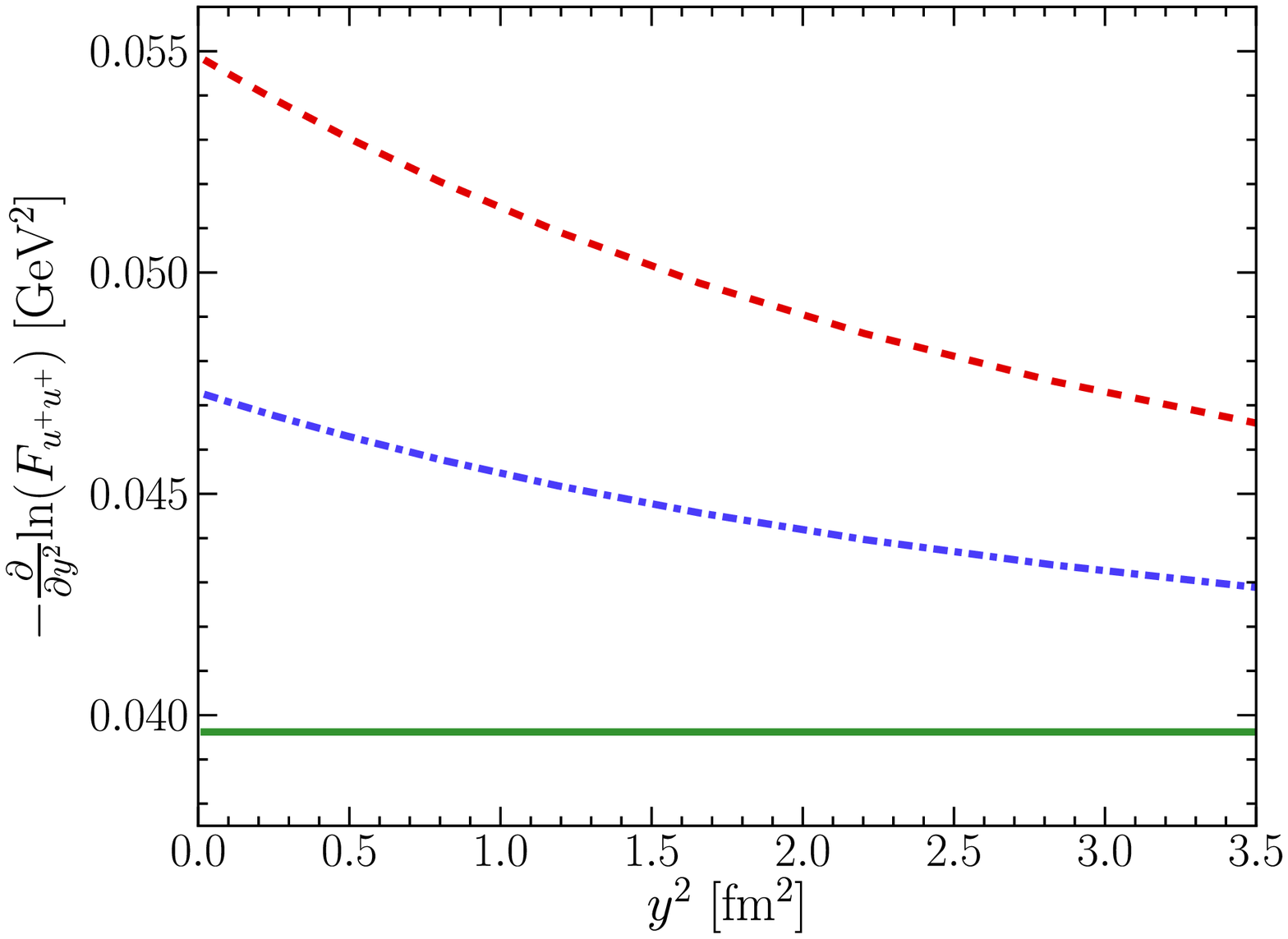}}

  \vspace{0.5em}

  \subfloat[]{\includegraphics[width=0.470\textwidth]{%
      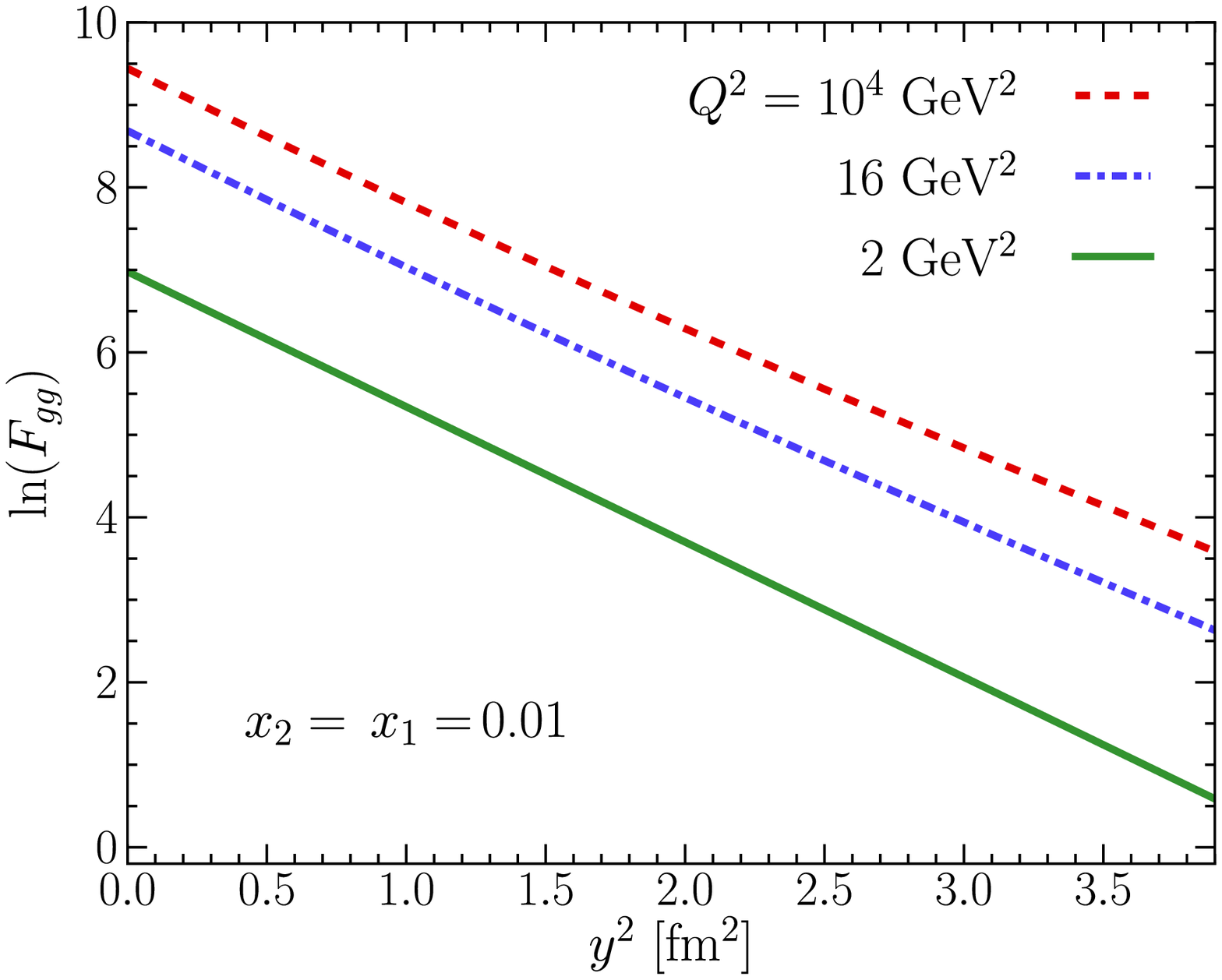}}
  \subfloat[]{\includegraphics[width=0.509\textwidth]{%
      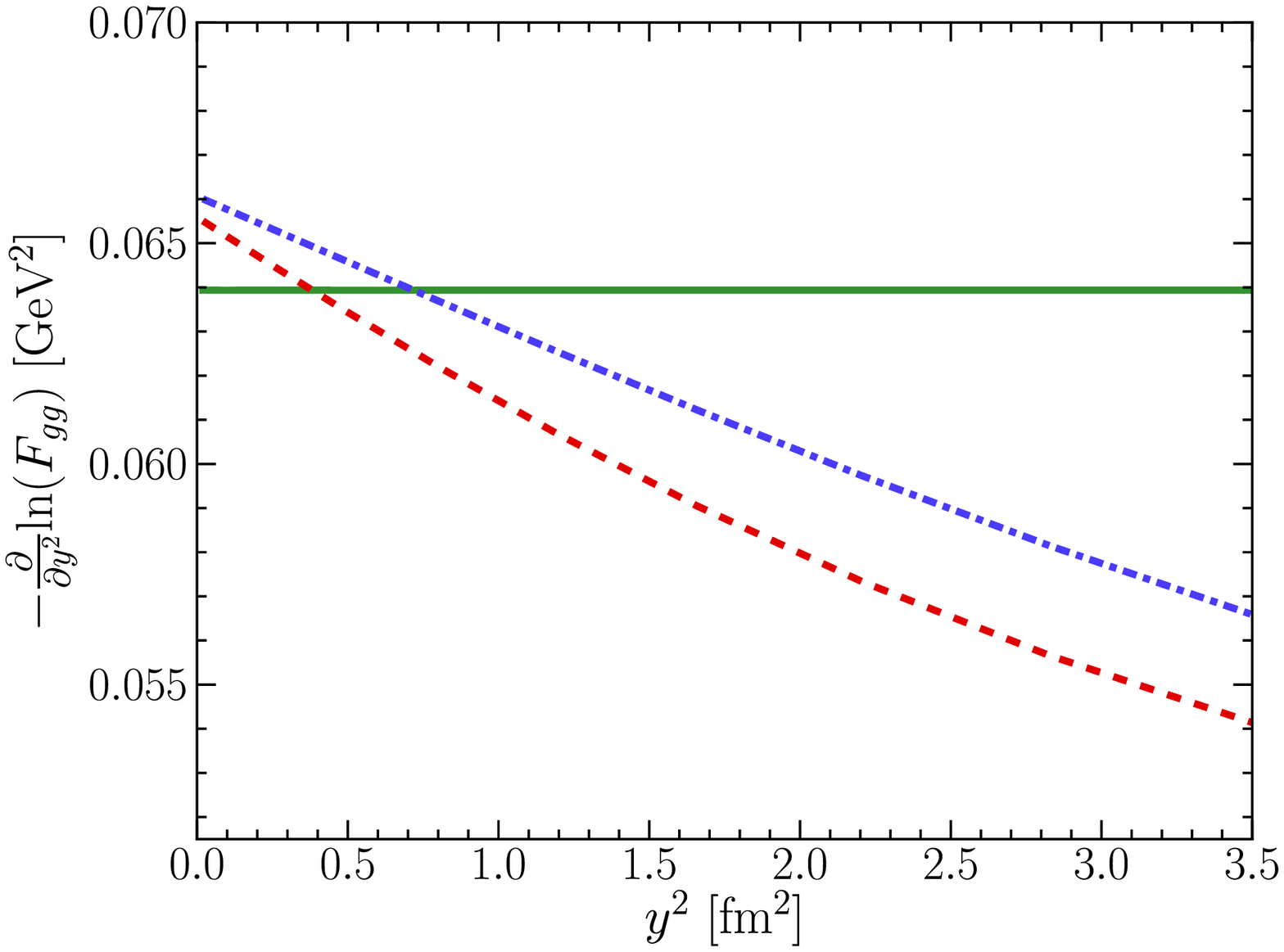}}
  \caption{\label{fig:ydep} $y^2$ dependence of the DPD for two $u^-$
    (top), two $u^+$ (center) and two gluons (bottom).  The left panels
    show the natural logarithm of the DPD and the right panels the
    corresponding slope in $y^2$.  Longitudinal momentum fractions are
    fixed at $x_1 = x_2 = 0.01$.}
\end{figure}

\afterpage{\clearpage} 

Let us first see to which extent the Gaussian $y$ dependence of our
starting condition \eqref{eq:g-ansatz} is changed by evolution.  To this
end, figure \ref{fig:ydep}(a) shows $\ln F_{u^-u^-}$ as a function of
$y^2$ for different values of the scale.  A Gaussian $y$ dependence of the
DPD translates into a straight line in this plot.  We see that the shape
remains approximately Gaussian even up to the high scale of $Q^2 = 10^4
\gev^2$, and that the slope in $y^2$ becomes steeper with $Q^2$,
corresponding to a narrowing of the Gaussian profile.  This is seen more
quantitatively in figure~\ref{fig:ydep}(b), where we show the slope of
the curves as a function of $y^2$, multiplied with an overall minus sign.
The departure from a Gaussian $y$ dependence after evolution is reflected
in a slow decrease of the slope with $y$, but overall the effect is rather
mild.

The corresponding plots for the DPDs for two $u^+$ or two gluons are shown
in figure~\ref{fig:ydep}(c) to (f).  For two $u^+$ we observe a similar
trend as for two $u^-$, with a slight departure from a Gaussian behavior
and an overall narrowing of the $y$ profile at higher scales.  For two
gluons, the $y$ dependence also remains approximately Gaussian after
evolution, but the local $y$ slope shown in figure~\ref{fig:ydep}(f) shows
a different behavior than for quarks, with a tiny increase at low $y$ and
a weak decrease at higher $y$.  The overall size of the effect is,
however, quite small.

Let us now take a closer look at the evolution of the width of the $y$
dependence.  We quantify this by taking the difference quotient
\begin{align}
  \label{eq:diff-quot}
\frac{ \ln F_{aa}(x,x,y) - \ln F_{aa}(x,x,0) }{y^2} \,\bigg|_{y = 0.4\fm}
 & = -\frac{1}{4\ms h_{aa}^{\text{eff}}(x,x)}
\end{align}
between $y=0$ and $y= 0.4\fm$, a region where according to
figure~\ref{fig:ydep} the approximation of a linear $y^2$ dependence of
$\ln F_{aa}(x,x,y)$ works very well.  The function
$h_{aa}^{\text{eff}}(x,x)$ thus defined may be regarded as an effective
Gaussian width.

\begin{figure}[tb]
  \centering
  \subfloat[]{\includegraphics[width=0.49\textwidth]{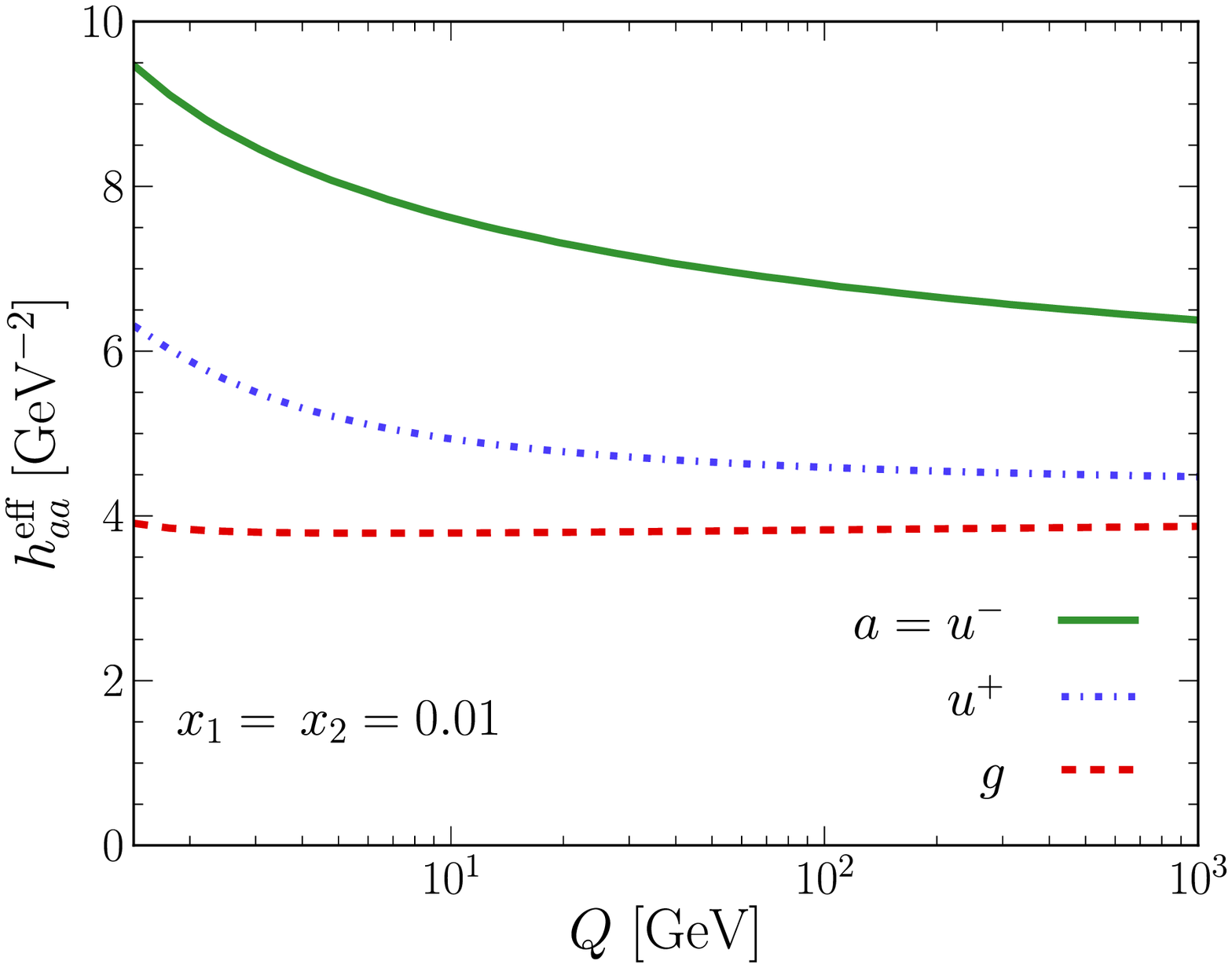}}
  \subfloat[]{\includegraphics[width=0.493\textwidth]{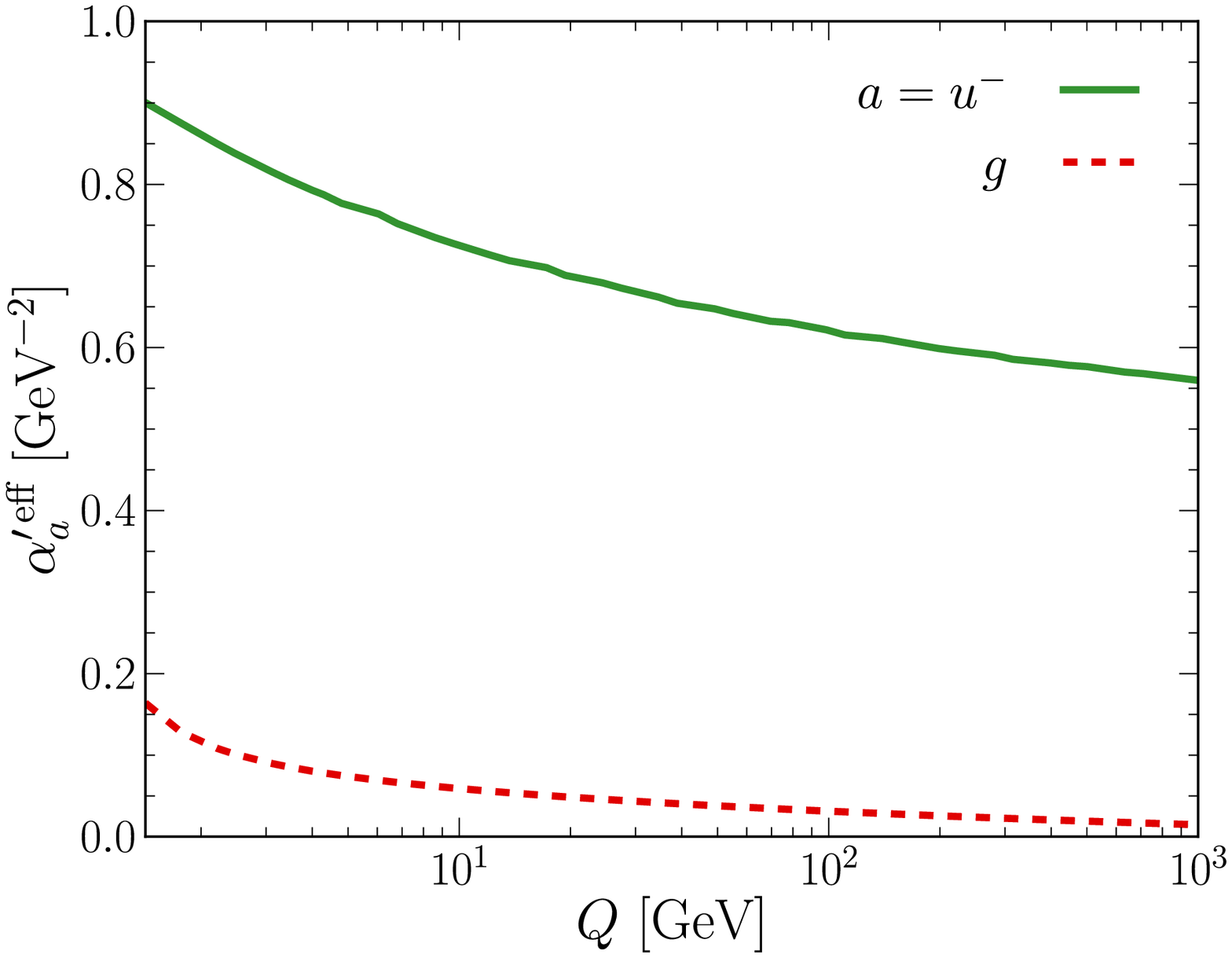}}
  \caption{\label{fig:fit-result-params} (a): Evolution of the effective
    Gaussian width $h_{aa}^{\text{eff}}(x,x)$ defined by
    \protect\eqref{eq:diff-quot} and evaluated at $x = 0.01$ for $a=u^+$,
    $u^-$ and $g$.  (b): Evolution of the effective shrinkage parameter
    $\alpha'^{\,\text{eff}}_a$ obtained by fitting
    $h_{aa}^{\text{eff}}(x,x)$ to \eqref{eq:fitfunc} in the range $0.004
    \le x \le 0.04$ for $a = u^-$ and $g$.}
\end{figure}

Figure~\ref{fig:fit-result-params}(a) shows the evolution of
$h_{aa}^{\text{eff}}(x,x)$ at $x = 0.01$ for $a=u^-$, $u^+$ and $g$.  The
width for the double $u^-$ distribution starts at a larger value than for
the other partons, while the starting value of $h_{gg}^{\text{eff}}$ is
the smallest and $h_{u^+u^+}^{\text{eff}}$ is found almost half way in
between.  As we already saw in figure~\ref{fig:ydep}, the effective
Gaussian width decreases under evolution for both $u^-$ and $u^+$, whereas
it barely changes for the gluon.  Note that $h_{u^-u^-}^{\text{eff}}$
strongly decreases with $x$ for our choice of parameters in
\eqref{eq:trans-param}.  As the valence combination $u^-$ evolves to
higher scales, partons migrate from higher to lower $x$ values by
radiating gluons.  For partons at given $x$ and $Q$, the width of the $y$
distribution is thus influenced by the smaller values of this width for
partons with higher $x$ at lower $Q$.  This explains the decrease of
$h_{u^-u^-}^{\text{eff}}$ with $Q$ in
figure~\ref{fig:fit-result-params}(a), which can also be derived
analytically by adapting the argument for the transverse distribution of a
single quark given in section~2 of \cite{Diehl:2004cx}.
Turning to the double $u^+$ distribution, which mixes with gluons, we
observe that $h_{u^+u^+}^{\text{eff}}$ approaches $h_{gg}^{\text{eff}}$
with increasing scale, although it does so rather slowly.  The difference
between the transverse distribution of gluons and quarks, which we have
assumed at $Q_0$, thus persists over a wide range of scales.

\begin{figure}[tb]
  \centering
  \subfloat[]{\includegraphics[width=0.49\textwidth]{%
      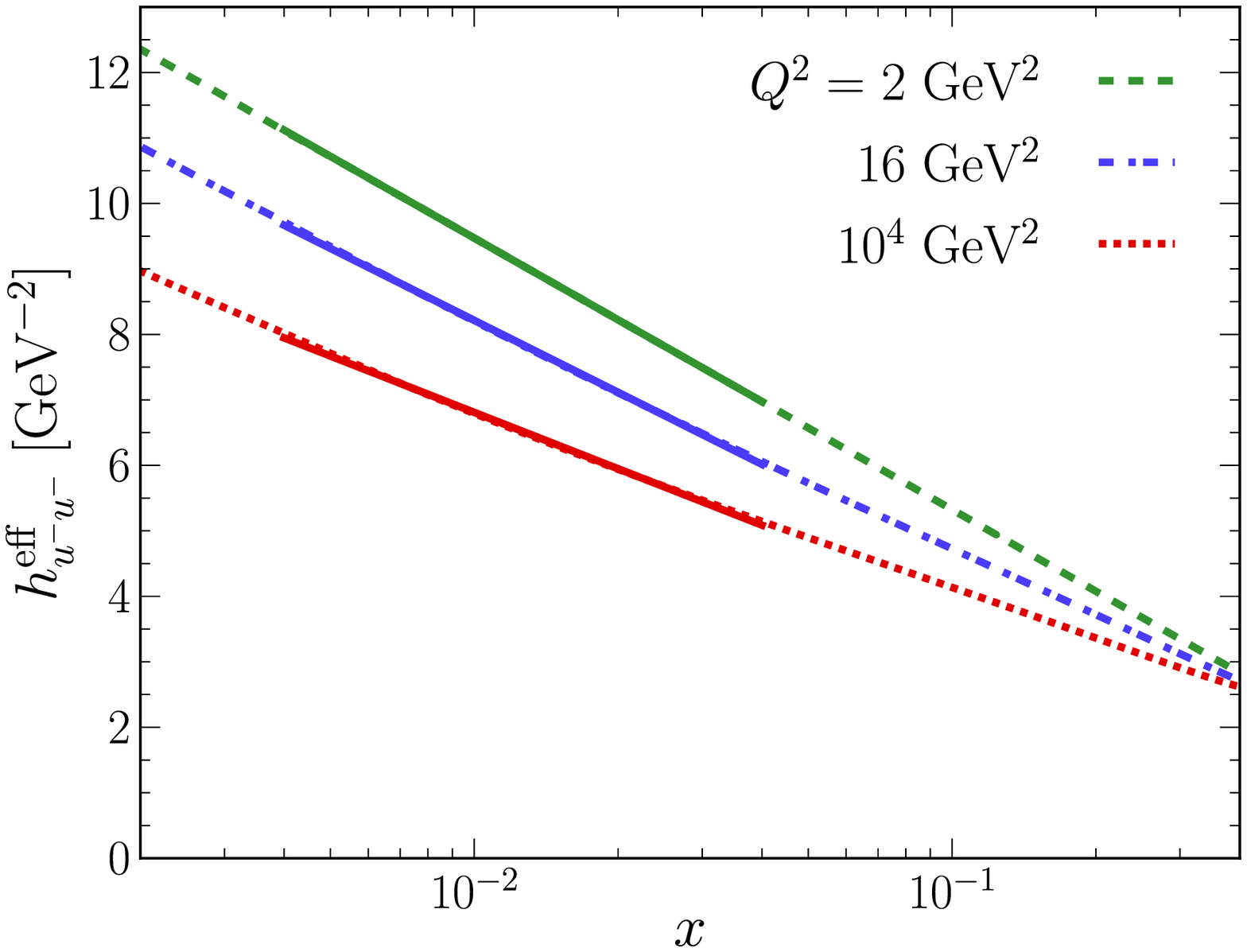}}
  \subfloat[]{\includegraphics[width=0.49\textwidth]{%
      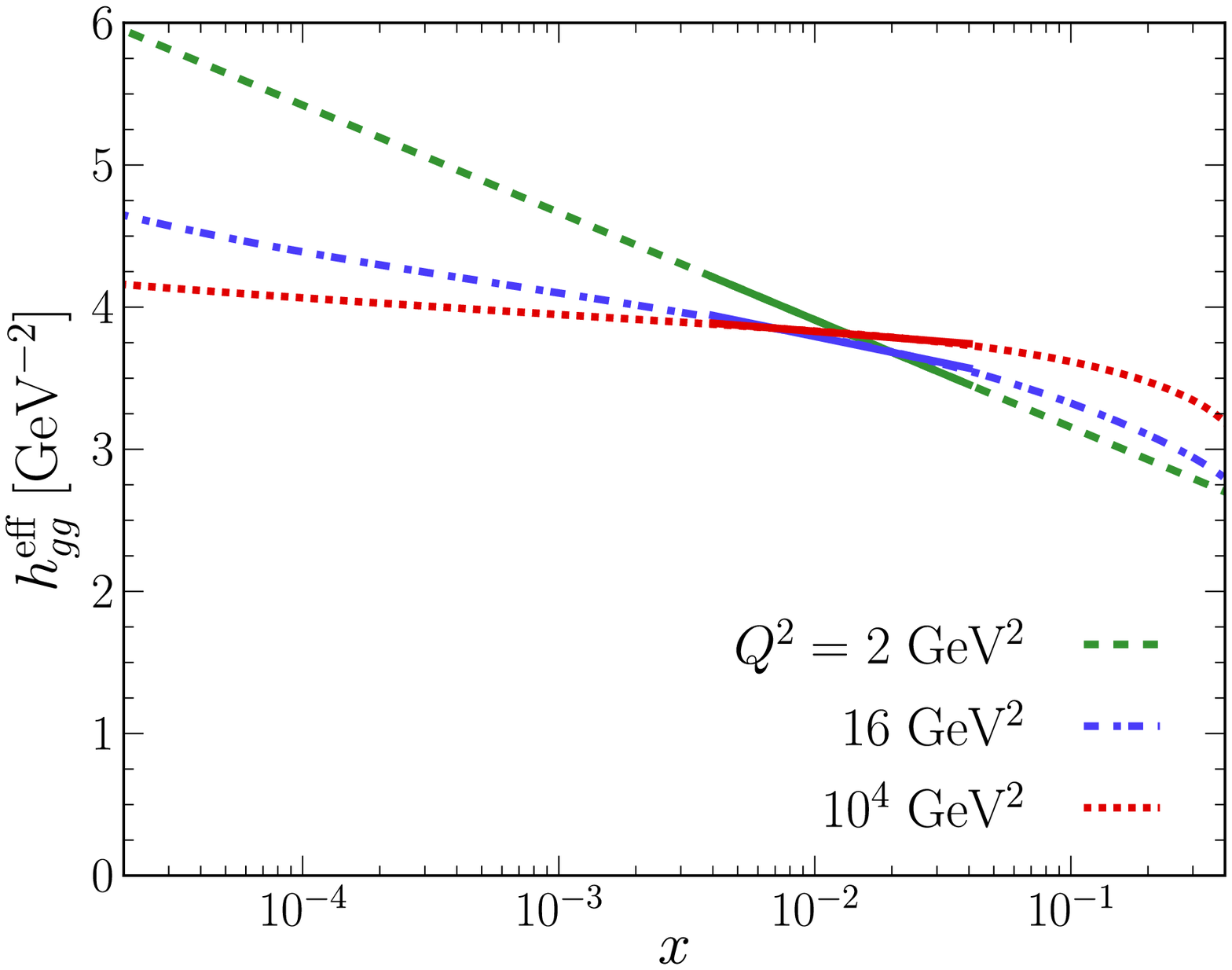}}

  \subfloat[]{\includegraphics[width=0.49\textwidth]{%
      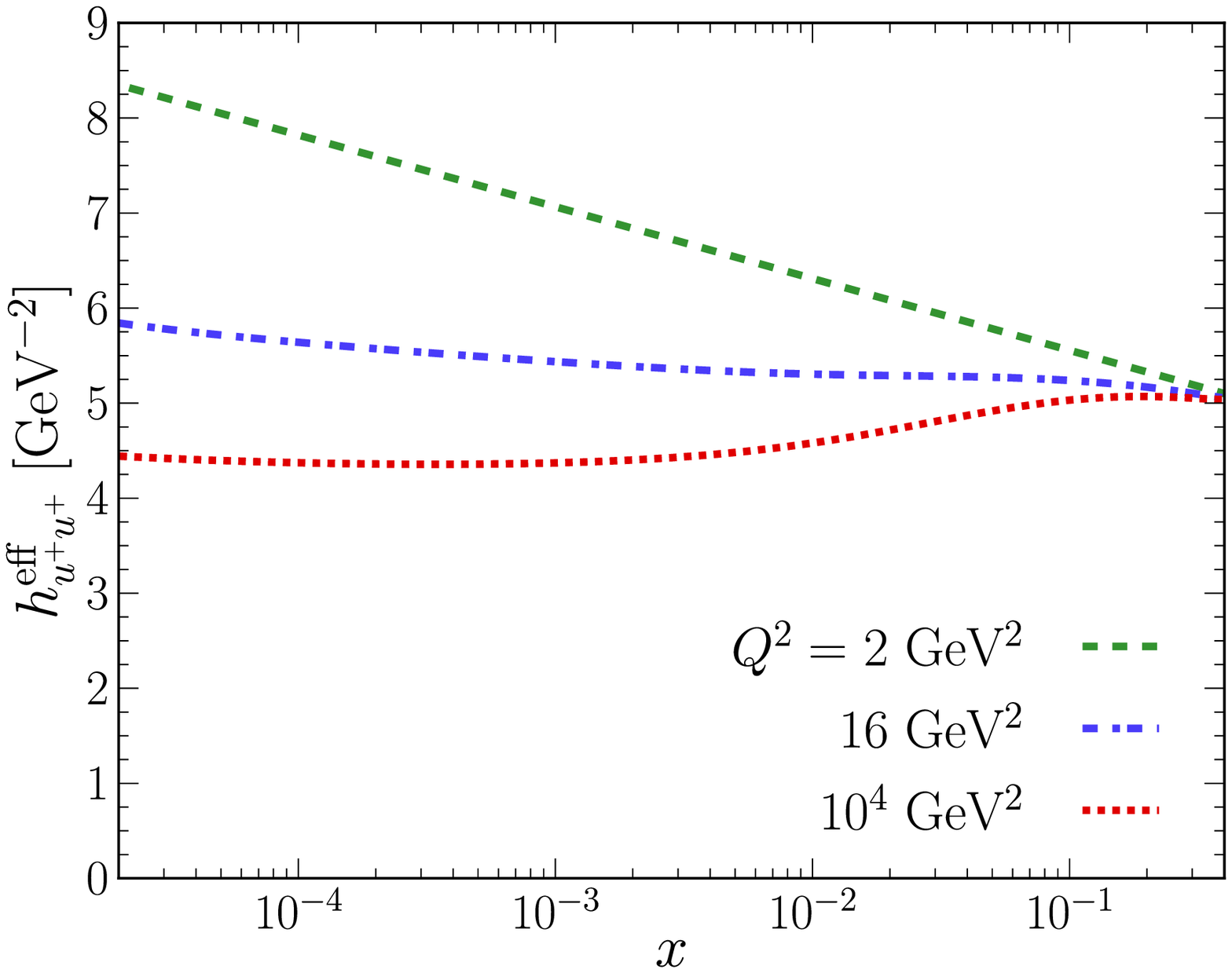}}
  \subfloat[]{\includegraphics[width=0.49\textwidth]{%
      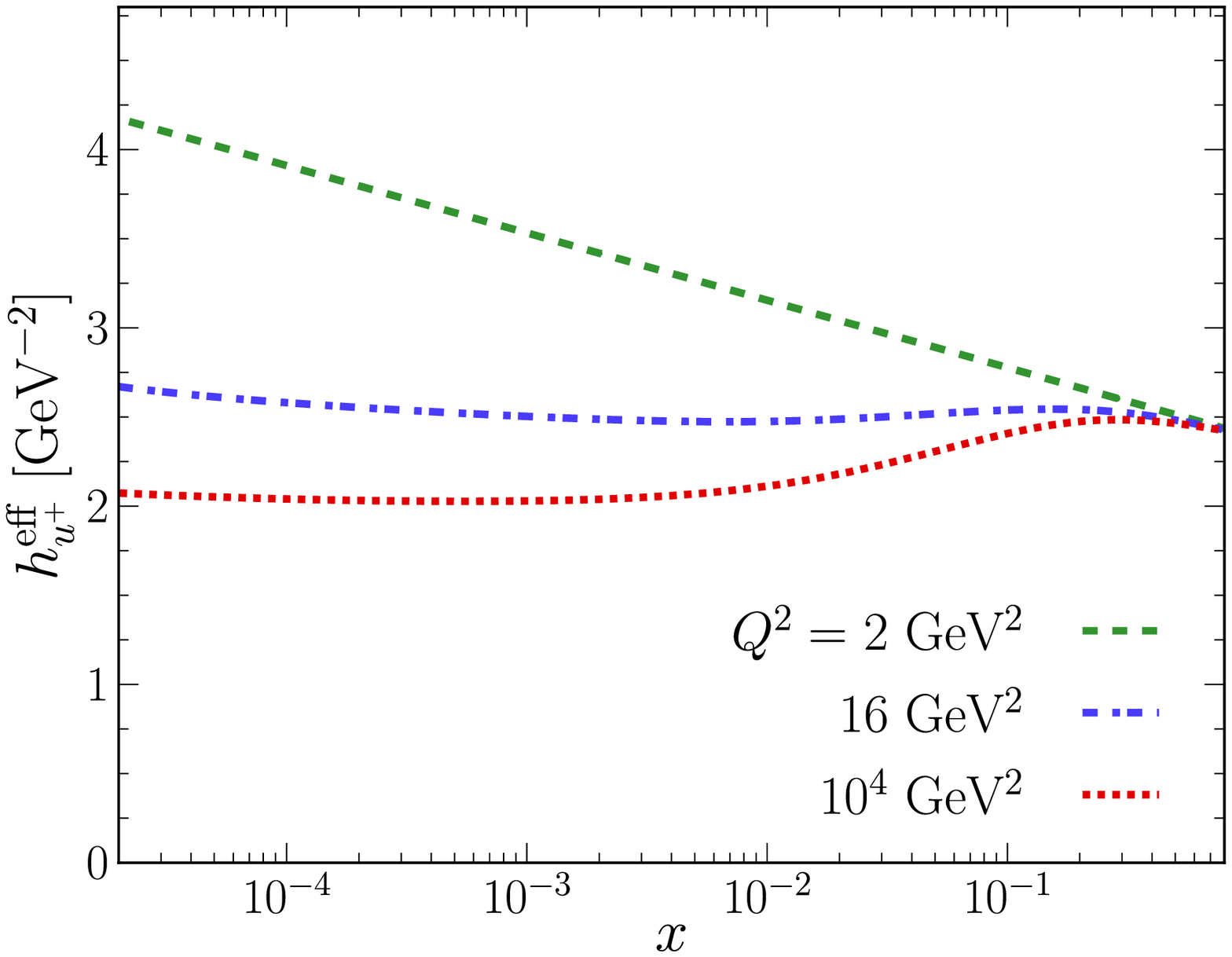}}
  \caption{\label{fig:fit-test} (a), (b), (c): Dependence of
    $h_{aa}^{\text{eff}}(x,x)$ on $x$.  The solid sections of the curves
    in panels (a) and (b) represent fits to \protect\eqref{eq:fitfunc} in
    the range $0.004 \le x \le 0.04$.  (d): Evolution of the effective
    Gaussian width $h_{u^+}^{\text{eff}}(x)$ of the impact-parameter
    dependent single parton distribution $f_{u^+}(x,\vek{b})$, determined
    in analogy to $h_{aa}^{\text{eff}}(x,x)$ as explained in the text.}
\end{figure}

The dependence of $h_{aa}^{\text{eff}}(x,x)$ on $x$ is shown in
figure~\ref{fig:fit-test}(a), (b) and (c) for the different parton types.
We see that evolution is faster at small momentum fractions $x$ than at
large ones.  At low $x$, there is a rapid decrease of
$h_{aa}^{\text{eff}}(x,x)$ with $Q^2$ for all parton types.  For $u^+$
this results in a region of intermediate $x$ where
$h_{u^+u^+}^{\text{eff}}(x,x)$ increases with $x$ at high $Q^2$.  For
$u^-$ and $g$ the curves for $h_{aa}^{\text{eff}}(x,x)$ are approximately
linear in $\ln(x)$ as long as we stay away from the large-$x$ region.
This allows us to extract an effective shrinkage parameter
$\alpha'^{\,\text{eff}}_a$ by fitting the effective Gaussian width to
\begin{align}
  \label{eq:fitfunc}
h_{aa}^{\text{eff}}(x,x) &=
  2\alpha'^{\,\text{eff}}_a \ln\frac{1}{x} + 2 B_a^{\text{eff}}
\end{align}
in an appropriate region of $x$, which we choose as $0.004 \leq x \leq
0.04$.  At the starting scale of evolution, we recover of course the value
of $\alpha'_a$ in our original ansatz \eqref{eq:dpd-trans} for
$h_{aa}(x_1,x_2)$ at $x_1=x_2$.  The scale dependence of
$\alpha'^{\,\text{eff}}_a$ is shown in
figure~\ref{fig:fit-result-params}(b).  We find that
$\alpha'^{\,\text{eff}}_a$ decreases quite rapidly for $a=g$ and more
gently for $a=u^-$.

Turning our attention to the double $u^+$ distribution, we see in
figure~\ref{fig:fit-test} that at large $Q^2$ a local fit of
$h_{u^+u^+}^{\text{eff}}(x,x)$ to the form \eqref{eq:fitfunc} would result
in a negative $\alpha'^{\,\text{eff}}_{u^-}$ whose value would strongly
depend on the chosen range of $x$.  To see whether this is a particular
feature of DPDs, we have investigated the evolution of the
impact-parameter dependent single parton distributions $f_{a}(x,\vek{b})$,
which proceeds according to the usual DGLAP equations at each value of
$\vek{b}$.  Using the program QCDNUM \cite{Botje:2010ay} to perform the
evolution, we find that the initial Gaussian $b$ dependence in our ansatz
\eqref{eq:tran-gau} approximately persists at higher scales, so that we
can extract an effective Gaussian width $h_a^{\text{eff}}(x)$ from the
difference quotient of $\ln f_{a}(x,\vek{b})$ between $b=0$ and $b=0.4\fm$
in full analogy to \eqref{eq:diff-quot}.  The scale dependence of
$h_{u^+}^{\text{eff}}(x)$ determined in this way is shown in
figure~\ref{fig:fit-test}(d) and shows the same qualitative behavior as
$h_{u^+u^+}^{\text{eff}}(x,x)$.  We conclude that an increase with $x$ of
the effective Gaussian width is not special to the evolution of DPDs.

A natural explanation of this increase is that, as $x$ decreases, the
evolution of quark distributions is more and more driven by their mixing
with gluons.  We recall that with the parameters \eqref{eq:trans-param}
for the initial conditions, gluons have a more narrow spatial distribution
than the sum $q^+$ of quarks and antiquarks.  Under evolution, the
effective Gaussian width for $q^+$ tends towards the one for gluons, and
this tendency is stronger at smaller $x$, where the gluon distribution is
larger.  To corroborate this explanation, we have repeated our study of
the DPDs for two alternative choices for the parameters in
\eqref{eq:trans-param}, taking either equal values $B_g = B_{q^+} = 1.2
\gev^{-2}$ or $B_g = 2.4 \gev^{-2}$ and $B_{q^+} = 1.2 \gev^{-2}$.  In
line with our expectation, no increase of $h_{u^+u^+}^{\text{eff}}(x,x)$
with $x$ is seen after evolution in these cases.

Our studies described so far have been done with the MSTW 2008 parton
distribution in the initial conditions, and it is natural to ask how much
our findings depend on this choice.  In \app{ap:pdfs} we show a selection
of recent LO PDF sets at the scale $Q_0^2 = 1 \gev^2$ and find that among
the sets suitable for our purposes (namely those that are positive at that
scale) the parameterizations of MSTW 2008 and GJR 08 \cite{Gluck:2007ck}
represent two extreme choices, with a very slow or a very fast increase of
the gluon at small $x$, respectively.  We have therefore repeated the
studies reported in this section by replacing MSTW 2008 with GJR 08 in our
ansatz \eqref{eq:tran-gau} at the scale $Q_0^2 = 2 \gev^2$.  We obtain
similar results, regarding both the qualitative effects of evolution
(including the behavior in figure~\ref{fig:fit-test}(c) and (d)) and the
rate of change of the parameters describing the $y$ dependence of the
DPDs.

%% file: EvoPol.tex
\section{Evolution of polarized double parton distributions}
\label{sec:pol}

We now investigate the evolution of spin correlations between
two partons inside a proton.  For simplicity we assume in this section a
multiplicative $y$ dependence of the DPDs,
\begin{align}
  \label{eq:y-factorization}
f_{p_1 p_2}(x_1,x_2,\y; Q) &=
   \tilde{f}_{p_1 p_2}(x_1,x_2; Q) \,G(\y) \,,
\end{align}
which is stable under scale evolution.  Since our focus is on the degree
of parton polarization rather than on the absolute size of the DPDs, we
set the $y$ dependent factor $G(\y)=1$ in all plots and omit the tilde in
$\tilde{f}_{p_1 p_2}(x_1,x_2; Q)$ from now on.

For the unpolarized DPDs we content ourselves with a simple factorizing
ansatz at the starting scale,
\begin{align}
  \label{eq:x-factorization}
{f}_{ab}(x_1,x_2; Q_0) &= f_a(x_1; Q_0)\, f_b(x_2; Q_0) \,,
\end{align}
which we take as $Q_0^2 = 1 \gev^2$ unless specified otherwise.  For the
single parton densities in \eqref{eq:x-factorization} we consider the two
LO sets MSTW 2008 and GJR 08, hereafter referred to as MSTW and GJR for
brevity.  This ansatz is clearly unsatisfactory close to the kinematic
limit $x_1+x_2 = 1$, where the DPDs are expected to vanish, but since we
are not particularly interested in that region we have refrained from
taking a more sophisticated form.  As discussed in \sect{sec:int-lim},
evolution to higher scales approximately conserves the factorized form
\eqref{eq:x-factorization} for sufficiently low $x_1$ and $x_2$.

\begin{figure}[tb]
  \centering
  \subfloat[MSTW]{\includegraphics[width=0.49\textwidth]{%
      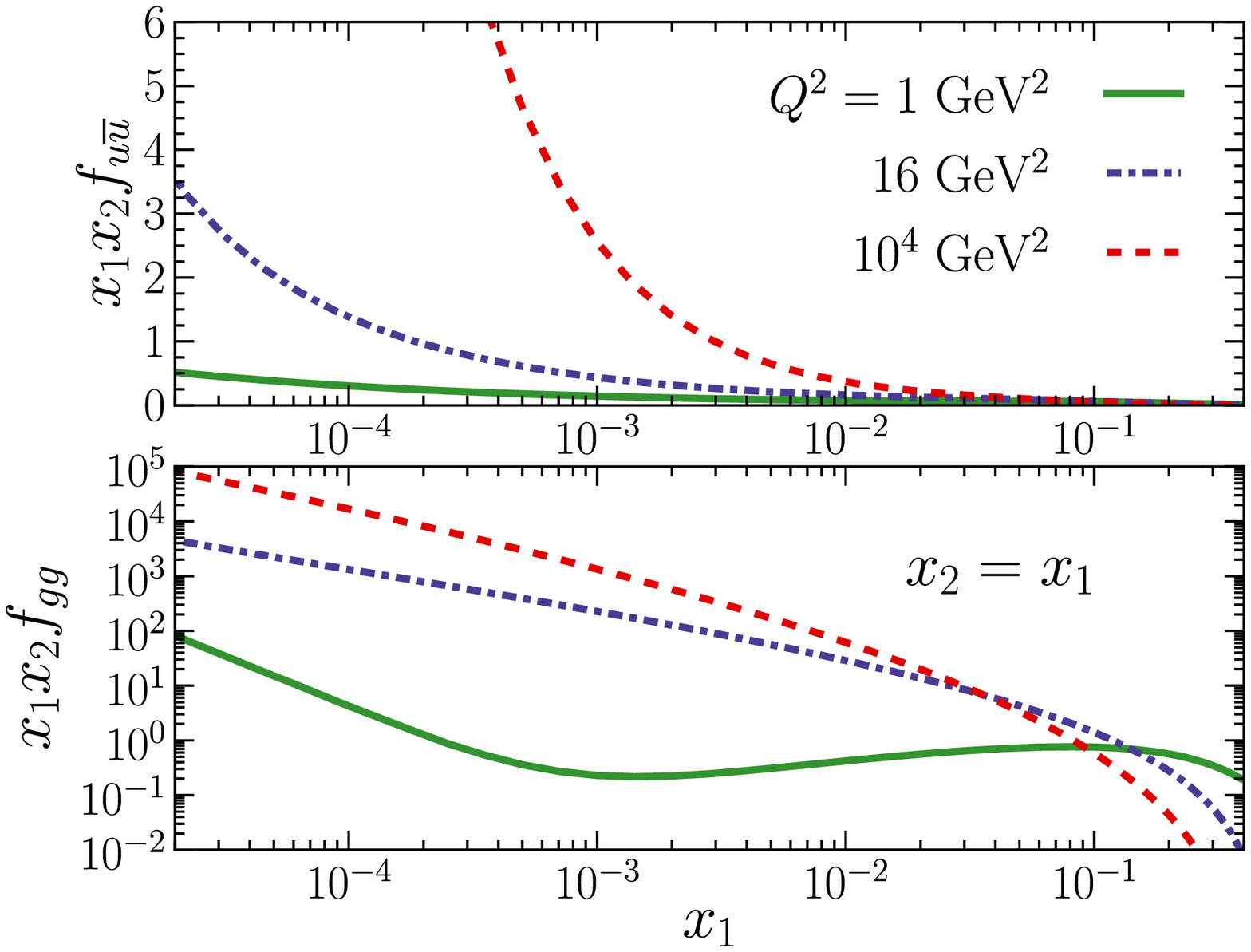}}
  \subfloat[GJR]{\includegraphics[width=0.49\textwidth]{%
      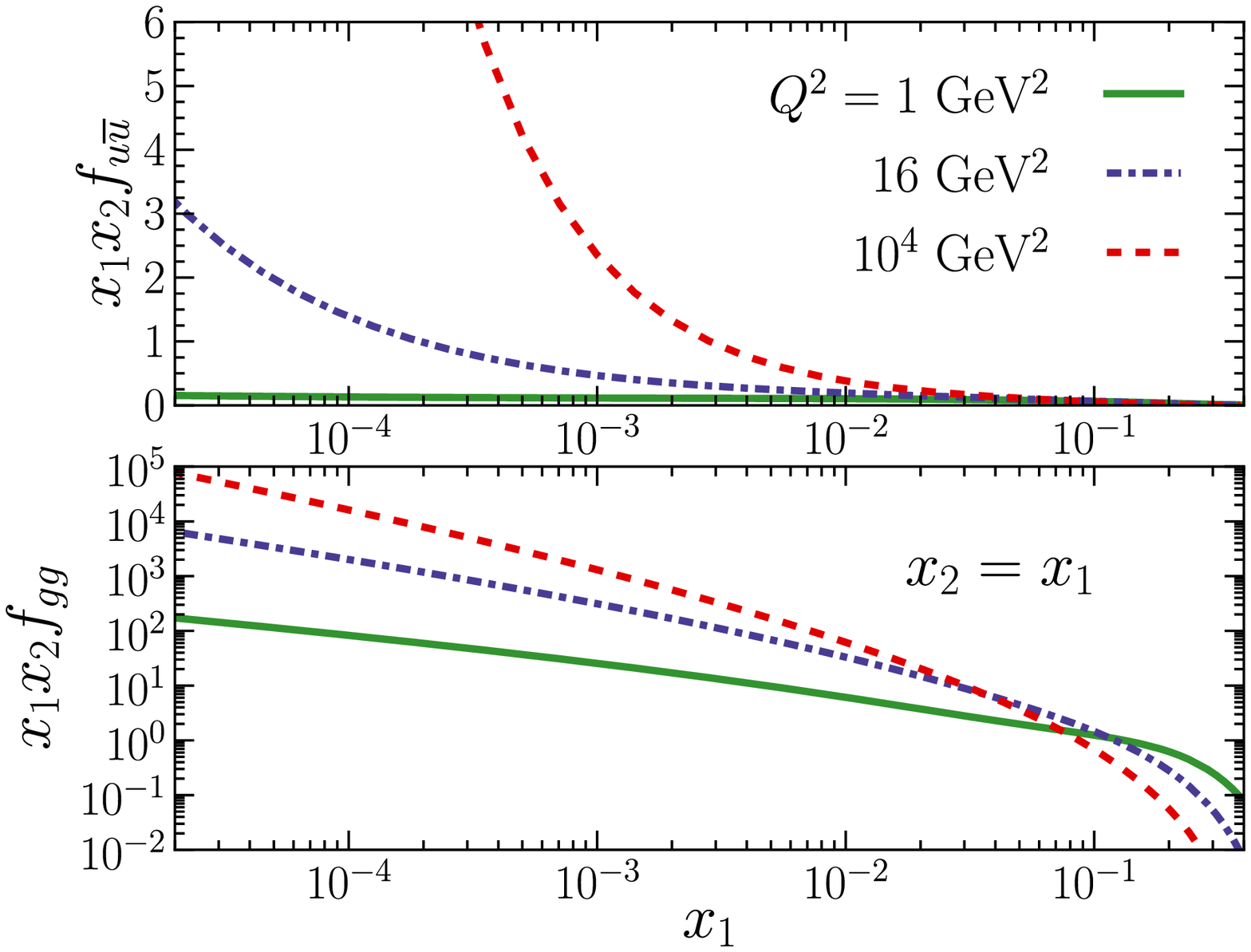}}
  \caption{\label{fig:unpol-v1} Unpolarized double parton distributions,
    constructed from the ansatz \protect\eqref{eq:y-factorization} and
    \protect\eqref{eq:x-factorization} with the LO PDFs of MSTW (a) or GJR
    (b).  The factor $G(\y)$ in \protect\eqref{eq:y-factorization} has
    been set to $1$ for simplicity.  The vertical scales have been chosen
    to facilitate comparison with polarized distributions in subsequent
    plots.}

  \vspace{0.5em}

  \subfloat[MSTW]{\includegraphics[width=0.49\textwidth]{%
      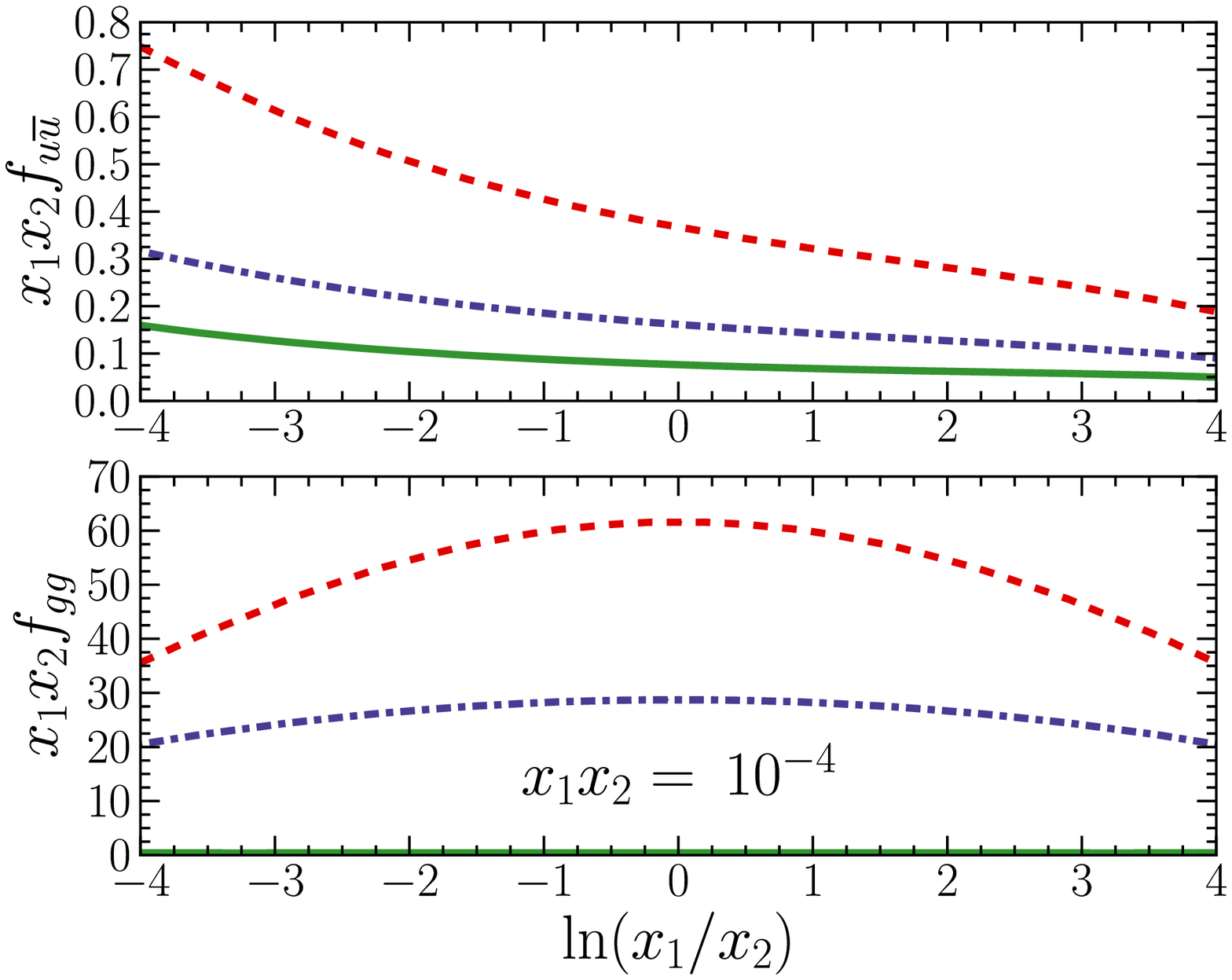}}
  \subfloat[GJR]{\includegraphics[width=0.49\textwidth]{%
      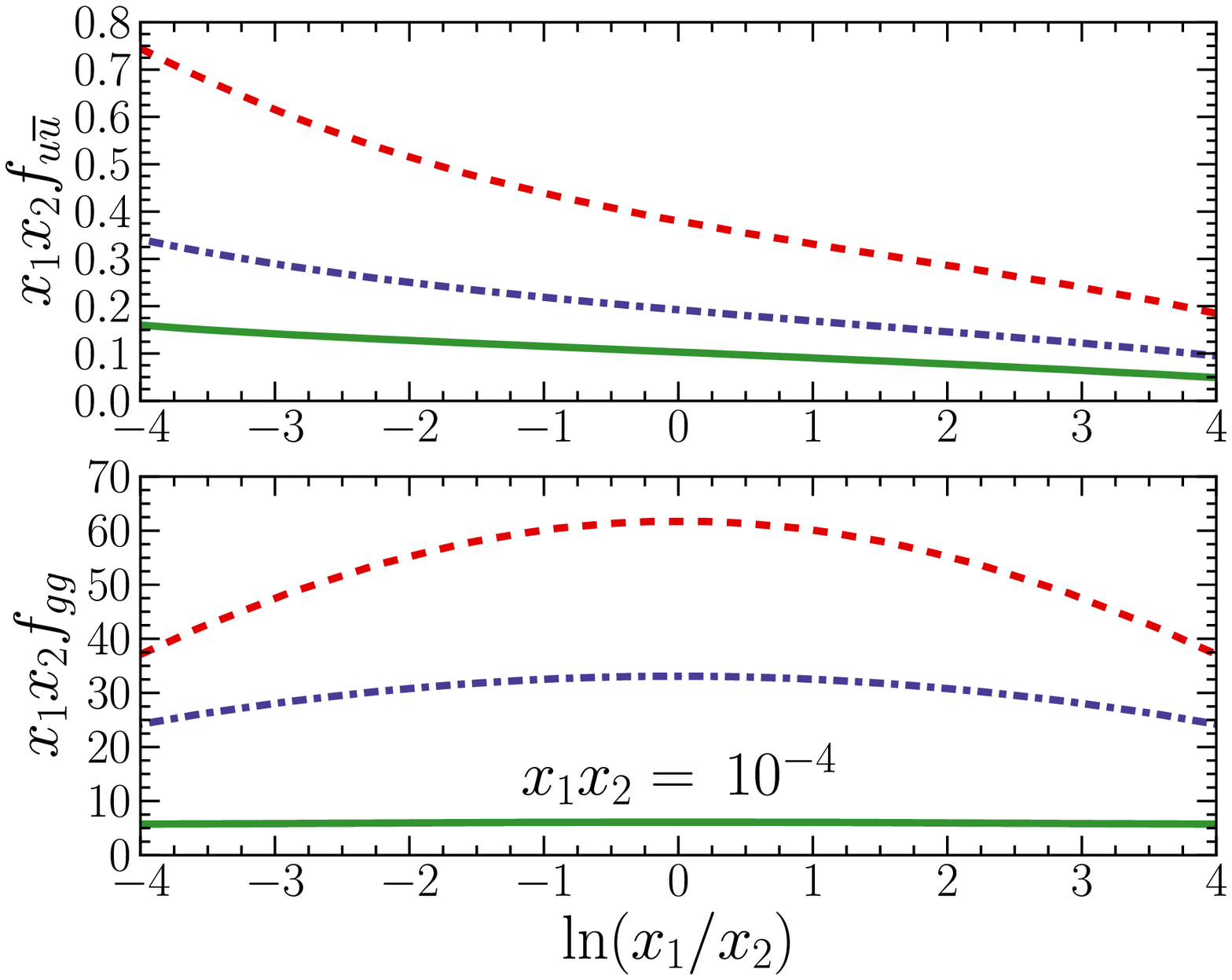}}
  \caption{\label{fig:unpol-v2} The same distributions as in
    figure~\protect\ref{fig:unpol-v1}, plotted against $\ln(x_1/x_2)$ at
    fixed $x_1\ms x_2$.}
\end{figure}

In figures~\ref{fig:unpol-v1} and \ref{fig:unpol-v2} we show our
unpolarized model DPDs for a $u\bar{u}$ pair and for two gluons, either as
functions of $x_1$ at $x_2=x_1$ or as functions of $\ln(x_1/x_2)$ at fixed
$x_1\ms x_2 = 10^4$.  Notice that $\ln(x_1/x_2)$ is related to the
rapidity difference between the systems of particles produced in the two
hard-scattering subprocesses.  Denoting the total four-momenta of the
final states in the two subprocesses by $q_1$ and $q_2$, we have
\begin{align}
\Delta Y = Y_1 - Y_2 &= \frac{1}{2}\,
   \biggl[\ms \ln\frac{q_1^+}{q_1^-} - \ln\frac{q_2^+}{q_2^-} \ms\biggr]
 = \ln\frac{x_1}{x_2} + \frac{1}{2} \ln\frac{q_2^2}{q_1^2} \,.
\end{align}
For equal c.m.\ energies of the two subprocesses we simply have $\Delta Y
= \ln(x_1/x_2)$.  In that case, rapidity differences in the range $-4 \le
\Delta Y \le 4$ correspond to longitudinal momentum fractions $0.0014 \leq
x_i \leq 0.074$ for $x_1\ms x_2=10^{-4}$.  In figure~\ref{fig:unpol-v1} we
see that at low scales the double gluon distributions constructed from
MSTW and GJR PDFs strongly differ in size and shape.  Only for larger
scales do the two sets approach each other, which is plausible since at
high scales the single parton densities are more directly constrained by
data than at low scales.

To model the polarized DPDs is much more difficult.  There is no reason to
believe that the single parton distributions for a polarized parton in a
polarized proton should be suitable even as a starting point to describe
the DPDs for two polarized partons in an unpolarized proton.  In other
words, it is far from obvious how to connect the spin correlations between
one parton and the proton with the spin correlations between two partons,
and we will not try to do so.

Instead, we pursue two scenarios.  In the first one, which we call the
``max scenario'', we make use of the positivity bounds for DPDs derived in
\cite{Diehl:2013mla}.  At the starting scale $Q_0$ of evolution, we
maximize each polarized DPD individually with respect to its unpolarized
counterpart.  For the combinations we will investigate, this gives
\begin{align}
  \label{eq:pos-saturated}
  |f_{\Delta a \Delta b}| &\le f_{ab} \,,
&
  |f_{\delta a \delta b}| &\le f_{ab} \,,
\end{align}
and
\begin{align}
  \label{eq:pos-sat-mixed}
  (y M)^2\, |f_{a\ms \delta g}| &\le f_{ag}
\end{align}
for $a,b = q,\bar{q},g$ at equal values of $x_1, x_2$ and $y$ on the left-
and right-hand sides.  As follows from equation~(4.6) in
\cite{Diehl:2013mla}, the bounds in \eqref{eq:pos-saturated} can be
satisfied simultaneously, as well as the bounds in
\eqref{eq:pos-sat-mixed}, but not the two sets together.  The distribution
$f^t_{\delta a \delta b}$ is subject to the same bound as $f_{\delta a
  \delta b}$ in \eqref{eq:pos-saturated}.  Since it also follows the same
evolution equation and does not mix with any other distribution, we will
not discuss it further.

As shown in \cite{Diehl:2013mla}, leading-order evolution to higher scales
preserves the above bounds.  By contrast, if the bounds are saturated at
some scale, they will in general be violated at lower scales.  For this
reason, we take a rather low value of $Q_0$ in this study.  Evolved to
high scales, results in the max scenario show how large polarization
effects can possibly be if one assumes that the density interpretation of
DPDs and thus their positivity holds down to the scale $Q_0$.

The sign of the distributions on the l.h.s.\ of \eqref{eq:pos-saturated}
and \eqref{eq:pos-sat-mixed} can be either positive or negative.  For DPDs
involving only transverse or linear polarization, such as $f_{\delta q
  \delta q}$ or $f_{\delta g \delta g}$, this is of no consequence for the
evolution behavior and we take the positive sign for definiteness.  For
polarized DPDs that mix with others under evolution, relative signs are
important.  In the following, we will always assume the positive sign for
all polarized distributions in the max scenario.  Other choices typically
yield lower polarization after evolution.  Exploring several combinations
for the signs of $f_{\Delta q \Delta\bar{q}}$, $f_{\Delta g \Delta g}$,
$f_{\Delta q \Delta g}$ and $f_{\Delta g \Delta\bar{q}}$, we find that
after evolution $f_{\Delta u\Delta \bar{u}}$ is suppressed by a factor
between $0.5$ and $1$ relative to the values shown in the figures of
\sect{sec:qqbar-pol}.  For distributions with one or two longitudinally
polarized gluons, the corresponding suppression is stronger in parts of
phase space.

Our second scenario, called the ``splitting scenario'', contains more
detailed dynamical input.  At small distances $y$, DPDs can be calculated
perturbatively in terms of single parton distributions as discussed in
\cite{Diehl:2011yj}.  We then have
\begin{align}
  \label{eq:pert-split}
  f_{q\bar{q}}(x_1,x_2,y) &=
  \frac{\alpha_s}{2\pi^2}\, \frac{1}{y^2}\,
    \frac{f_{g}(x_1+x_2)}{x_1+x_2}\,
       T_{g\to q\bar{q}}\Bigl( \frac{x_1}{x_1+x_2} \Bigr)
\end{align}
and analogous relations for the other DPDs, which are collected in
\app{ap:split}.  At large $y$ these relations will no longer hold.  In the
splitting scenario, we assume that at scale $Q_0$ the ratio of polarized
and unpolarized DPDs computed in the perturbative regime is valid up to
large values of $y$, even if the form \eqref{eq:pert-split} is not.  We
thus continue to use the factorized form \eqref{eq:y-factorization} and
\eqref{eq:x-factorization} for the unpolarized DPDs, while the polarized
ones at scale $Q_0$ are given by
\begin{align}
  \label{eq:split-coeff}
  f_{\Delta q \Delta \bar{q}} &= - f_{q\bar{q}} \,,
& 
  f_{\delta q \delta \bar{q}} &= - \frac{2z\bar{z}}{z^2+\bar{z}^2}\;
                                   f_{q\bar{q}} \,,
\nonumber\\[0.2em]
  f_{\Delta g \Delta g}       &=
  \frac{z\bar{z}\ms (2-z\bar{z})}{z^2+\bar{z}^2+z^2\bar{z}^2}\; f_{gg} \,,
&
  f_{\delta g \delta g}       &=  
  \frac{z^2\bar{z}^2}{z^2+\bar{z}^2+z^2\bar{z}^2}\; f_{gg} \,,
\nonumber\\[0.2em]
  f_{\Delta q \Delta g}       &= \frac{1-z^2}{1+z^2}\; f_{qg}
\intertext{and}
  \label{eq:split-coeff-single}
  (y M)^2 f_{q\ms \delta g}     &= \frac{2z}{1+z^2}\; f_{qg} \,,
&
  (y M)^2 f_{g\ms \delta g}     &=
  \frac{z^2}{z^2+\bar{z}^2+z^2\bar{z}^2}\; f_{g g} \,,
\end{align}
where $z=x_1/(x_1+x_2)$ and $\bar{z} = 1-z$.  Further non-zero
distributions are obtained by interchanging parton labels and momentum
fractions or by interchanging quarks with antiquarks; other combinations
such as $f_{qq}$ or $f_{\bar{q}\ms \delta q}$ vanish at $Q_0$ in this
scenario.  The resulting set of distributions saturates several of the
positivity bounds discussed in \cite{Diehl:2013mla}.  Specifically, one
obtains two vanishing eigenvalues in each of the spin-density
matrices~$\rho$ for quark-antiquark, quark-gluon and gluon-gluon DPDs,
given in eq.~(3.2) to (3.6) of \cite{Diehl:2013mla}.

We note that the appearance of the factors $(y M)^2$ in
\eqref{eq:pos-sat-mixed} and \eqref{eq:split-coeff-single} is a
consequence of the factors $\y^k \y^{k'} M^2$ in the DPD definitions
\eqref{eq:def-qg} and \eqref{eq:def-gg}.  In cross sections, the
distributions $f_{q\ms \delta g}$ and $f_{g\ms \delta g}$ always appear
multiplied with $(y M)^2$.  For convenience we will set this factor equal
to $1$ in our plots.


\subsection{Quark and antiquark distributions}
\label{sec:qqbar-pol}

We start our examination of spin correlations with the DPDs for
longitudinally or transversely polarized quarks and antiquarks.

We will show a series of figures which are all using the MSTW parton
distributions in the initial conditions, with curves for the three scales
$Q^2= 1, 16$ and $10^4 \gev^2$.  In each figure, the upper row shows the
polarized DPDs and the lower row shows the ratio between polarized and
unpolarized DPDs for the same parton type.  The latter ratio is restricted
to the range from $-1$ to $1$ according to \eqref{eq:pos-saturated} and
will be called the ``degree of polarization''.  It is the size of this
ratio that indicates how important spin correlations are in the cross
sections of DPS processes.  As we did for unpolarized DPDs in
figures~\ref{fig:unpol-v1} and \ref{fig:unpol-v2}, we show the polarized
distributions both as functions of $x_1$ at $x_1=x_2$ and as functions of
$\ln(x_1/x_2)$ for $x_1x_2=10^{-4}$.  Results will be given both for the
max scenario and for the splitting scenario in the initial conditions.

\begin{figure}[tb]
  \centering
  \subfloat[]{\includegraphics[width=0.49\textwidth]{%
      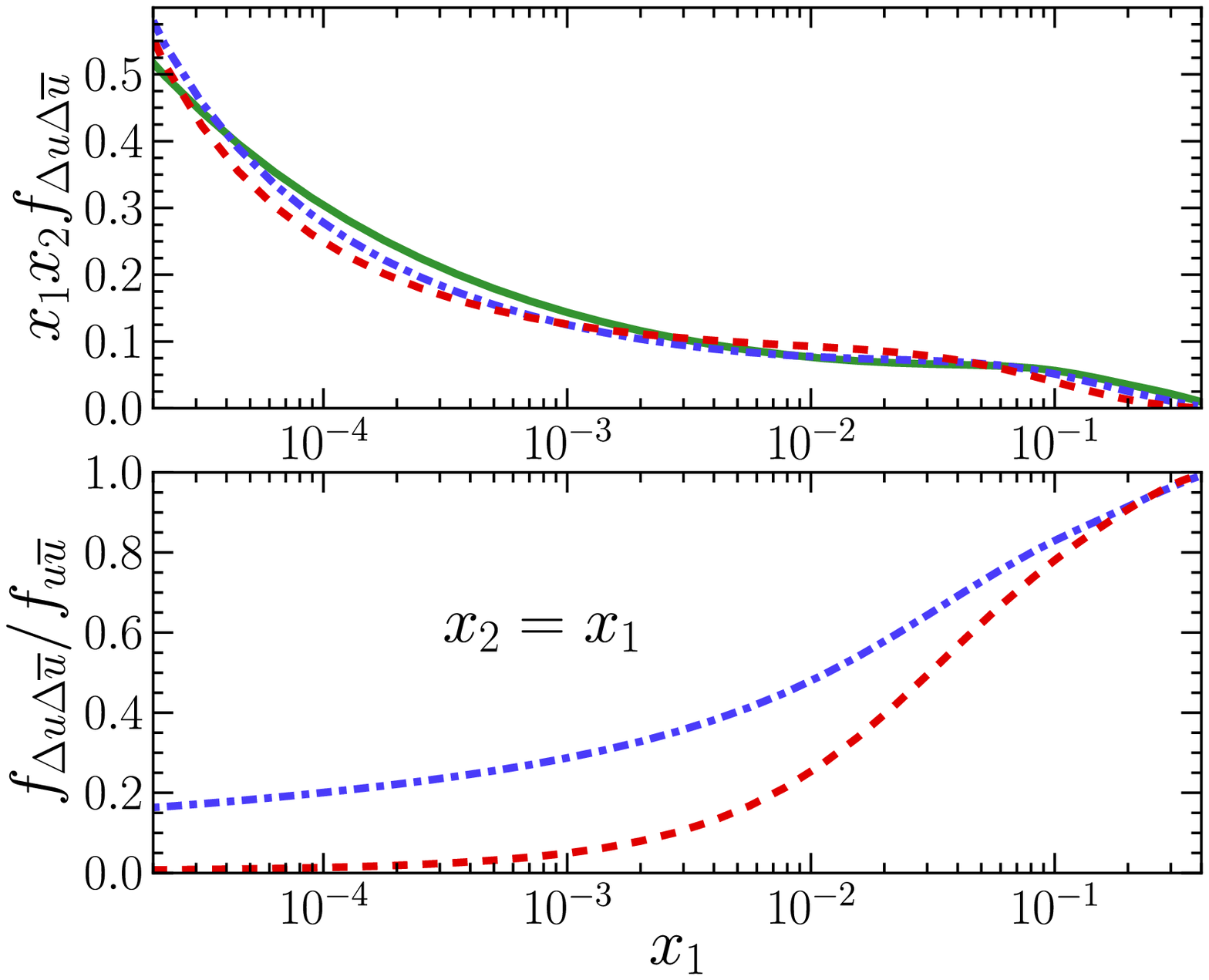}}
  \subfloat[]{\includegraphics[width=0.49\textwidth]{%
      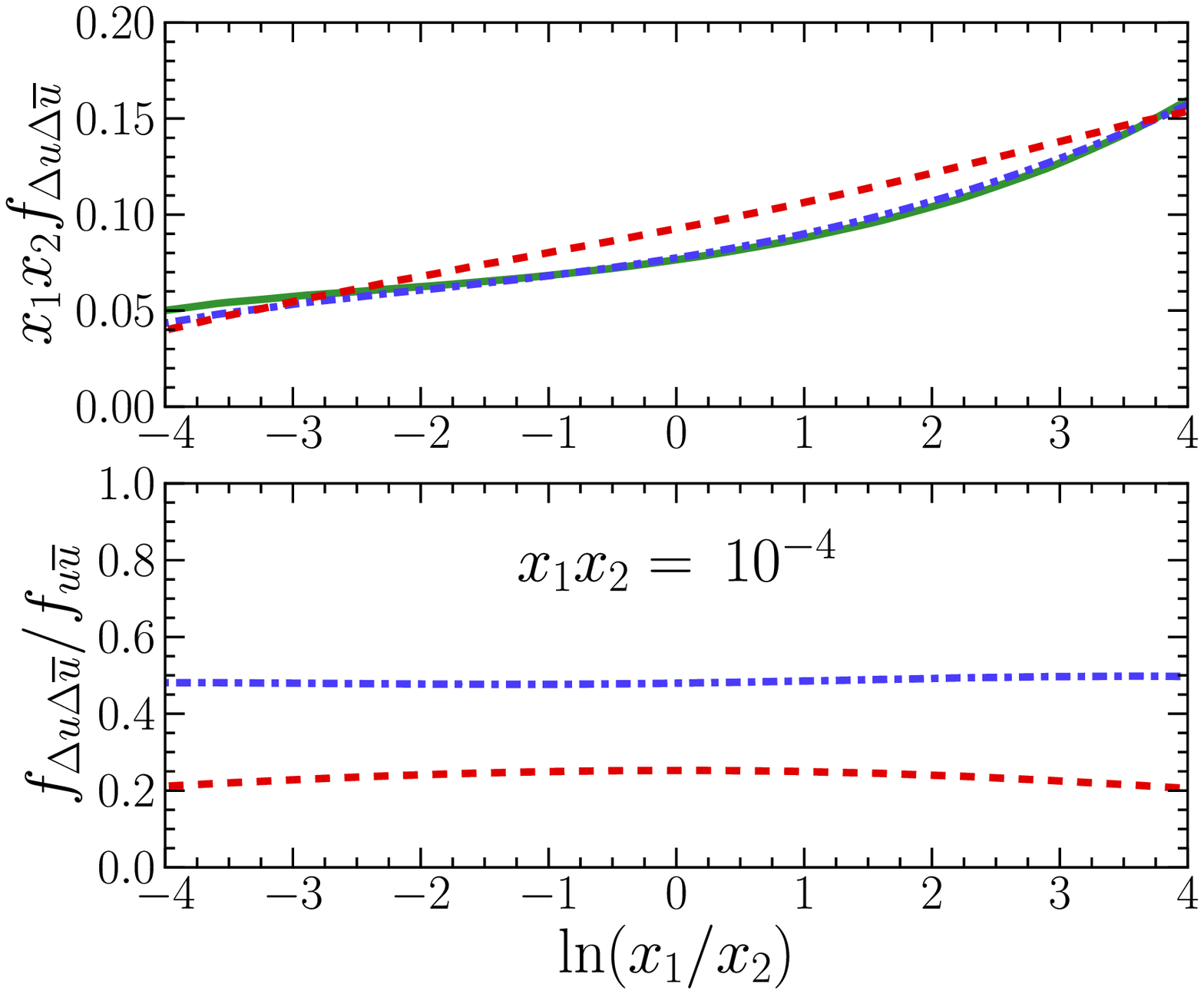}}
  \caption{\label{fig:long-u-max} Longitudinally polarized up quarks and
    antiquarks in the max scenario, with initial conditions using the MSTW
    PDFs. Here and in the following figures the upper row shows the
    polarized DPDs and the lower row the ratio between polarized and
    unpolarized DPDs.  In the max scenario, this ratio is $1$ at the
    starting scale by construction and will not be shown.  Color (line
    style) coding as in figure~\ref{fig:unpol-v1}.}
  \centering
  \subfloat[]{\includegraphics[width=0.49\textwidth]{%
      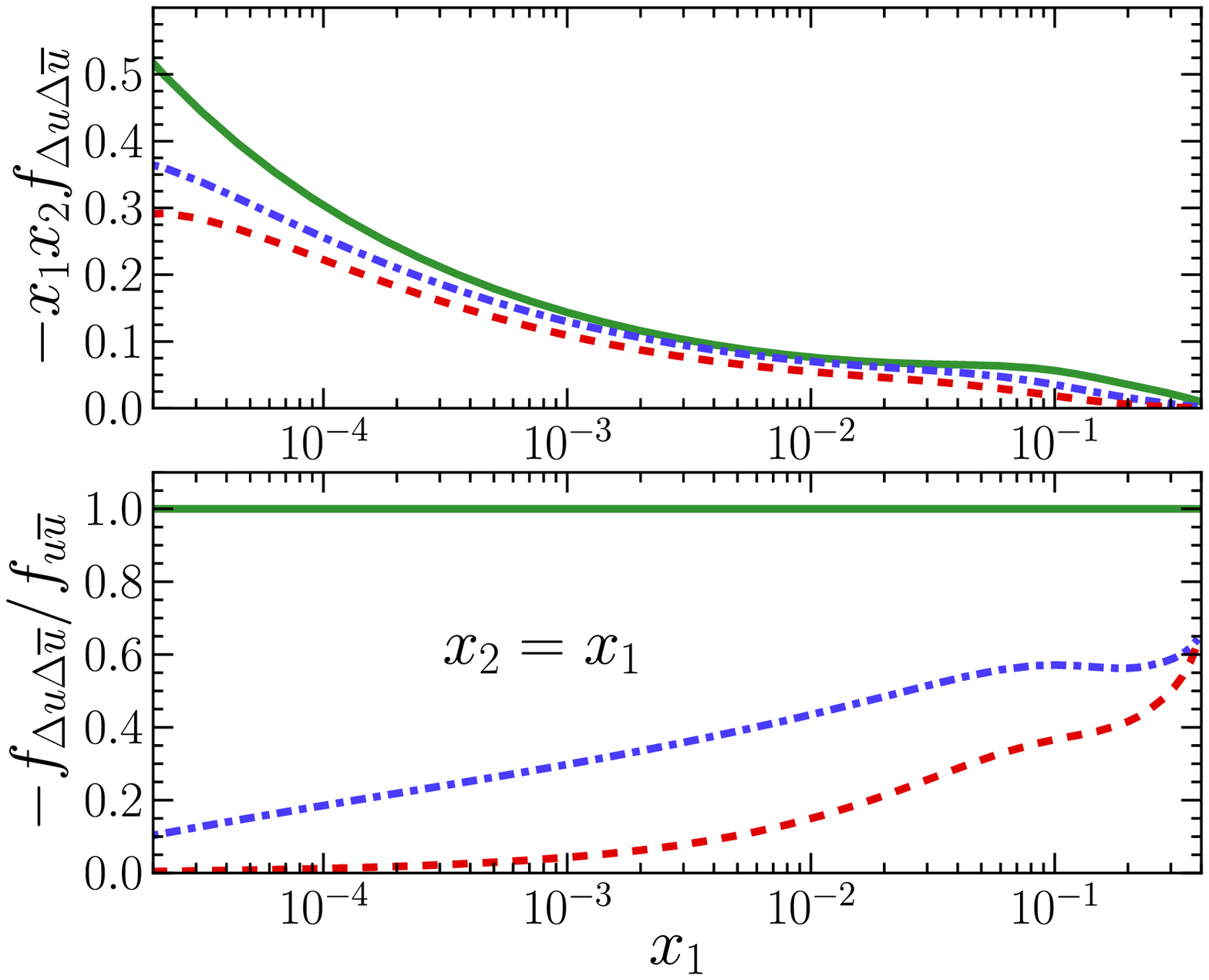}}
  \subfloat[]{\includegraphics[width=0.49\textwidth]{%
      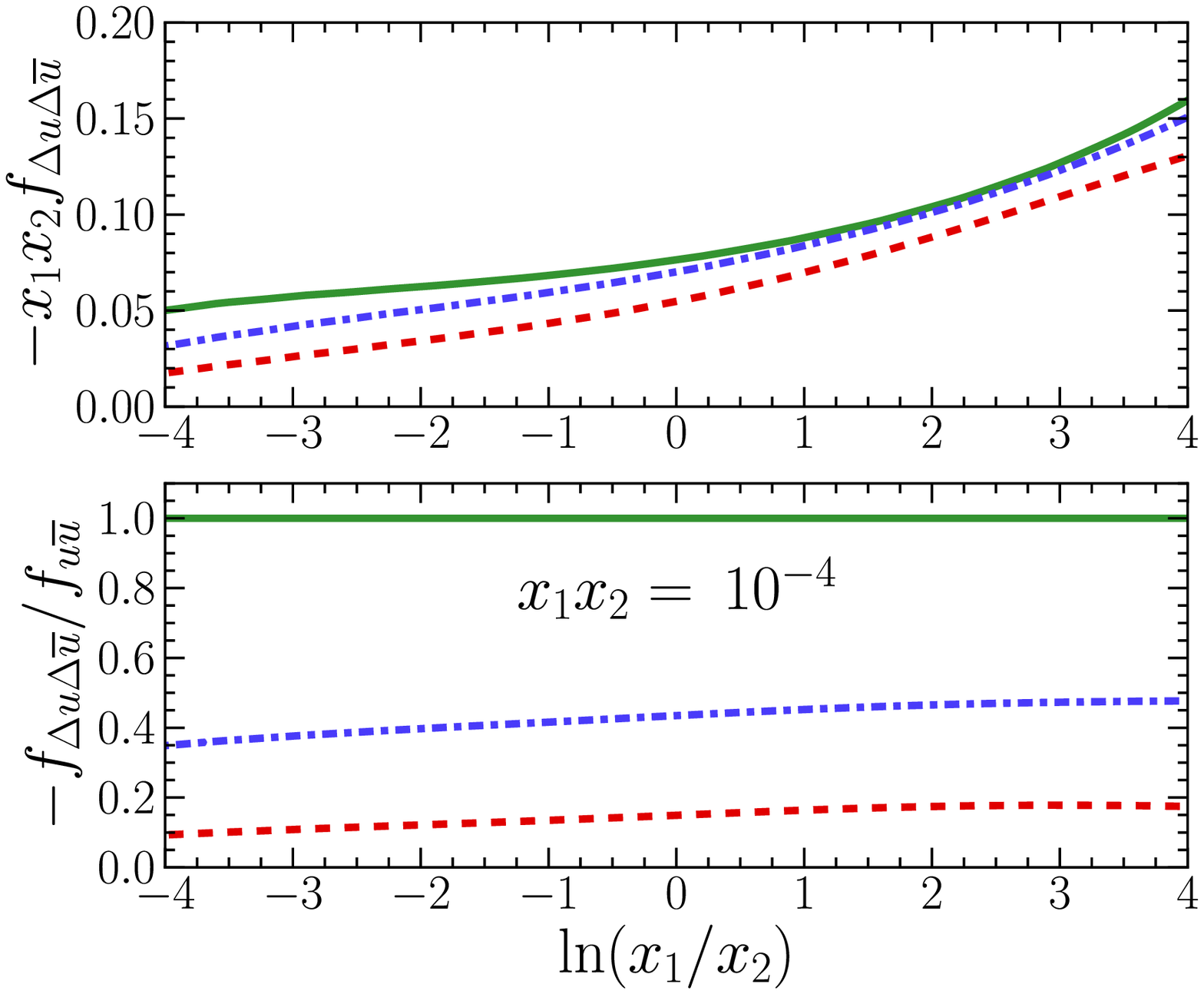}}
  \caption{\label{fig:long-u-mdl} As figure~\protect\ref{fig:long-u-max}
    but in the splitting scenario.  Note the minus sign on the vertical
    axes.}
\end{figure}

The distribution for longitudinally polarized up quarks and antiquarks in
the max scenario is shown in figure~\ref{fig:long-u-max}.  The polarized
distribution $f_{\Delta u\Delta\bar{u}}$ evolves very slowly and hardly
changes with $Q^2$.  However, the degree of polarization decreases with
the evolution scale.  This is due to the increase of the unpolarized DPDs,
which can be seen in figure~\ref{fig:unpol-v1}.  At low $x_i$ the degree
of longitudinal polarization decreases rapidly with $Q^2$, whereas at
intermediate and larger $x_i$ values it does so rather slowly.  We find a
degree of polarization around 50\% at $Q^2=16 \gev^2$ and above 20\% at
$Q^2=10^4 \gev^2$ for $x_1\ms x_2=10^{-4}$ and a wide range of
$\ln(x_1/x_2)$.

In the splitting scenario we have $f_{\Delta q \Delta q} = -f_{qq}$ for
all $x_1$ and $x_2$ at the starting scale.  In order to facilitate
comparison with the max scenario, we multiply the polarized distribution
with $-1$ in figure~\ref{fig:long-u-mdl}.  The mixing of $f_{\Delta q
  \Delta q}$ with distributions involving gluons and the zero starting
value of all distributions where the quark and antiquark have different
flavors induce some differences in evolution compared to the max scenario.
We observe a small decrease of $f_{\Delta u\Delta \bar{u}}$ with the
evolution scale and a somewhat lower degree of polarization.  At large
momentum fractions, the distribution does not quite reach 100\%
polarization in the $x_i$ range shown in the figure.  The dependence of
the degree of polarization on $\ln(x_1/x_2)$ is slightly tilted towards
larger polarization when the quark has a bigger momentum fraction than the
antiquark.

\begin{figure}[tb]
  \centering
  \subfloat[]{\includegraphics[width=0.49\textwidth]{%
      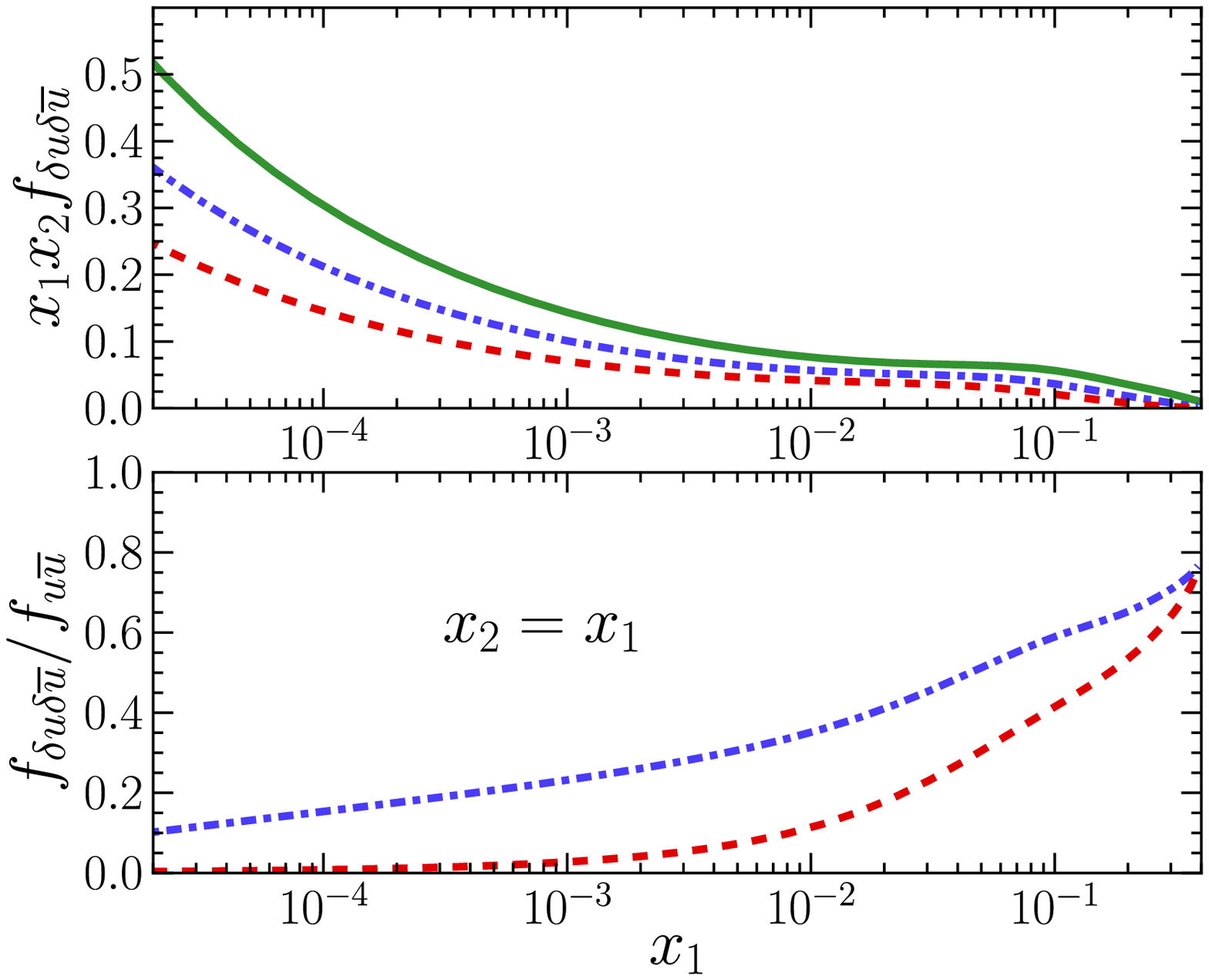}}
  \subfloat[]{\includegraphics[width=0.49\textwidth]{%
      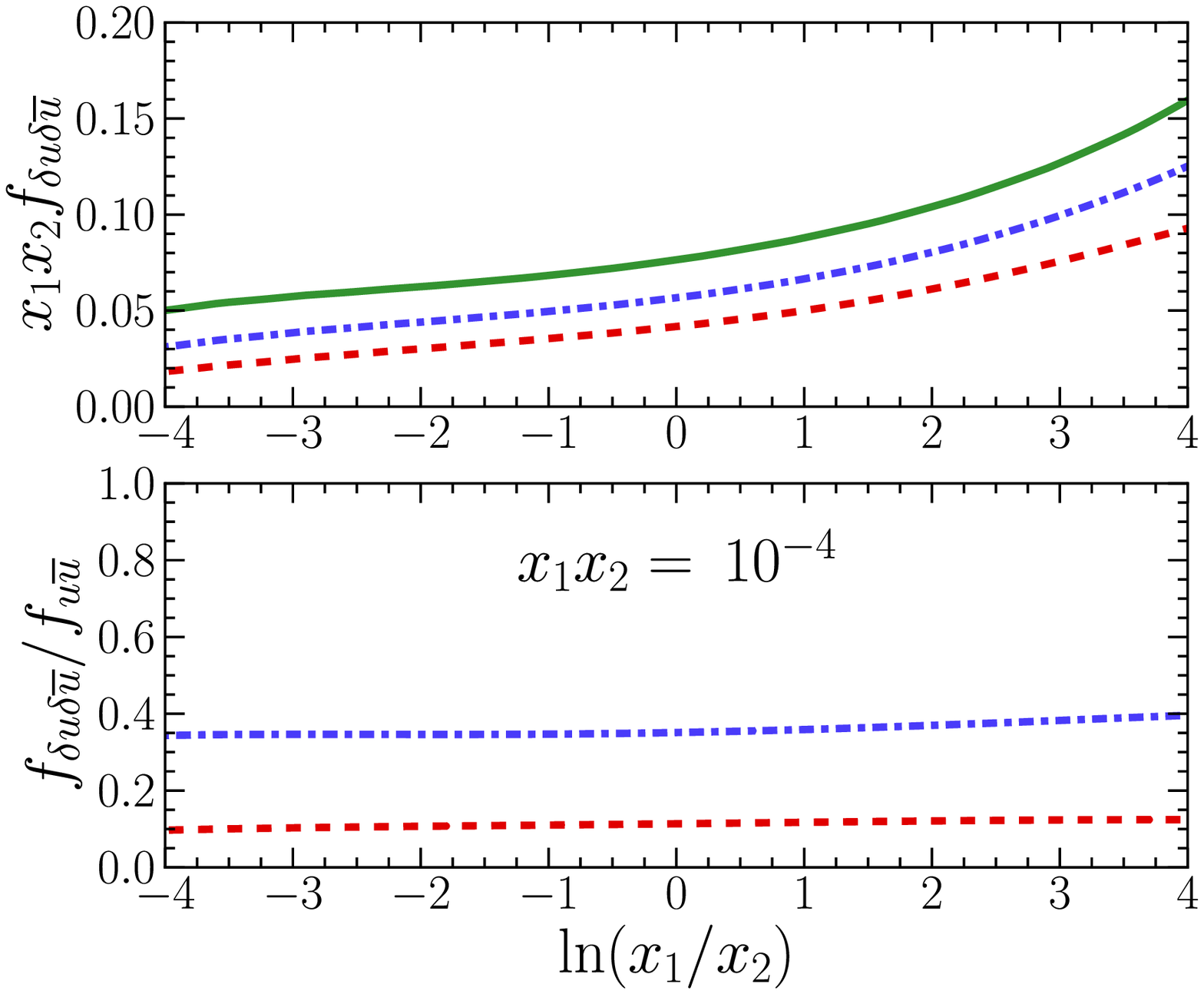}}
  \caption{\label{fig:trans-u-max} Transversely polarized up quarks and
    antiquarks in the max scenario, with initial conditions using the MSTW
    PDFs. Color (line style) coding as in figure~\ref{fig:unpol-v1}.}
  \centering
  \subfloat[]{\includegraphics[width=0.49\textwidth]{%
      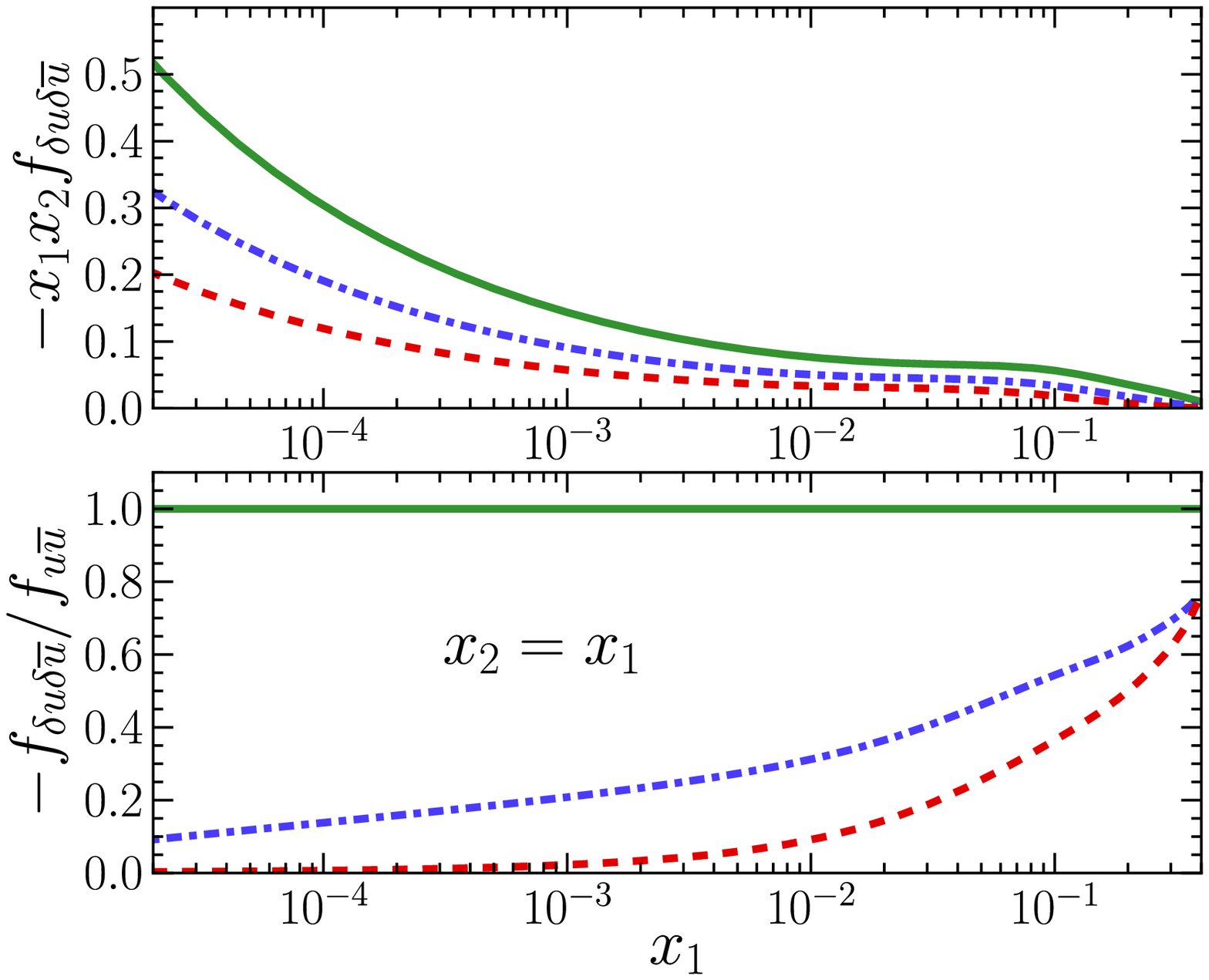}}
  \subfloat[]{\includegraphics[width=0.49\textwidth]{%
      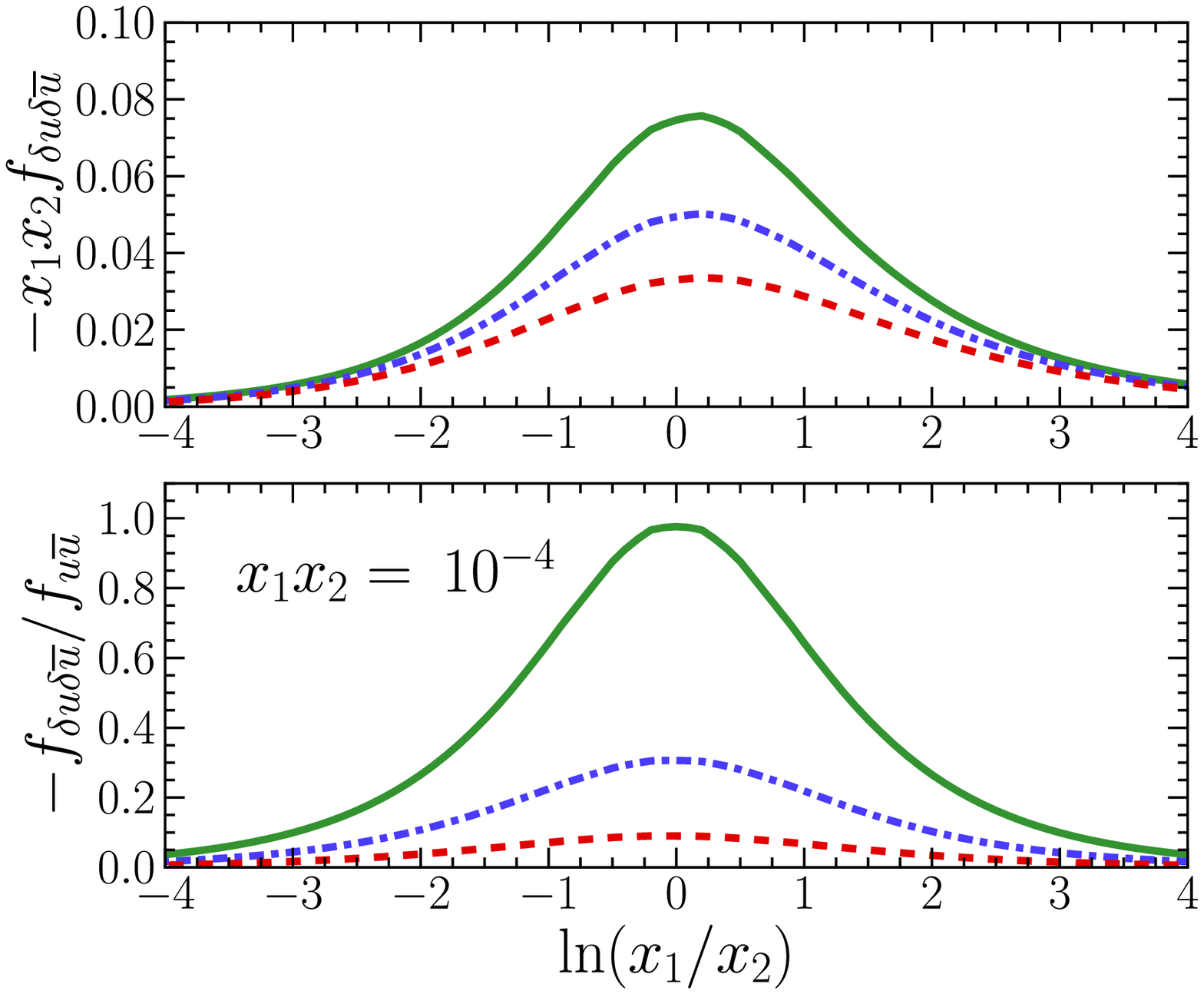}}
  \caption{\label{fig:trans-u-mdl} As figure~\protect\ref{fig:trans-u-max}
    but in the splitting scenario.  Note the minus sign on the vertical
    axes.}
\end{figure}

We now turn to transverse quark and antiquark polarization, which leads to
characteristic azimuthal correlations in the final state of DPD processes
\cite{Kasemets:2012pr}.  Transversely polarized quarks or antiquarks do
not mix with gluons under evolution, nor with quarks or antiquarks of
different flavors.  Figure~\ref{fig:trans-u-max} shows the DPD for
transversely polarized up quarks and antiquarks in the max scenario.
There is a slight decrease of the DPD with $Q^2$ over the entire $x_i$
range, but the suppression of the degree of polarization is mainly due to
the increase in the unpolarized distributions.  The evolution of the
degree of polarization is similar to the case of longitudinal polarization
in the max scenario, with a somewhat faster decrease.  At intermediate and
large $x_i$ values, the degree of polarization decreases slowly.  For
$x_1x_2=10^{-4}$ it amounts to 40\% at $Q^2=16 \gev^2$ and to 10\% at
$Q^2=10^4 \gev^2$ over a wide rapidity range.

In the splitting scenario we have maximal negative polarization $f_{\delta
  q \delta\bar{q}} = - f_{q\bar{q}}$ at the starting scale for $x_1=x_2$,
but the degree of polarization decreases when the two partons have
different momentum fractions and tends to zero for both $x_1\ll x_2$ and
$x_1\gg x_2$.  Figure~\ref{fig:trans-u-mdl} shows that the dependence of
the polarized DPD on $\ln(x_1/x_2)$ slightly flattens under evolution and
that its overall size at $x_1=x_2$ evolves in a similar way as in the max
scenario.  The same holds for the degree of polarization.

The polarization for other combinations of light quarks and antiquarks is
of similar size and shows a similar evolution behavior as for the case of
a $u\bar{u}$ pair just presented.  Generically, the evolved distributions
have a slightly larger degree of polarization for quarks compared with
antiquarks, and for up quarks compared with down quarks.  Obvious
exceptions in the splitting scenario are distributions that do not have a
quark and an antiquark of equal flavor.  These distributions start at
zero.  For transverse polarization they hence remain zero at all scales,
while for longitudinal polarization they become nonzero due to the mixing
with other distributions.  The ratio $f_{\Delta u \Delta u} /f_{uu}$ for
example can reach a few percent in the intermediate $x_i$ region for equal
momentum fractions after evolution.

In summary, we find that both longitudinal and transverse polarization
remains sizeable up to large scales for intermediate and large $x_i$
values, provided it is large at low scales.  In particular the
distributions for longitudinally polarized quarks, which enter linearly in
electroweak cross sections, can thus play a significant role.  In the
small $x_i$ region, however, both longitudinal and transverse polarization
are strongly suppressed at high scales.

We have repeated our study with the MSTW distributions replaced by those
of GJR in the initial conditions.  Naturally, this has some effect on the
quark distributions, but the differences on the degree of polarization are
comparably small and do not change the conclusions we have just drawn.


\subsection{Gluon distributions}

\begin{figure}[b]
  \centering
  \subfloat[]{\includegraphics[width=0.496\textwidth]{%
      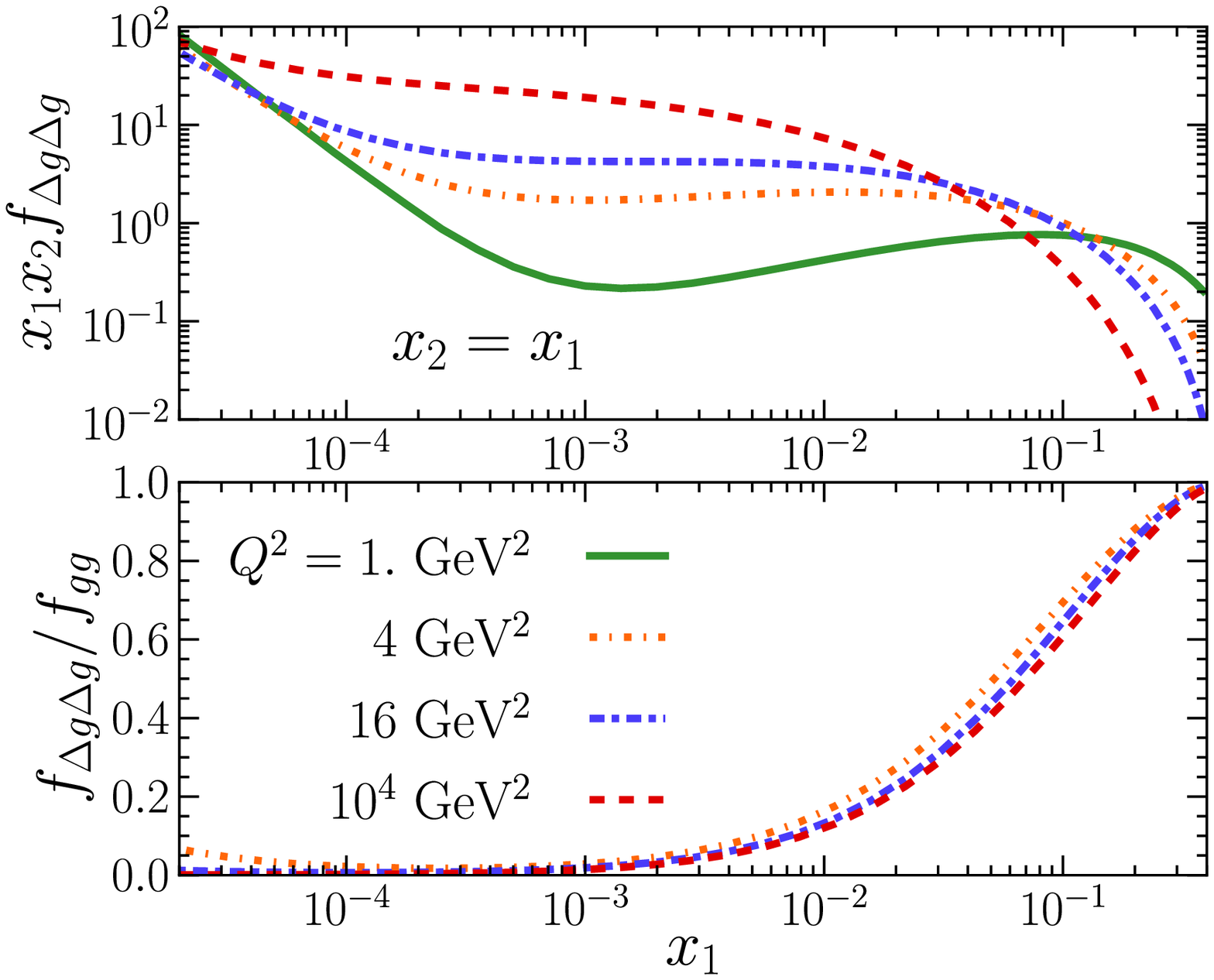}}
  \subfloat[]{\includegraphics[width=0.48\textwidth]{%
      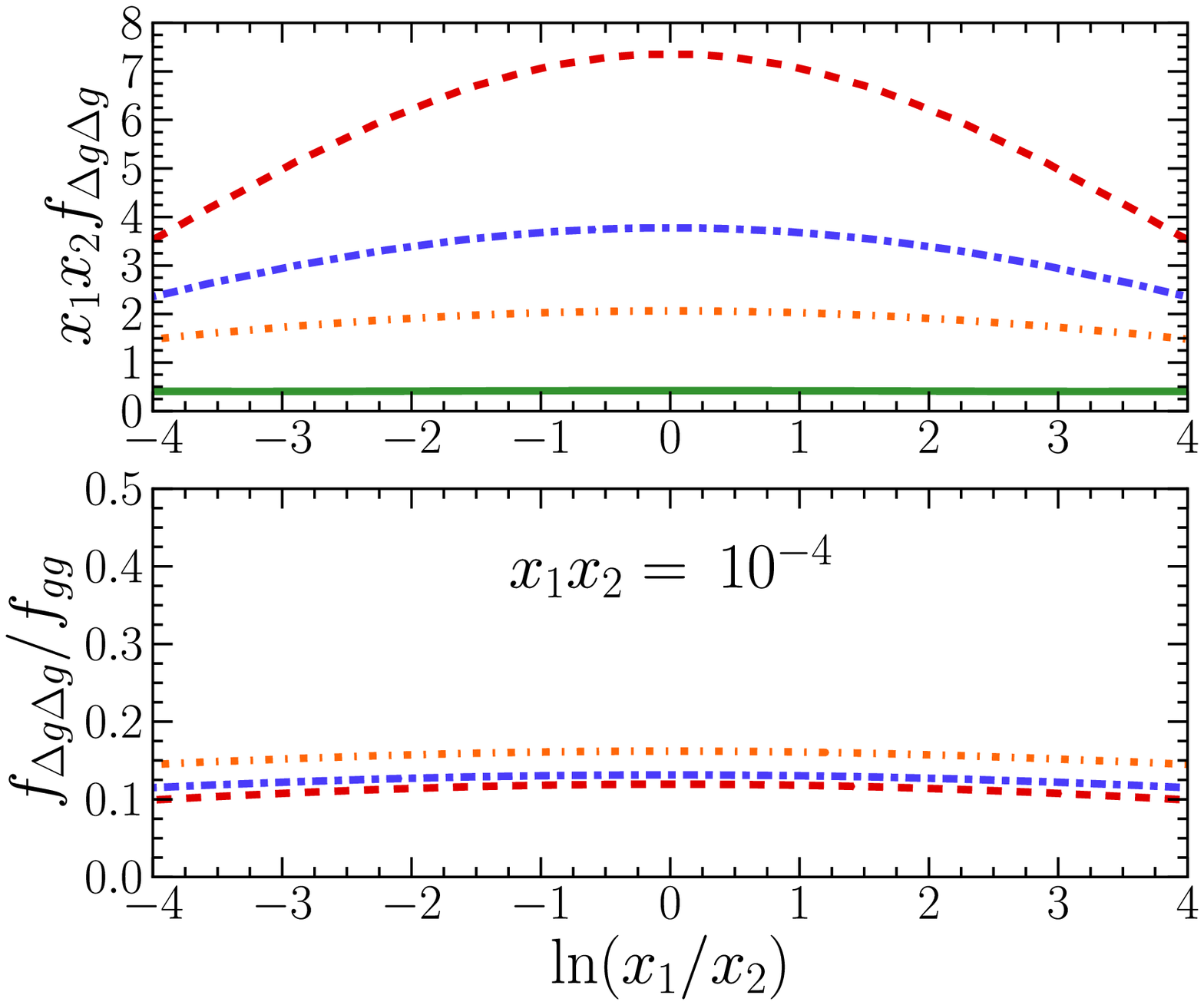}}
  \caption{\label{fig:long-gMSTW-max} Longitudinally polarized gluons in
    the max scenario, with initial conditions using the PDFs of MSTW.}
\end{figure}

Gluons can be polarized longitudinally or linearly.  As discussed in
\sect{sec:evo} the unpolarized (single or double) gluon density increases
rapidly at small momentum fractions due to the $1/x$ behavior of the gluon
splitting kernel.  The absence of this low-$x$ enhancement in the
polarized gluon splitting kernels lead us to expect that the degree of
gluon polarization will vanish rapidly in the small $x$ region.  As can be
seen in figure~\ref{fig:long-gMSTW-max} for longitudinally polarized
gluons, this is indeed the case.  The distribution $f_{\Delta g \Delta g}$
does increase with evolution scale, but at a much lower rate than
$f_{gg}$.  Evolution thus quickly suppresses the degree of longitudinal
gluon polarization in the small $x_i$ region.  In
figure~\ref{fig:long-gMSTW-max} we see that in the max scenario with MSTW
starting distributions this suppression stays rather constant between
$Q^2=4 \gev^2$ and $Q^2=10^4 \gev^2$.  For this range of scales, the
degree of longitudinal polarization is between 10\% and 15\% at $x_1\ms
x_2 = 10^{-4}$, with a very weak dependence on $\ln(x_1/x_2)$.

\begin{figure}[tb]
  \centering
  \subfloat[]{\includegraphics[width=0.496\textwidth]{%
      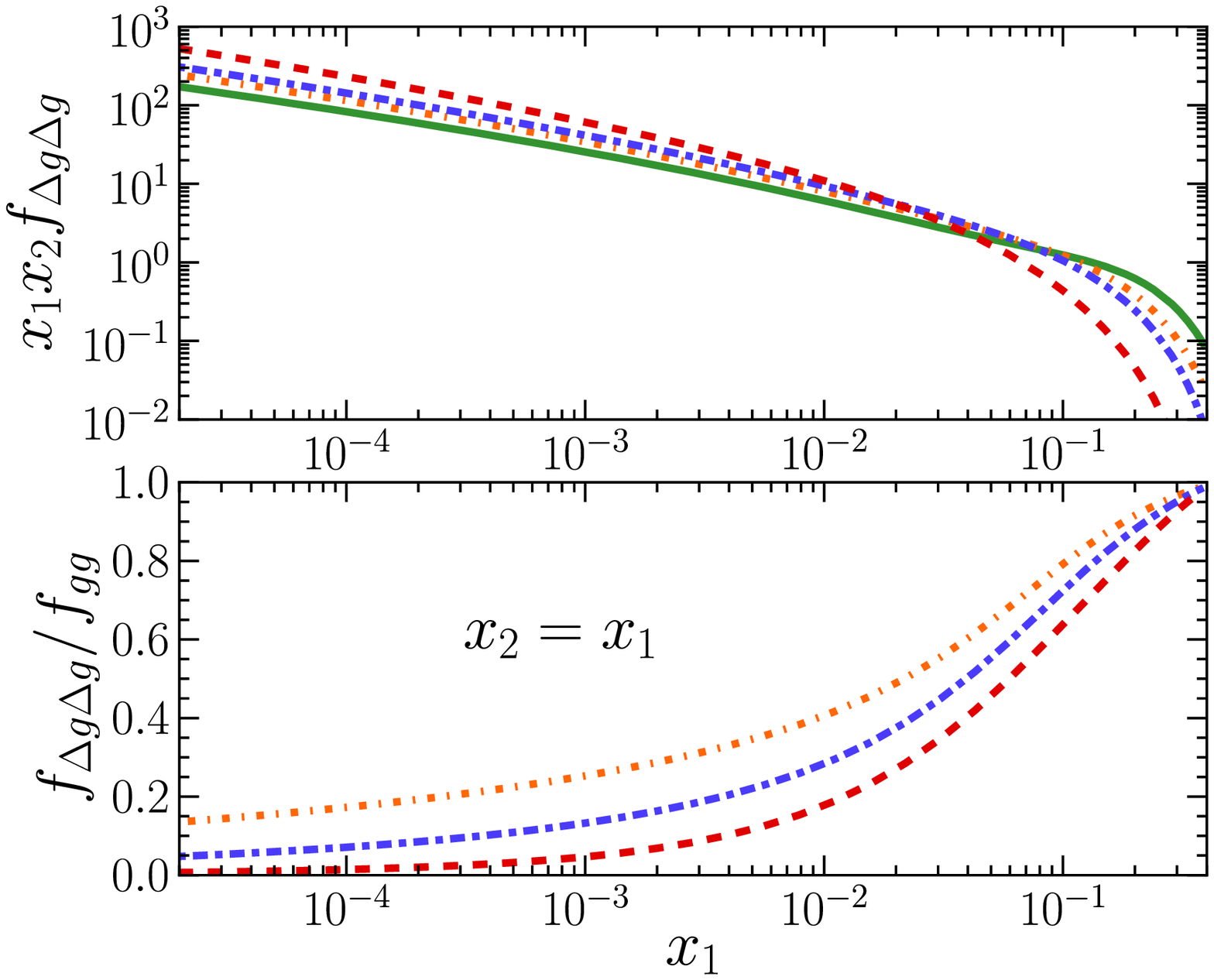}}
  \subfloat[]{\includegraphics[width=0.48\textwidth]{%
      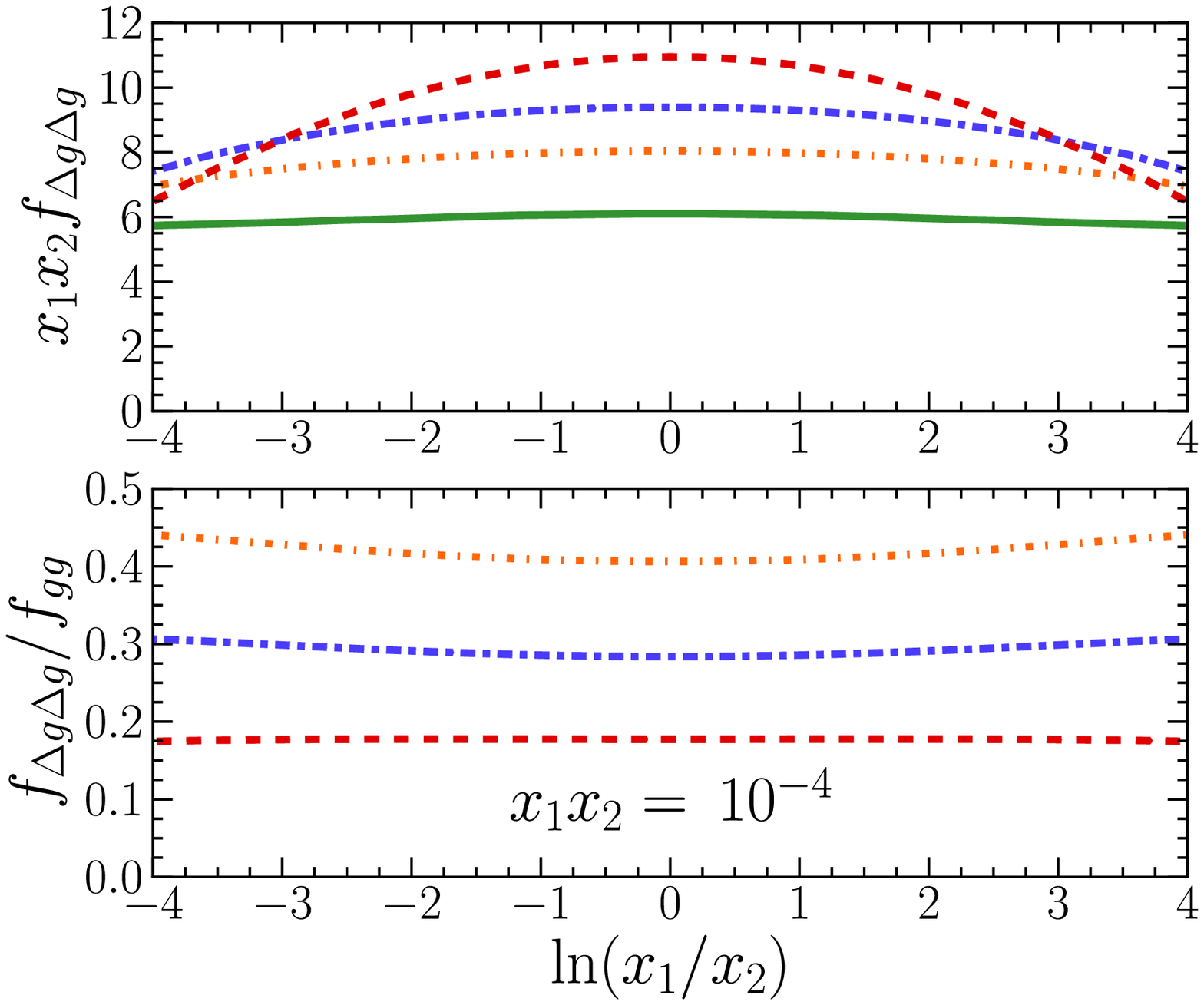}}
  \caption{\label{fig:long-gGJR-max} As
    figure~\protect\ref{fig:long-gMSTW-max} but with initial conditions
    using the PDFs of GJR.  Notice the different range of the vertical
    axes in the upper row compared with
    figure~\protect\ref{fig:long-gMSTW-max}.}
  \subfloat[]{\includegraphics[width=0.496\textwidth]%
    {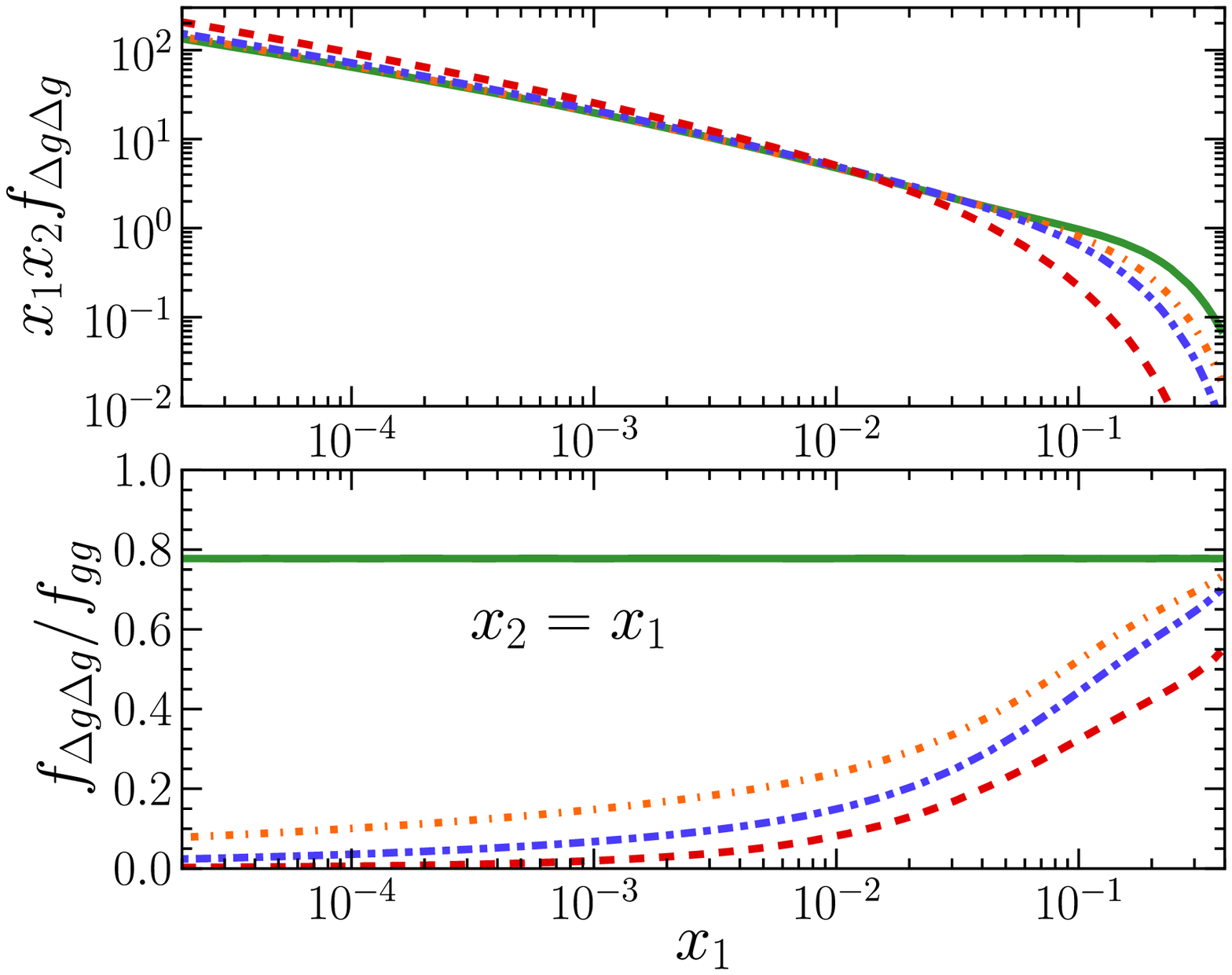}}
  \subfloat[]{\includegraphics[width=0.48\textwidth]{%
      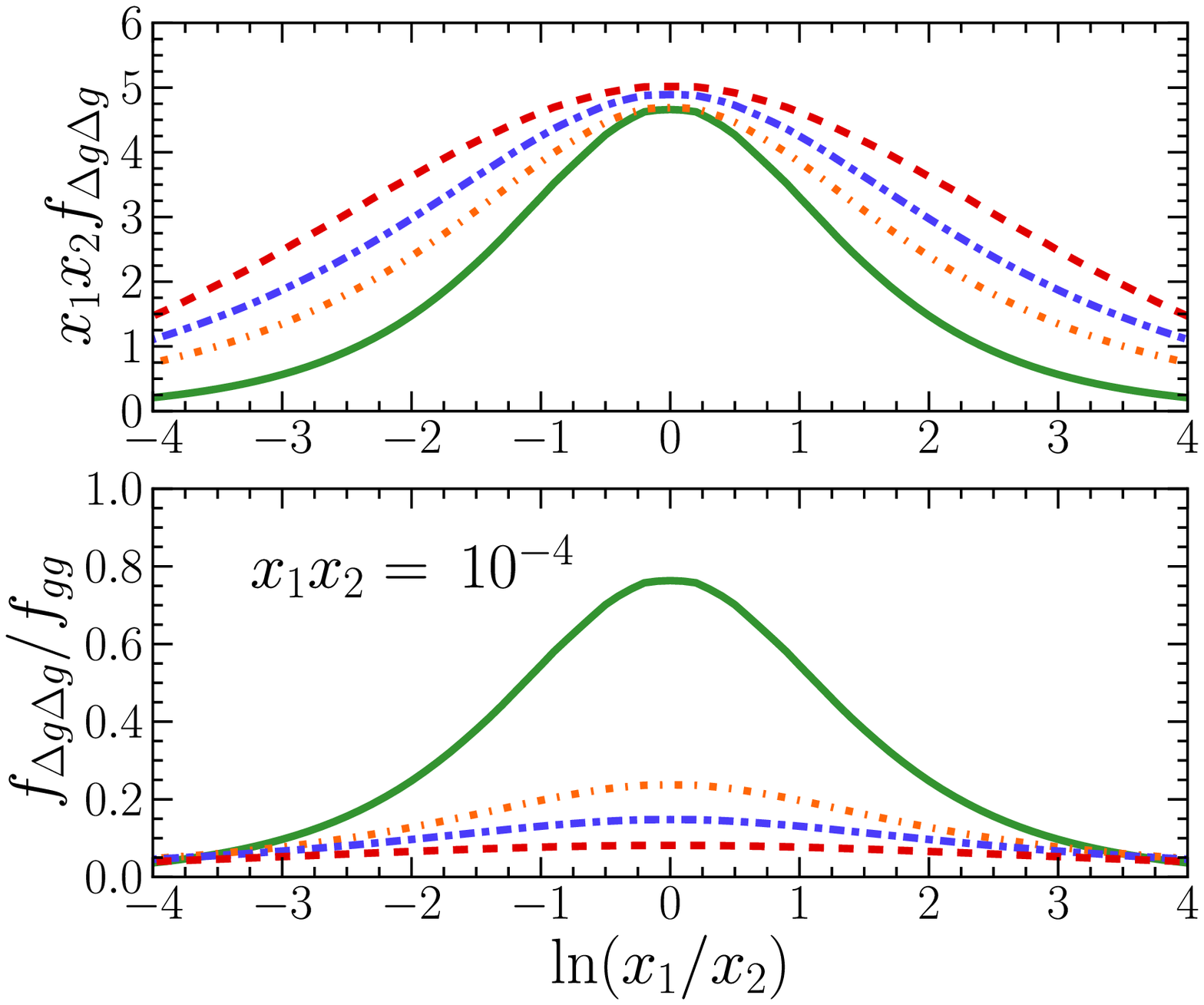}}
  \caption{\label{fig:long-g-mdl} Longitudinally polarized gluons in the
    splitting scenario, with initial conditions using the PDFs of GJR.
    Color (line style) coding as in
    figure~\protect\ref{fig:long-gMSTW-max}.}
\end{figure}

Our knowledge of the single gluon distribution at the low scale remains,
however, quite poor as is documented in \app{ap:pdfs}.  As an alternative
to the MSTW distributions used in figure~\ref{fig:long-gMSTW-max}, we show
in figure~\ref{fig:long-gGJR-max} the corresponding results obtained with
the GJR parton densities.  The degree of polarization at $Q^2=10^4 \gev^2$
is nearly twice as large as for MSTW and amounts to almost 20\% at $x_1\ms
x_2=10^{-4}$.  At $Q^2=16 \gev^2$ the difference is even more striking,
with a degree of polarization equal to 30\%, a factor of three larger than
for MSTW.  To understand this difference, we recall from
figures~\ref{fig:unpol-v1} and \ref{fig:unpol-v2} that at high scales the
unpolarized gluon DPDs obtained with the two PDF sets are relatively
similar.  The difference in the degree of longitudinal polarization
between the two cases is hence mainly due to the polarized DPDs.  At the
starting scale, these are much larger if we use the GJR set instead of
MSTW, and this large difference persists after evolution.

In the splitting scenario the differences between the results obtained
with the two PDF sets are similar to the differences we just described
for the max scenario.  We show in figure~\ref{fig:long-g-mdl} the
results obtained with the GJR set and note that the degree of polarization
obtained with MSTW distributions is significantly smaller.  At the
starting scale, the degree of longitudinal polarization has a maximum of
78\% for $x_1=x_2$ in the splitting scenario and quickly decreases when
the two gluons have different momentum fractions.  Evolution decreases the
degree of polarization in a similar manner as in the max scenario, leaving
us with polarization around 10\% for central rapidities and $Q^2$ between
$16$ and $10^4 \gev^2$.

\begin{figure}[tb]
  \centering
  \subfloat[]{\includegraphics[width=0.496\textwidth]{%
      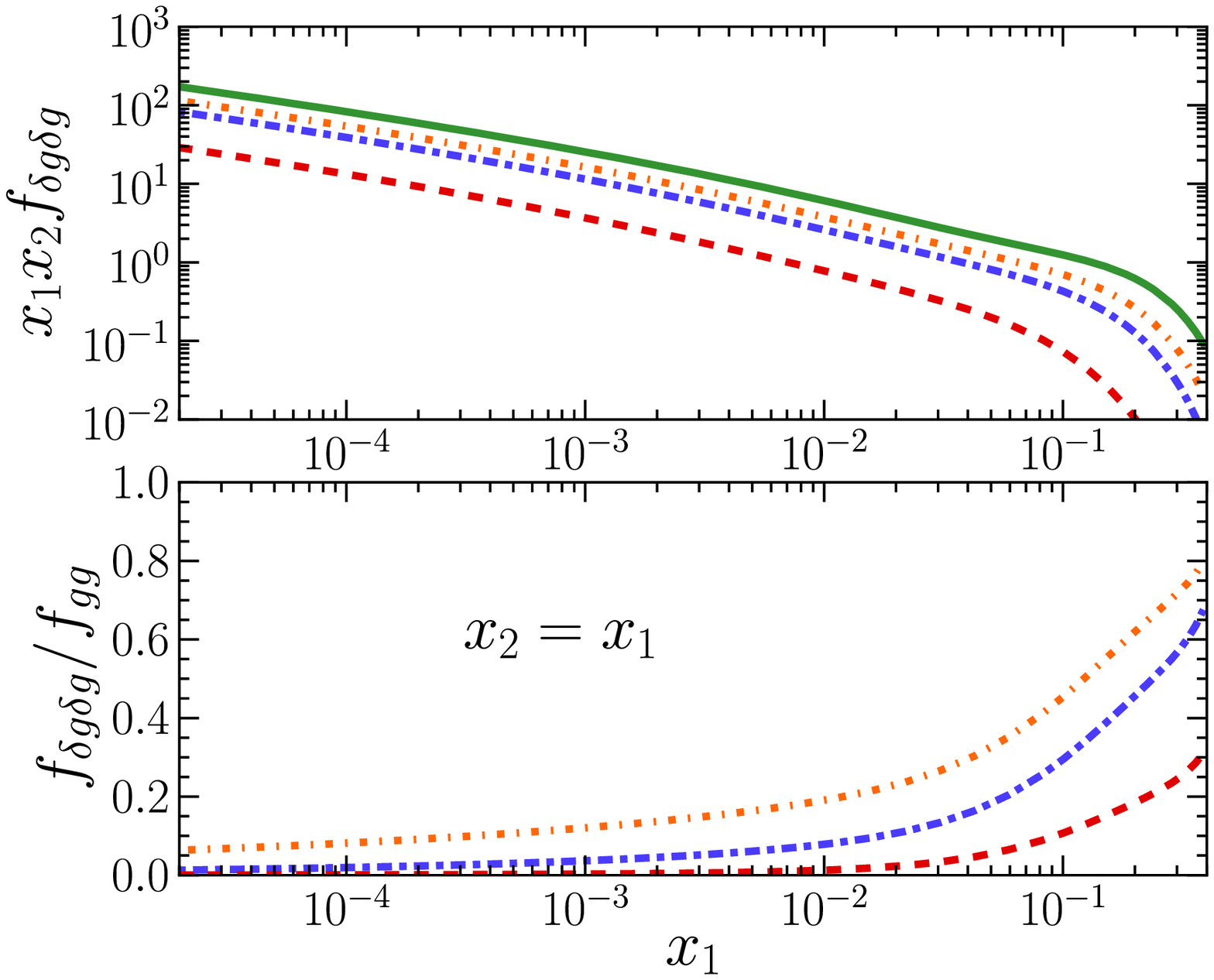}}
  \subfloat[]{\includegraphics[width=0.48\textwidth]{%
      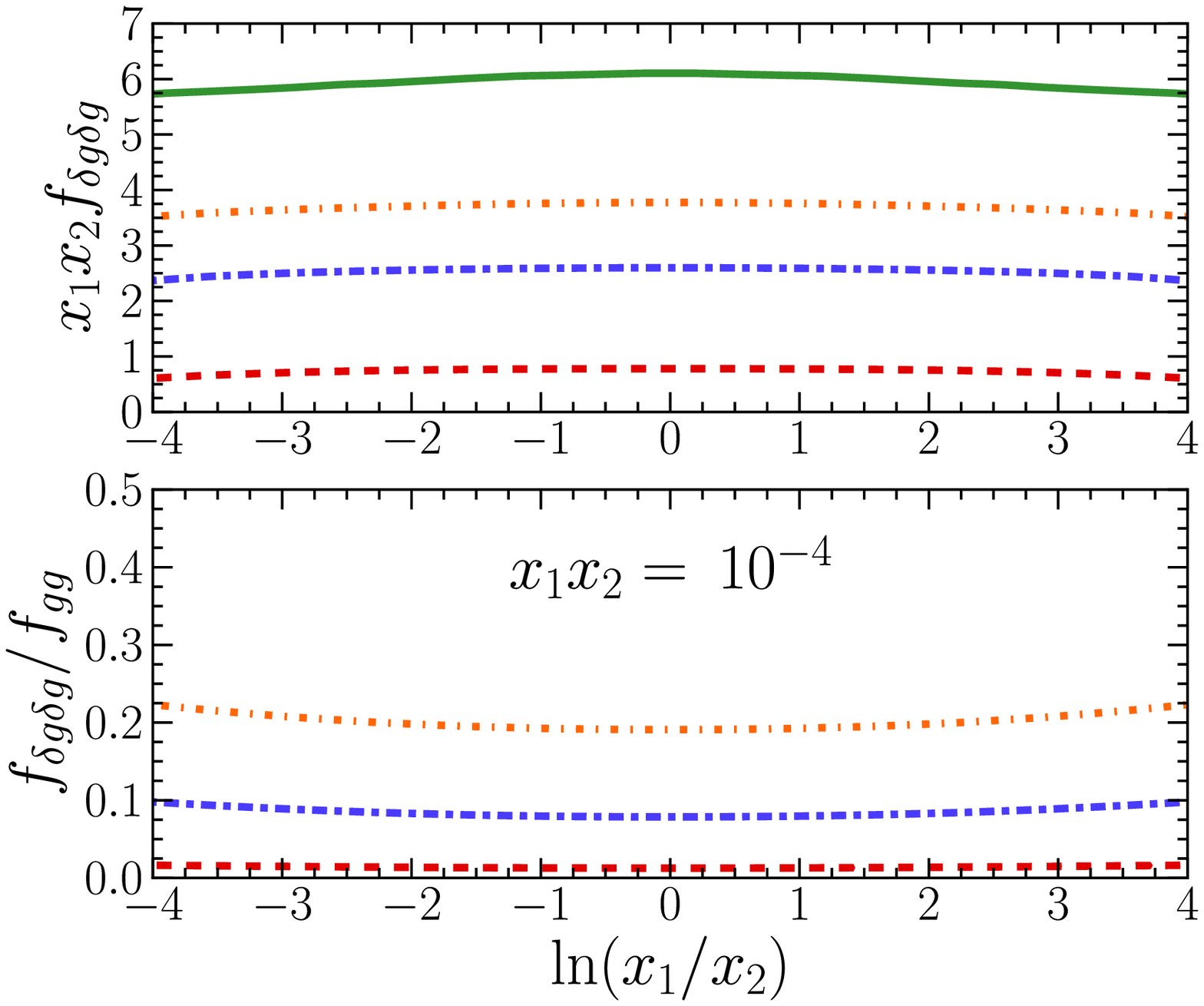}}
  \caption{\label{fig:lin-g-max} Distribution for two linearly polarized
    gluons in the max scenario, using the GJR PDFs in the initial
    conditions. Color (line style) coding as in
    figure~\protect\ref{fig:long-gMSTW-max}.  Here and in the following,
    the LO expression of the evolution kernel $P_{\delta g \delta g}$ is
    used, except in figure~\protect\ref{fig:linglu-nlo}.}
\end{figure}

Linearly polarized gluons give rise to azimuthal asymmetries in DPS cross
sections, in a similar way as transversely polarized quarks and
antiquarks.  The effect of evolution on the distribution of two linearly
polarized gluons in the max scenario is shown in
figure~\ref{fig:lin-g-max}.  We see that even the polarized distribution
$f_{\delta g \delta g}$ itself decreases with the scale.  Together with
the rapid increase of the unpolarized two-gluon DPD this results in a
rapid decrease of the degree of linear polarization, especially at small
$x_i$.  As in the case of longitudinal gluon polarization, using the MSTW
distributions at the starting scale (not shown here) results in an even
faster suppression.  In that case the degree of polarization is tiny
already at $Q^2=16 \gev^2$.  In the splitting scenario, the ratio
$f_{\delta g \delta g} / f_{gg}$ is at most 11\% at the starting scale,
which leads of course to even lower polarization after evolution.  We must
hence conclude that even in the most optimistic scenario shown in
figure~\ref{fig:lin-g-max}, the correlation between two linearly polarized
gluons is quickly washed out by evolution and can only be appreciable at
rather large $x_i$ or rather low scales.

So far we have only considered the case when both partons have the same
type of polarization.  There is however the possibility to have an
unpolarized gluon and a linearly polarized one, whose polarization
direction is correlated with the interparton distance $\y$.  In the max
scenario, the corresponding DPD at $x=x_1=x_2$ is well approximated by
\begin{align}
  \label{eq:pol-approx}
(y M)^2 f_{g\ms \delta g}(x,x; Q) \approx
  \sqrt{f_{gg}(x,x; Q)\, f_{\delta g\ms \delta g}(x,x; Q)} \,,
\end{align}
for not too large $x$.  At the starting scale, this is trivial, and at
higher scales it reflects the fact that double DGLAP evolution proceeds
approximately independently for the two partons, as long as $x_1+x_2$ is
not close to $1$.  In the splitting scenario, \eqref{eq:pol-approx} does
not hold even at the starting scale.

\begin{figure}[tb]
  \centering
  \subfloat[]{\includegraphics[width=0.496\textwidth]{%
      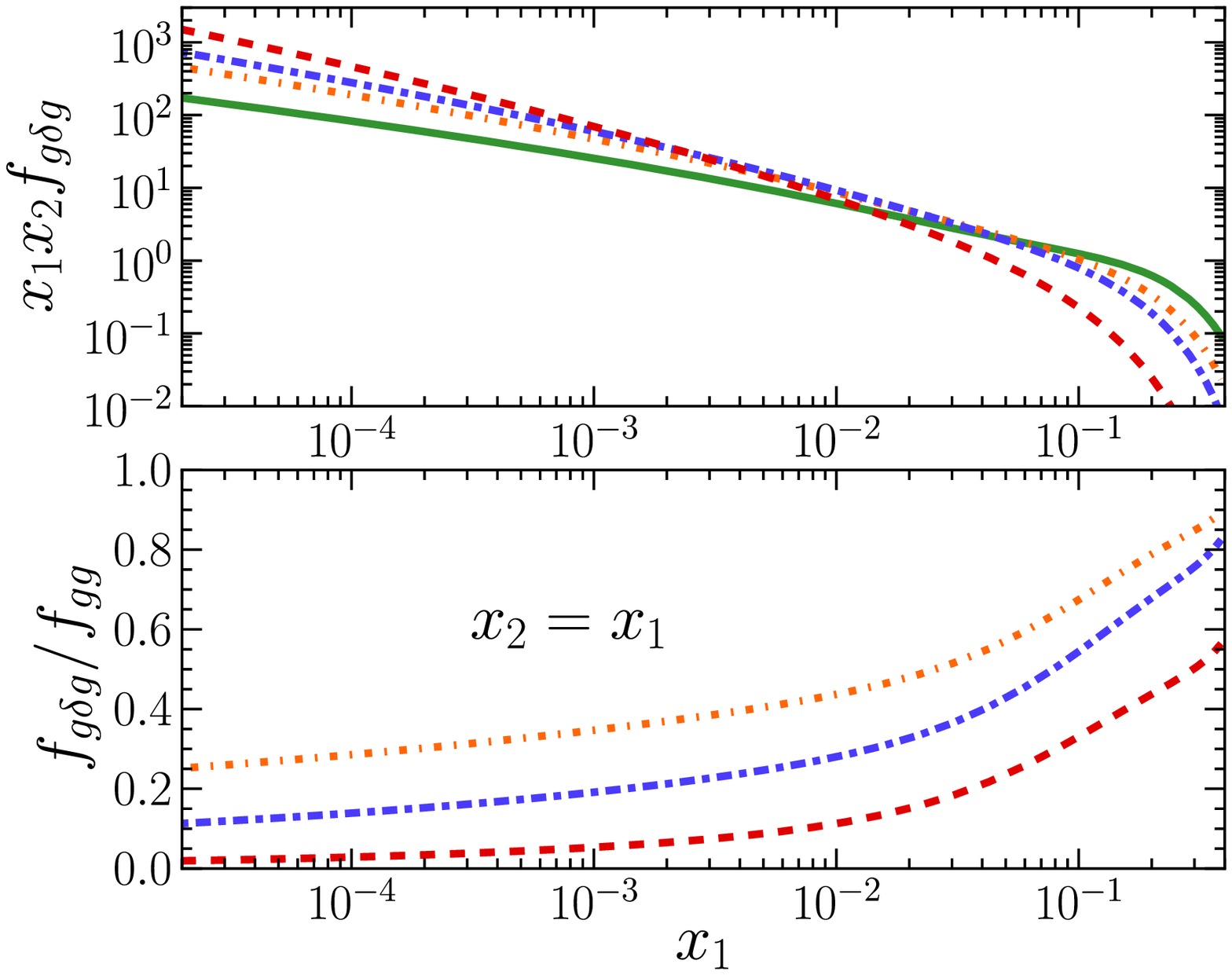}}
  \subfloat[]{\includegraphics[width=0.48\textwidth]{%
      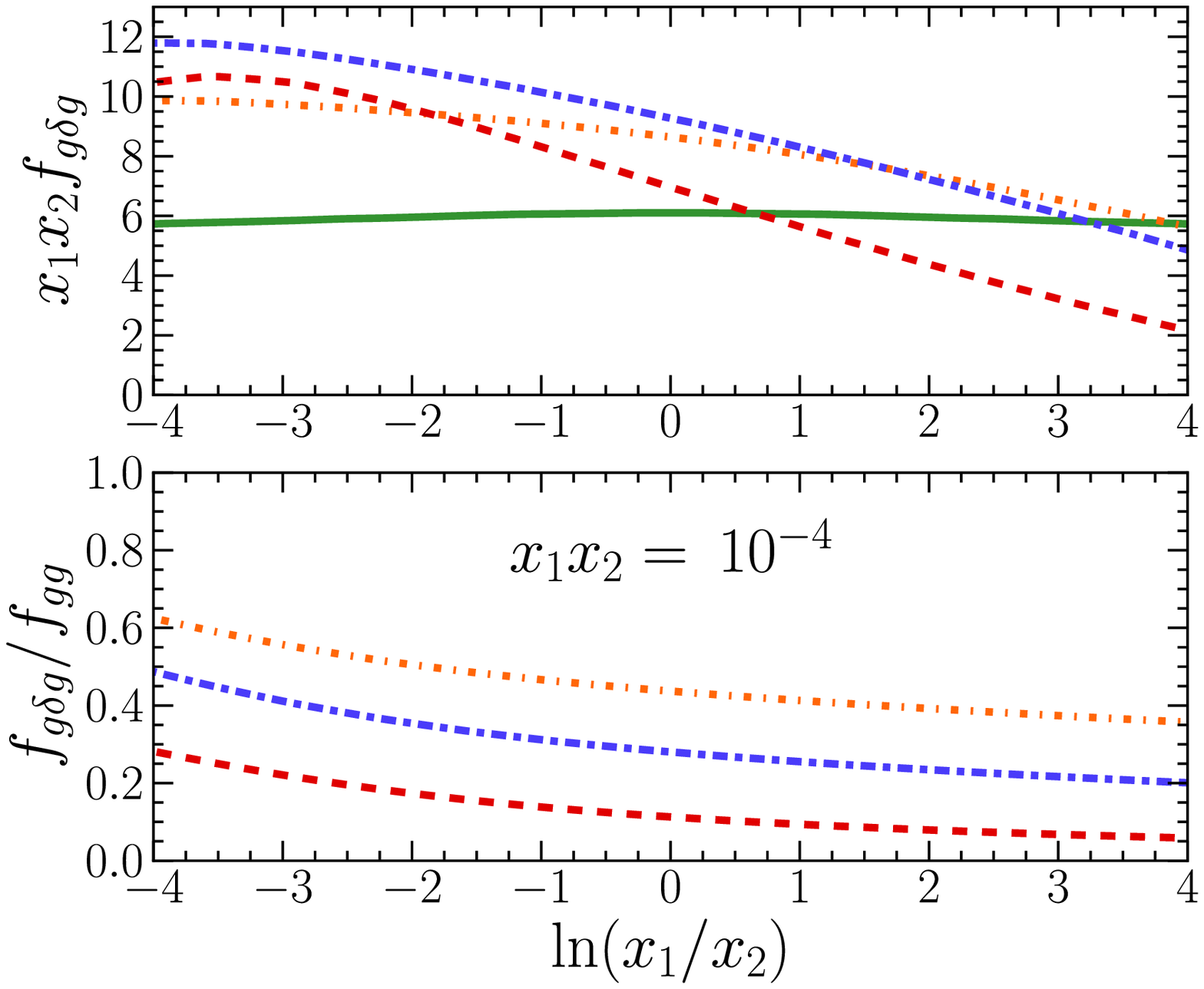}}
  \caption{\label{fig:g-ling-max} Two-gluon distribution with one linearly
    polarized gluon in the max scenario, with GJR PDFs used at the
    starting scale.  The factor $(y M)^2$ in the starting conditions
    \eqref{eq:pos-sat-mixed} has been set to $1$ for simplicity.  Color
    (line style) coding as in figure~\ref{fig:long-gMSTW-max}.}
 \subfloat[]{\includegraphics[width=0.496\textwidth]{%
     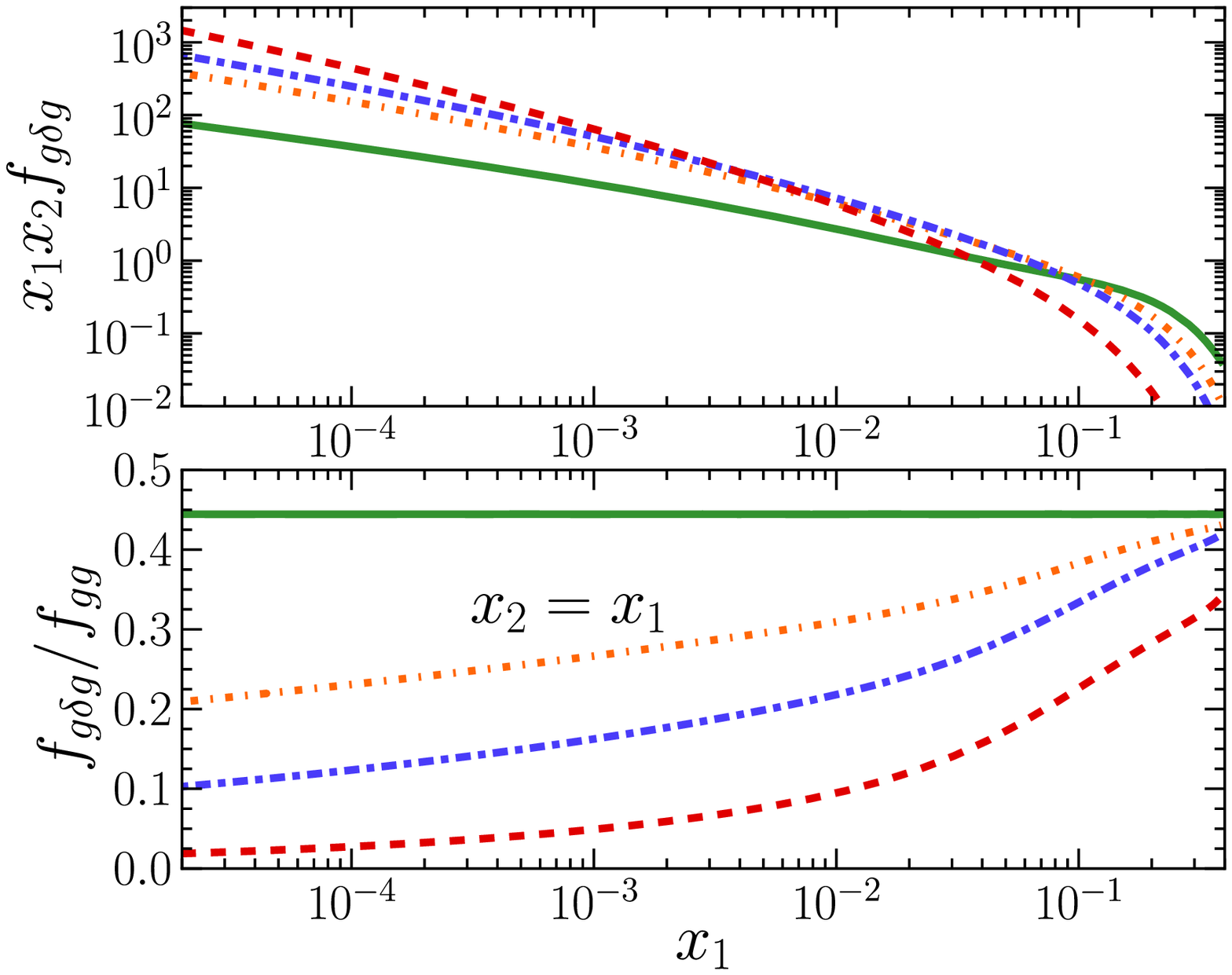}}
  \subfloat[]{\includegraphics[width=0.48\textwidth]{%
      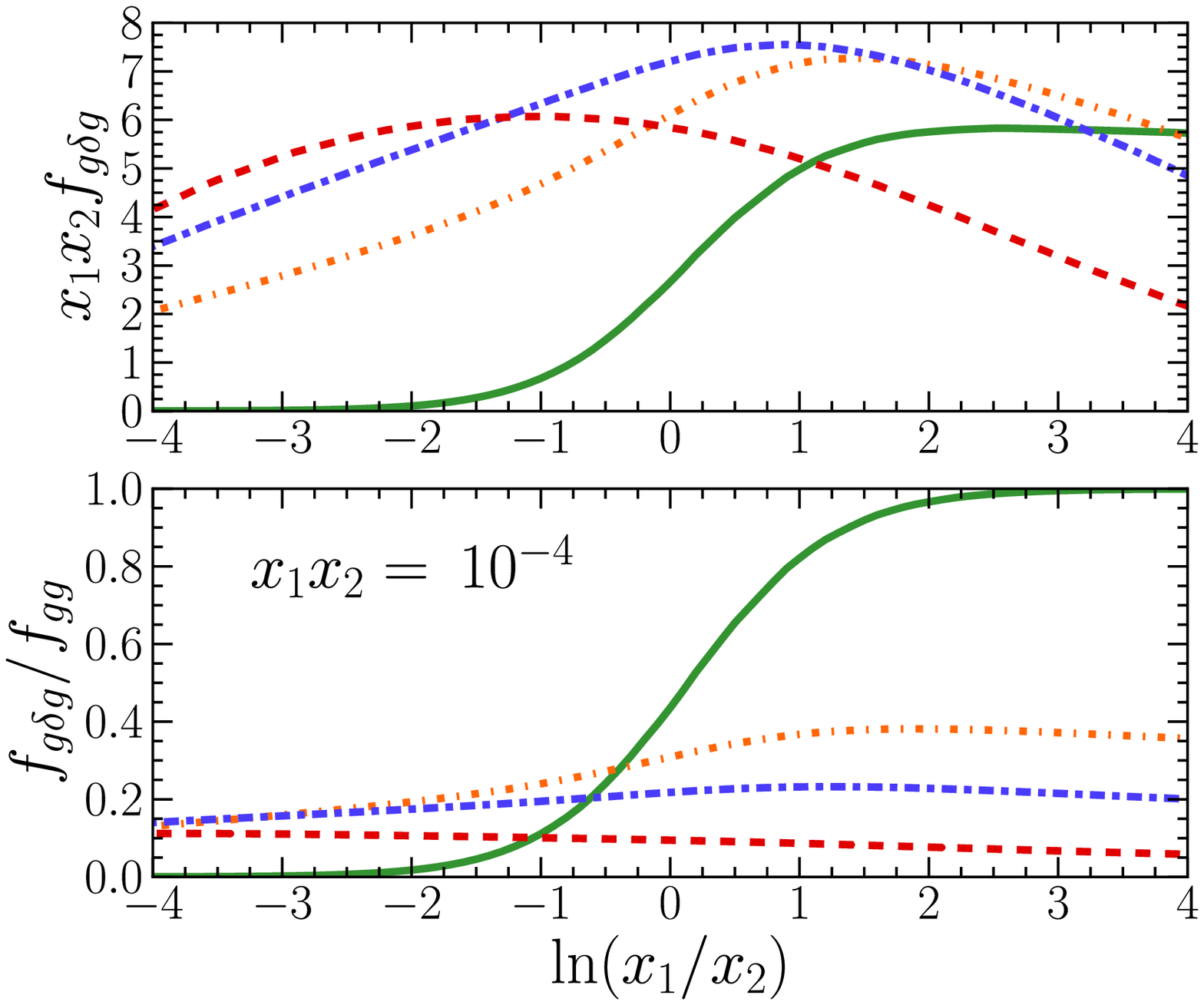}}

\vspace{-0.4em}

  \caption{\label{fig:g-ling-mdl} As figure~\protect\ref{fig:g-ling-max}
    but in the splitting scenario.  The factor $(y M)^2$ in the starting
    conditions \eqref{eq:split-coeff-single} has been set to $1$ for
    simplicity.}
\end{figure}

Figure~\ref{fig:g-ling-max} shows $f_{g\ms \delta g}$ in the max scenario.
The presence of one unpolarized gluon increases the distribution for small
$x_i$ values and results in a significantly larger degree of polarization.
However, for very large $Q^2$ linear polarization is still strongly
suppressed at small $x_i$.  For unequal $x_i$ we observe that the degree
of polarization is enhanced when the unpolarized gluon has the smaller
momentum fraction, reaching 30\% for $\ln(x_1/x_2) = -4$ even at the high
scale $Q^2=10^4 \gev^2$.  A significant degree of polarization for one
linearly polarized gluon is also found in the splitting scenario, as shown
in figure~\ref{fig:g-ling-mdl}.  The main difference to the max scenario
appears for unequal $x_i$.  According to \eqref{eq:split-coeff-single} the
splitting $g\to gg$ is such that for very asymmetric kinematics $x_1 \gg
x_2$ the slow gluon carries maximal linear polarization.  As is seen in
the figure, evolution weakens this trend, and at very high scales we find
again a higher degree of polarization if $x_1 < x_2$ rather than $x_1 >
x_2$.


\subsubsection*{Linear polarization: effect of NLO corrections}

As we noted in \sect{sec:evo}, the evolution kernel for linearly polarized
gluons has a qualitatively different small-$x$ behavior at leading and
next-to-leading order in $\alpha_s$.  To examine how this impacts the
fraction of linearly polarized gluons in DPDs at low $x_i$, we have
incorporated the leading low-$x$ term in the NLO kernel as given in
\eqref{eq:lin-split}.

\begin{figure}[ht]
  \centering
  \subfloat[MSTW]{\includegraphics[width=0.48\textwidth]{%
      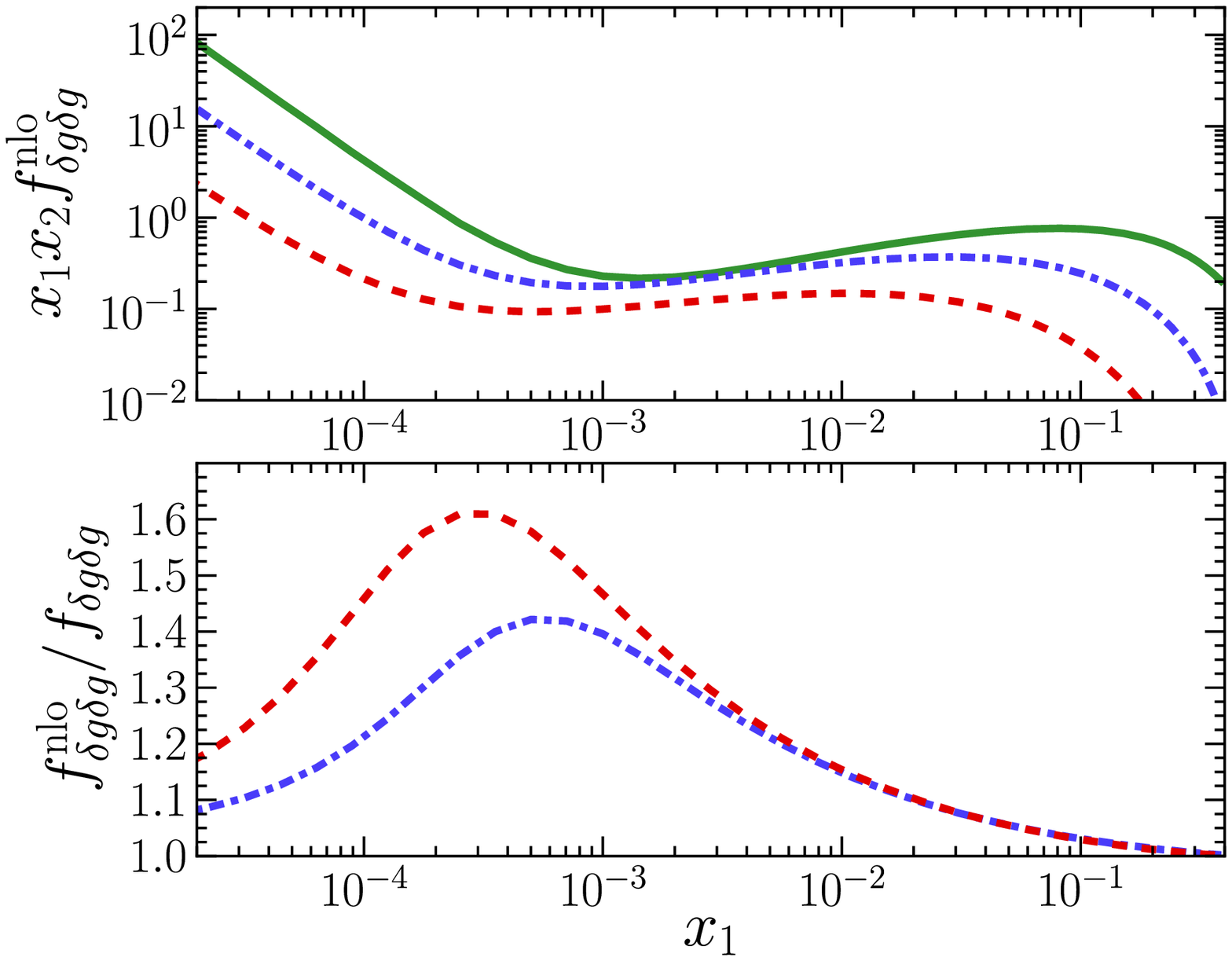}}
  \subfloat[GJR]{\includegraphics[width=0.48\textwidth]{%
      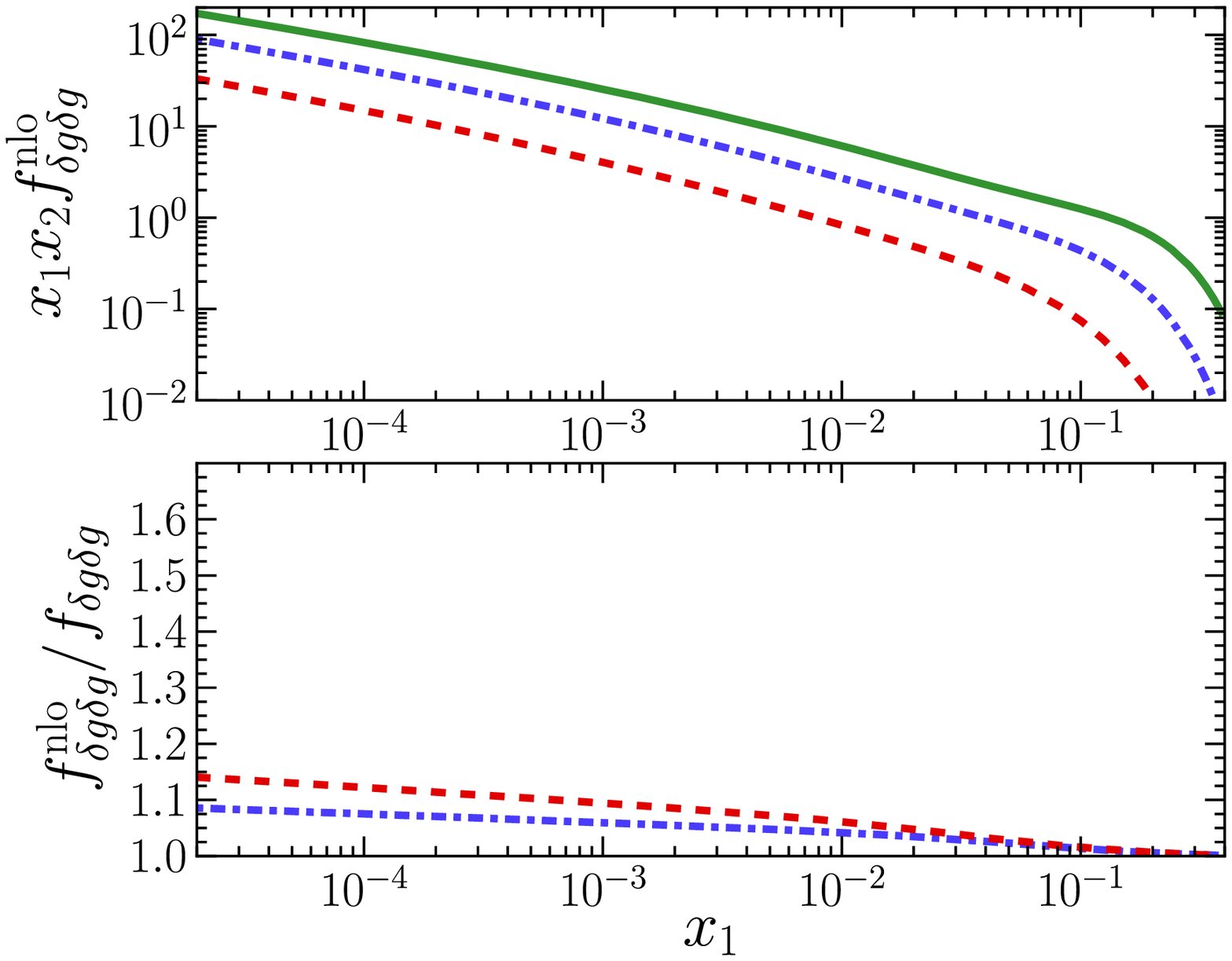}}
  \caption{\label{fig:linglu-nlo} Comparison of the DPD for two linearly
    polarized gluons evolved at LO $(f_{\delta g \delta g})$ or including
    the leading low-$x$ part of the NLO kernel $(f_{\delta g \delta
      g}^{\text{nlo}})$ given in equation~\protect\eqref{eq:lin-split}.
    For the initial conditions we use the max scenario with the PDFs of
    MSTW (a) or of GJR (b).  Color (line style) coding as in
    figure~\ref{fig:unpol-v1}.}
\end{figure}

The results in the splitting scenario with either MSTW or GJR input
distributions are shown in figure~\ref{fig:linglu-nlo}.  Although the NLO
corrections increase the polarization as one may expect, they do not
significantly change the overall picture.  The NLO enhancement is largest
for the case of MSTW distributions around $x_1=x_2 = 10^{-3}$, where the
polarized DPD has a dip and hence the increased migration of partons from
larger to smaller $x_i$ is most important.


\subsection{Quark-gluon distributions}

In the two previous subsections we have seen that in general gluon
polarization is washed out under evolution at a faster pace than
the polarization of quarks.  In this section we will see that the
corresponding decrease of polarization for quark-gluon distributions is in
between the pure quark and gluon cases.  For the unpolarized DPDs, our
factorized ansatz \eqref{eq:x-factorization} results in the approximate
relation
\begin{align}
f_{qg}(x,x; Q) \approx \sqrt{
  f_{qq}(x,x; Q)\, f_{gg}(x,x; Q)}
\end{align}
as long as $x$ is not too large.

\begin{figure}[tb]
  \centering
  \subfloat[]{\includegraphics[width=0.496\textwidth]{%
      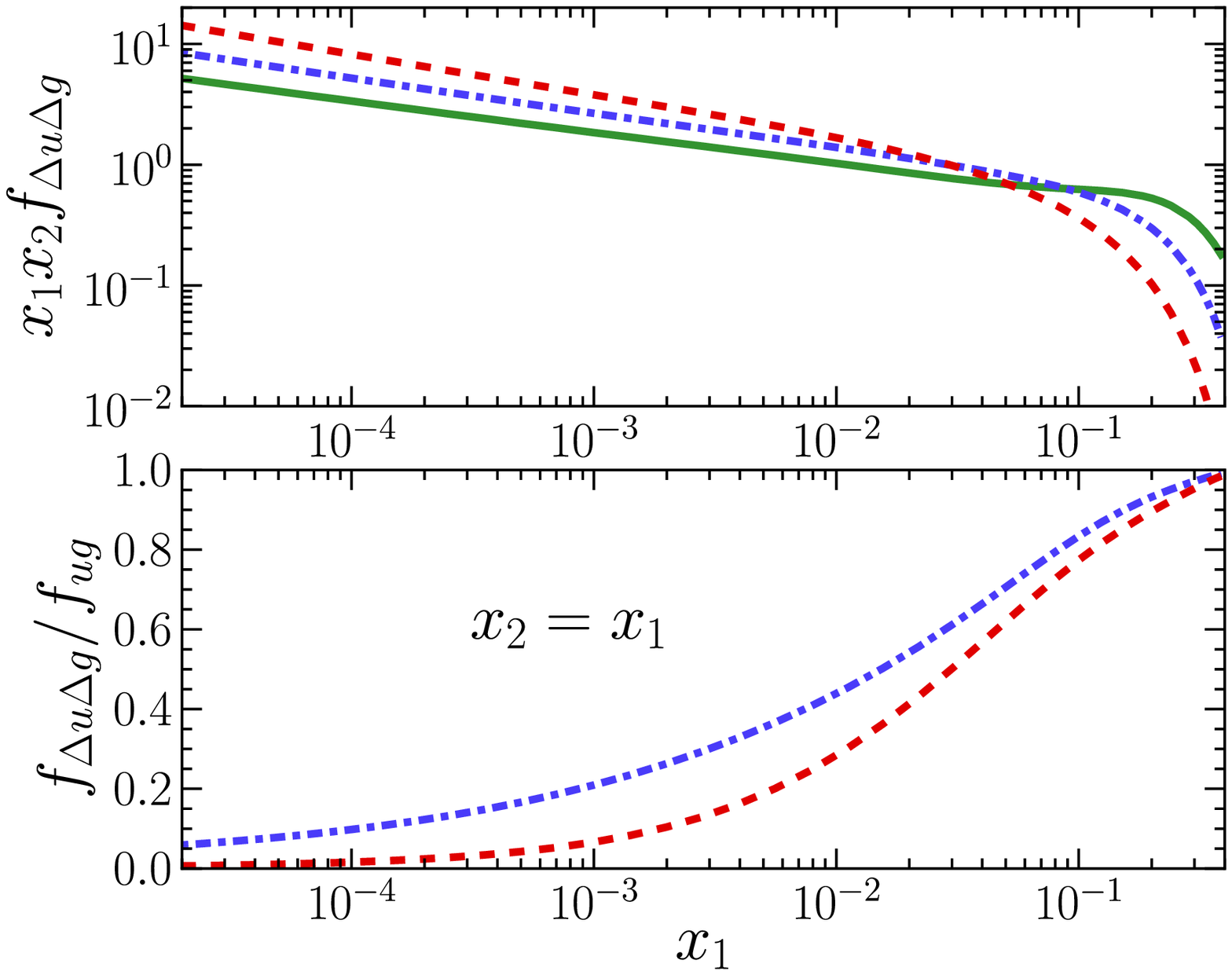}}
  \subfloat[]{\includegraphics[width=0.48\textwidth]{%
      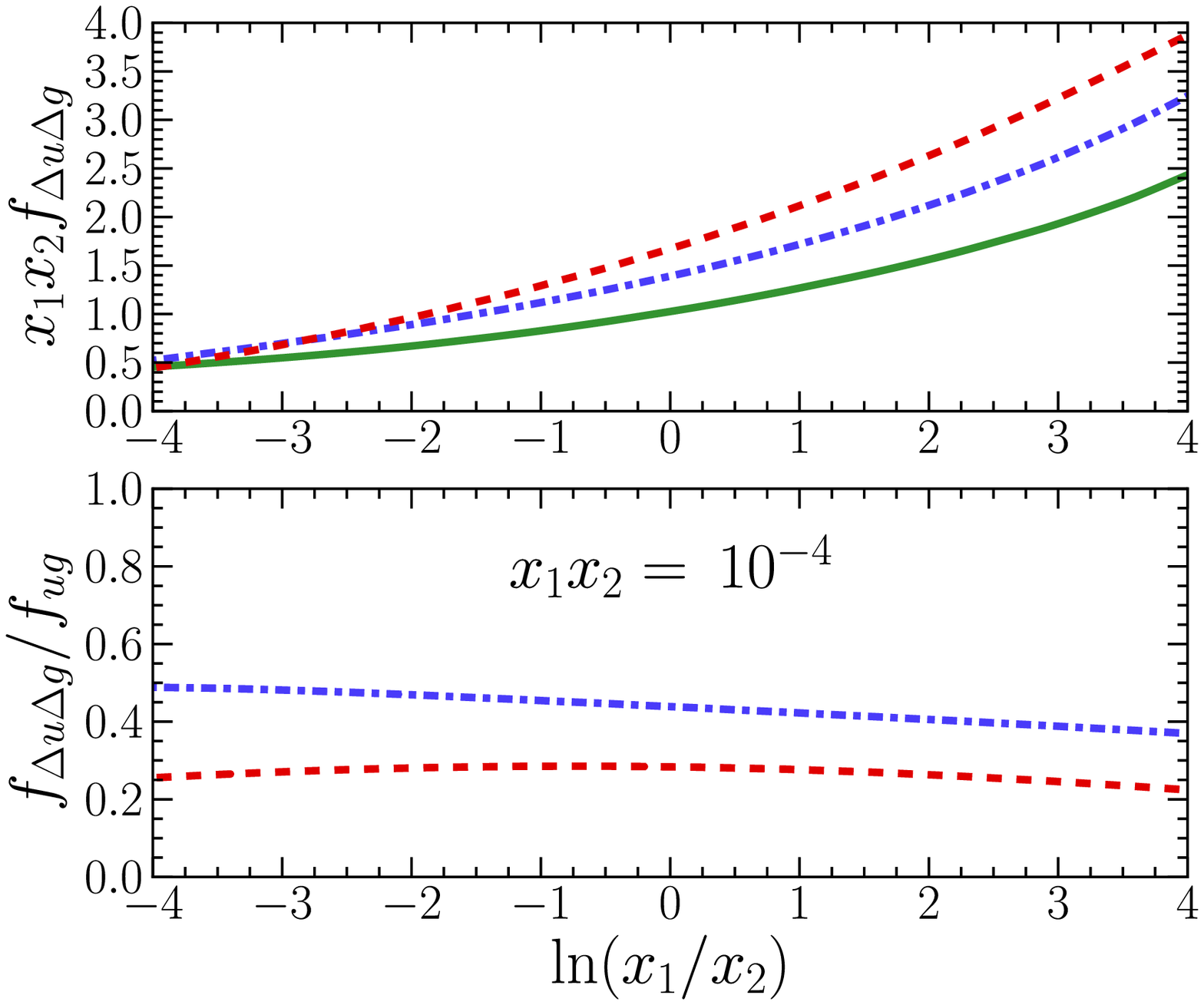}}
  \caption{\label{fig:long-mix-max} Longitudinally polarized distribution
    for an up quark and a gluon in the max scenario, with initial
    conditions using the GJR PDFs.  Color (line style) coding as in
    figure~\ref{fig:unpol-v1}.}
  \subfloat[]{\includegraphics[width=0.496\textwidth]{%
      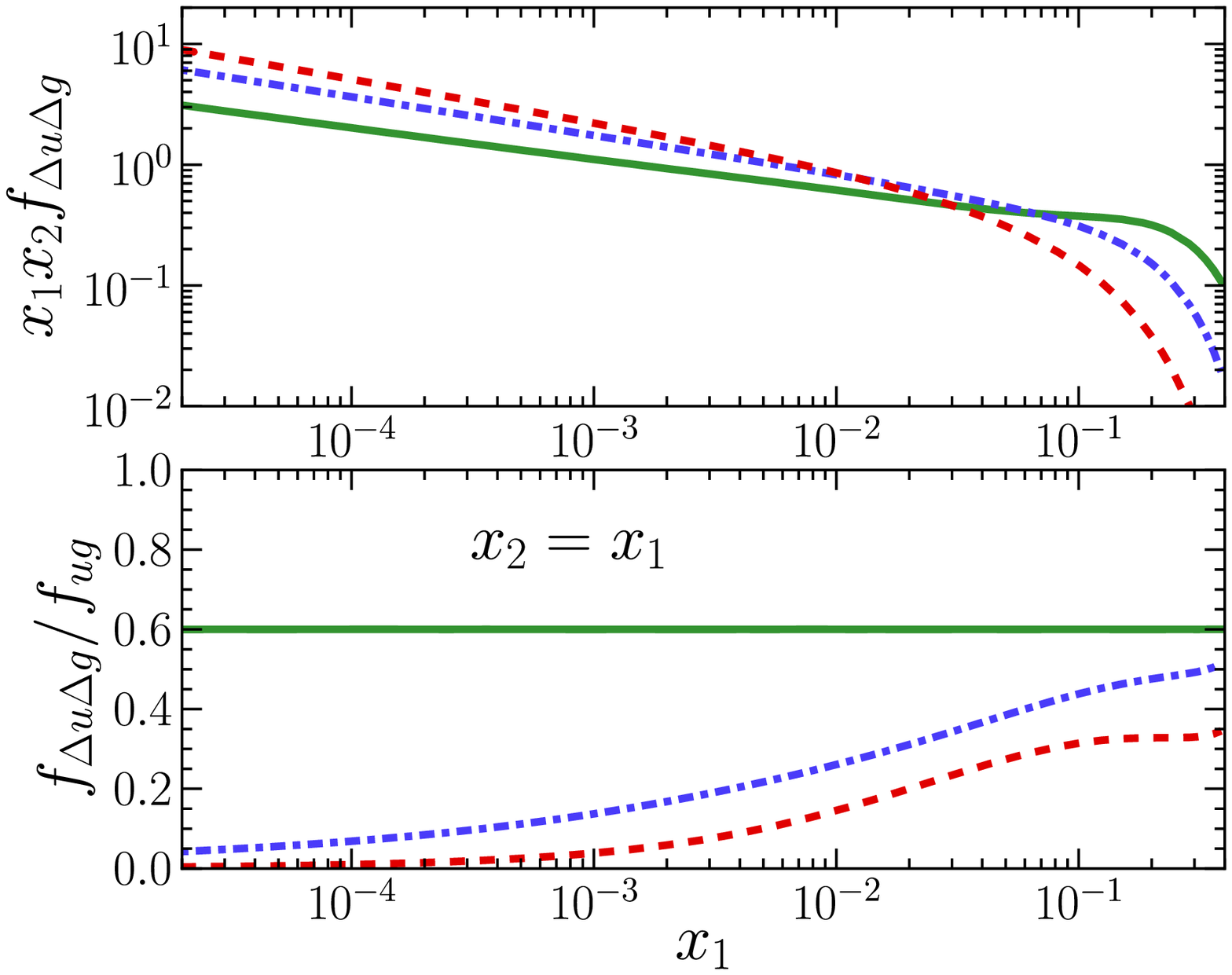}}
  \subfloat[]{\includegraphics[width=0.48\textwidth]{%
      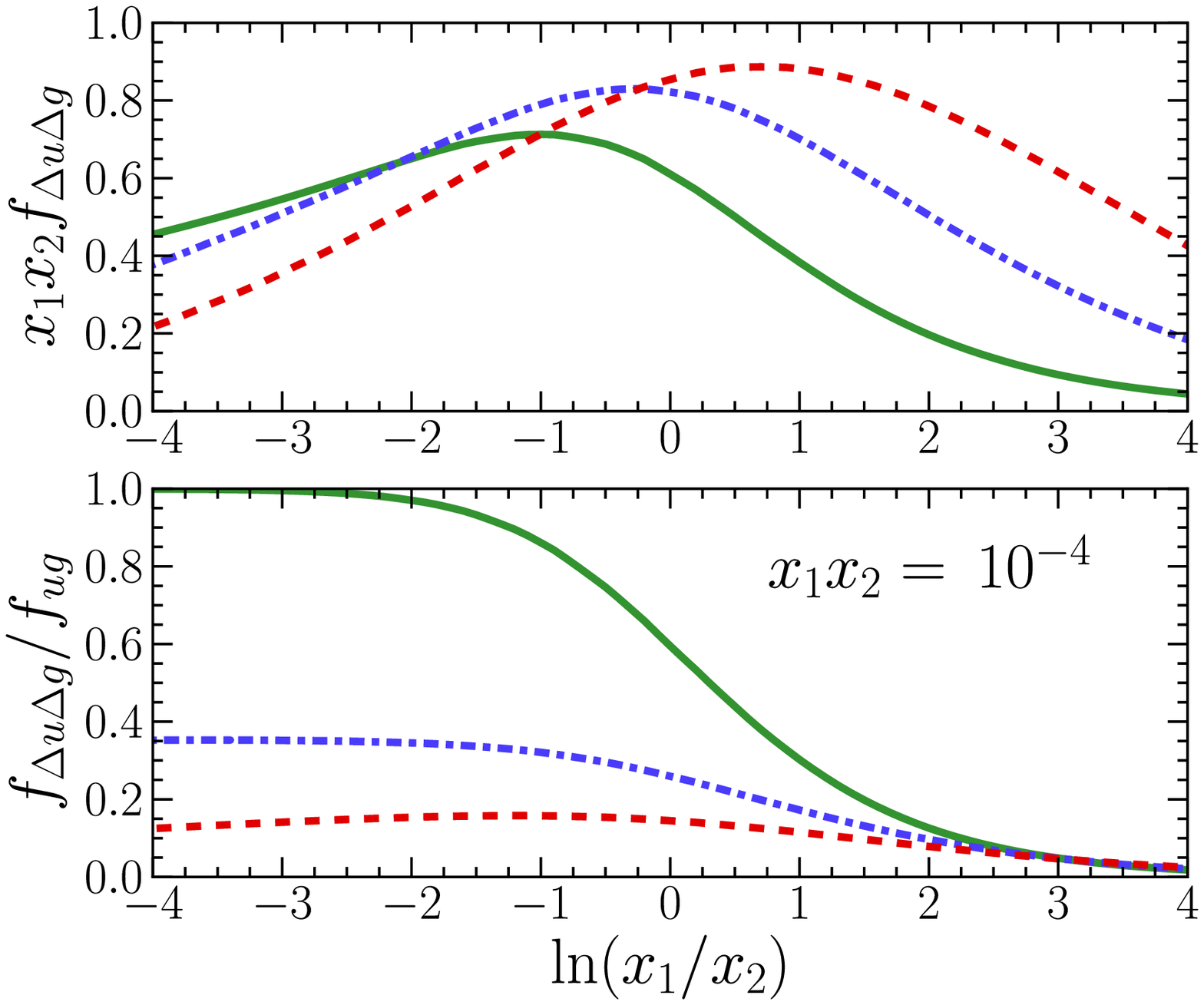}}
  \caption{\label{fig:long-mix-mdl} As
    figure~\protect\ref{fig:long-mix-max} but in the splitting scenario.}
\end{figure}

The mixed quark-gluon distribution for longitudinal polarization is shown
in figure ~\ref{fig:long-mix-max} for the max scenario using GJR
distributions as input.  There degree of polarization remains large up to
high scales except for very small $x_i$, and in
figure~\ref{fig:long-mix-max}(b) we find more than 20\% polarization for
$Q^2=10^4 \gev^2$ and $x_1\ms x_2 = 10^{-4}$.  The corresponding results
for the splitting scenario are shown in figure~\ref{fig:long-mix-mdl}.
For $x_1=x_2$ the degree of polarization is 60\% at the starting scale and
decreases moderately fast as long as $x_i$ is not too small.  In
asymmetric kinematics, the polarization remains sizeable up to high scales
if $x_1 \ll x_2$, while it is small at all scales for $x_1 \gg x_2$.

\begin{figure}[t]
  \centering
  \subfloat[]{\includegraphics[width=0.496\textwidth]{%
      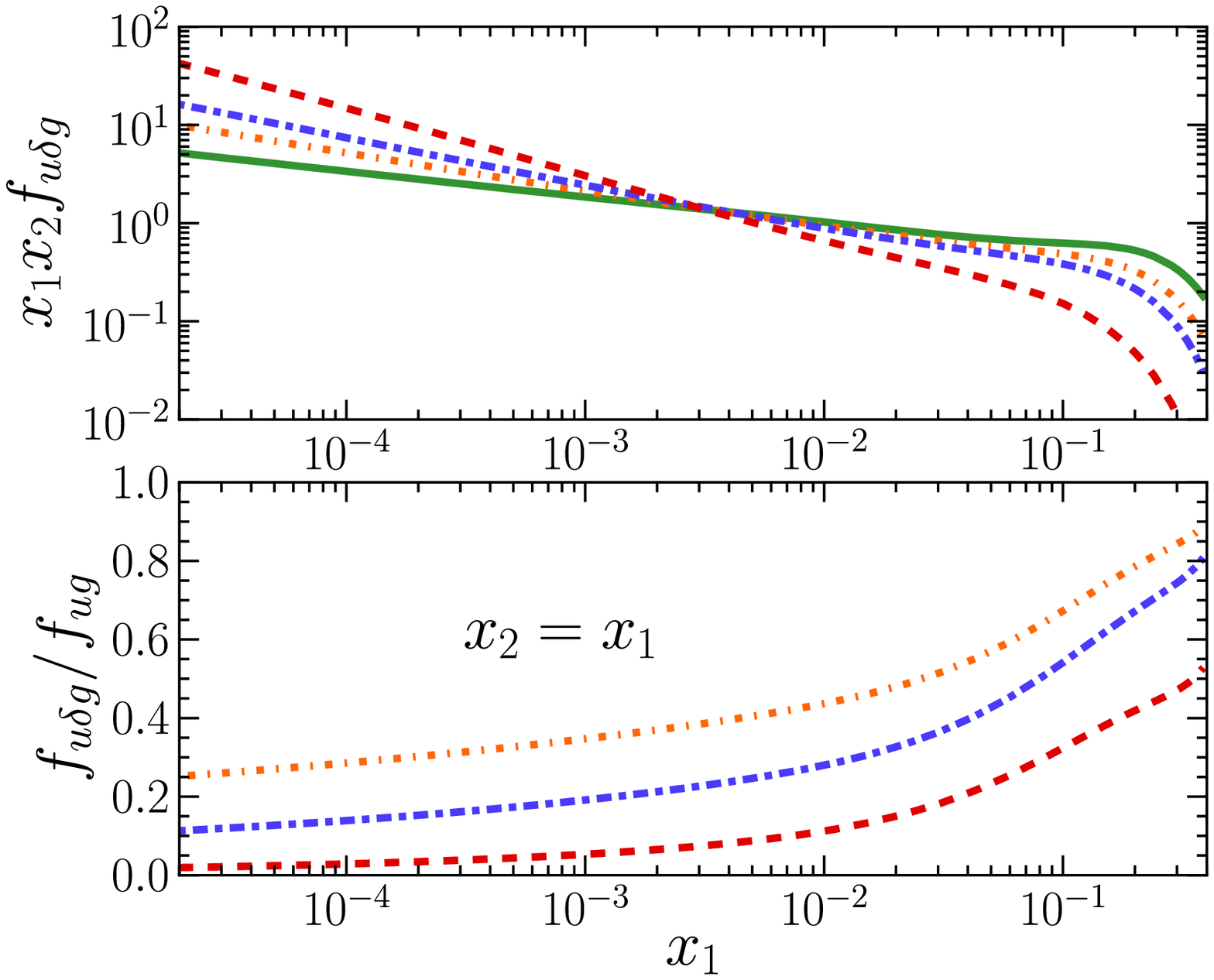}}
  \subfloat[]{\includegraphics[width=0.48\textwidth]{%
      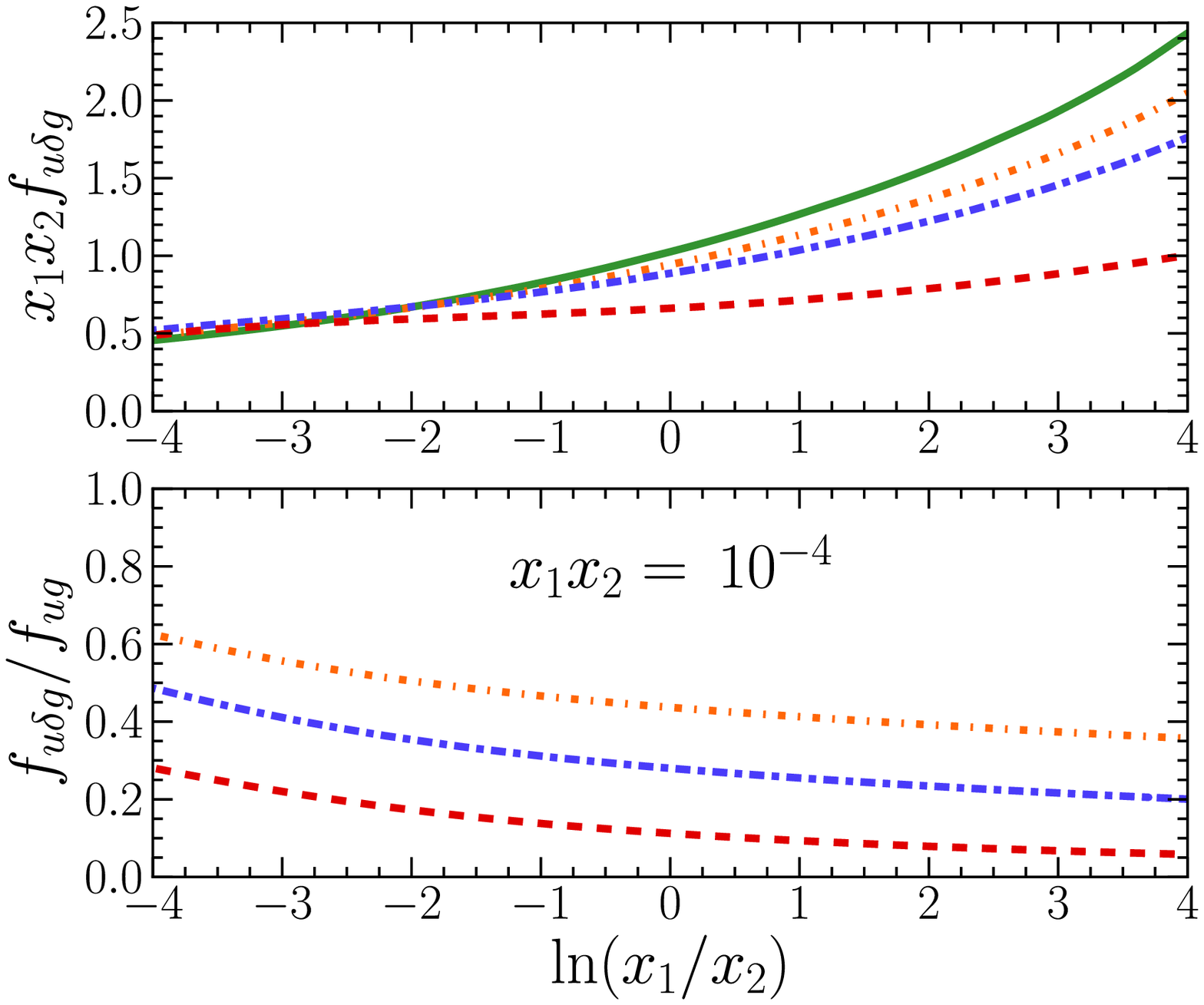}}
  \caption{\label{fig:u-ling-max} Distribution for an unpolarized up quark
    and a linearly polarized gluon in the max scenario, with GJR PDFs used
    at the starting scale.  The factor $(y M)^2$ in the starting
    conditions \eqref{eq:pos-sat-mixed} has been set to $1$ for
    simplicity.  Color (line style) coding as in
    figure~\ref{fig:long-gMSTW-max}.}
  \subfloat[]{\includegraphics[width=0.496\textwidth]{%
      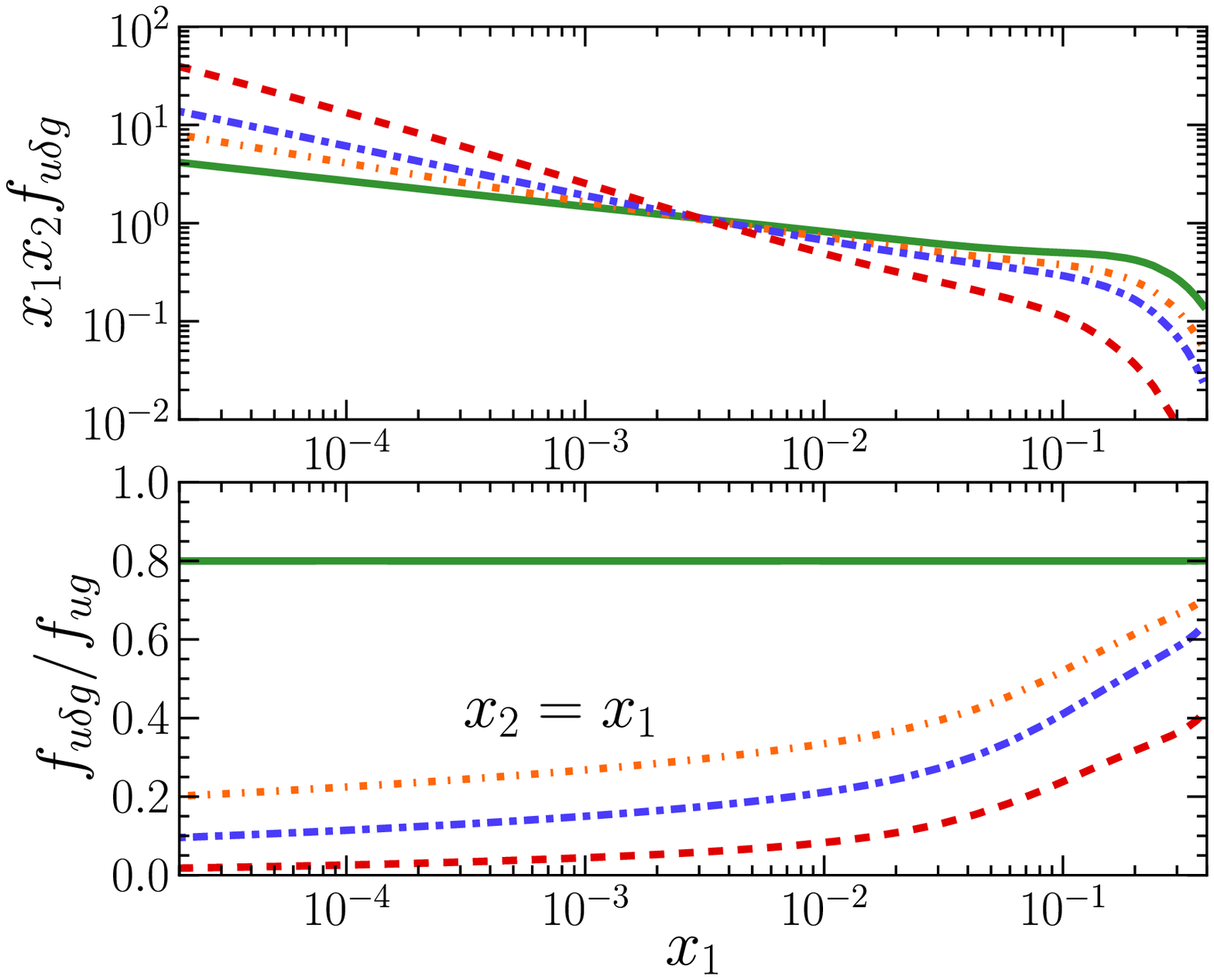}}
  \subfloat[]{\includegraphics[width=0.48\textwidth]{%
      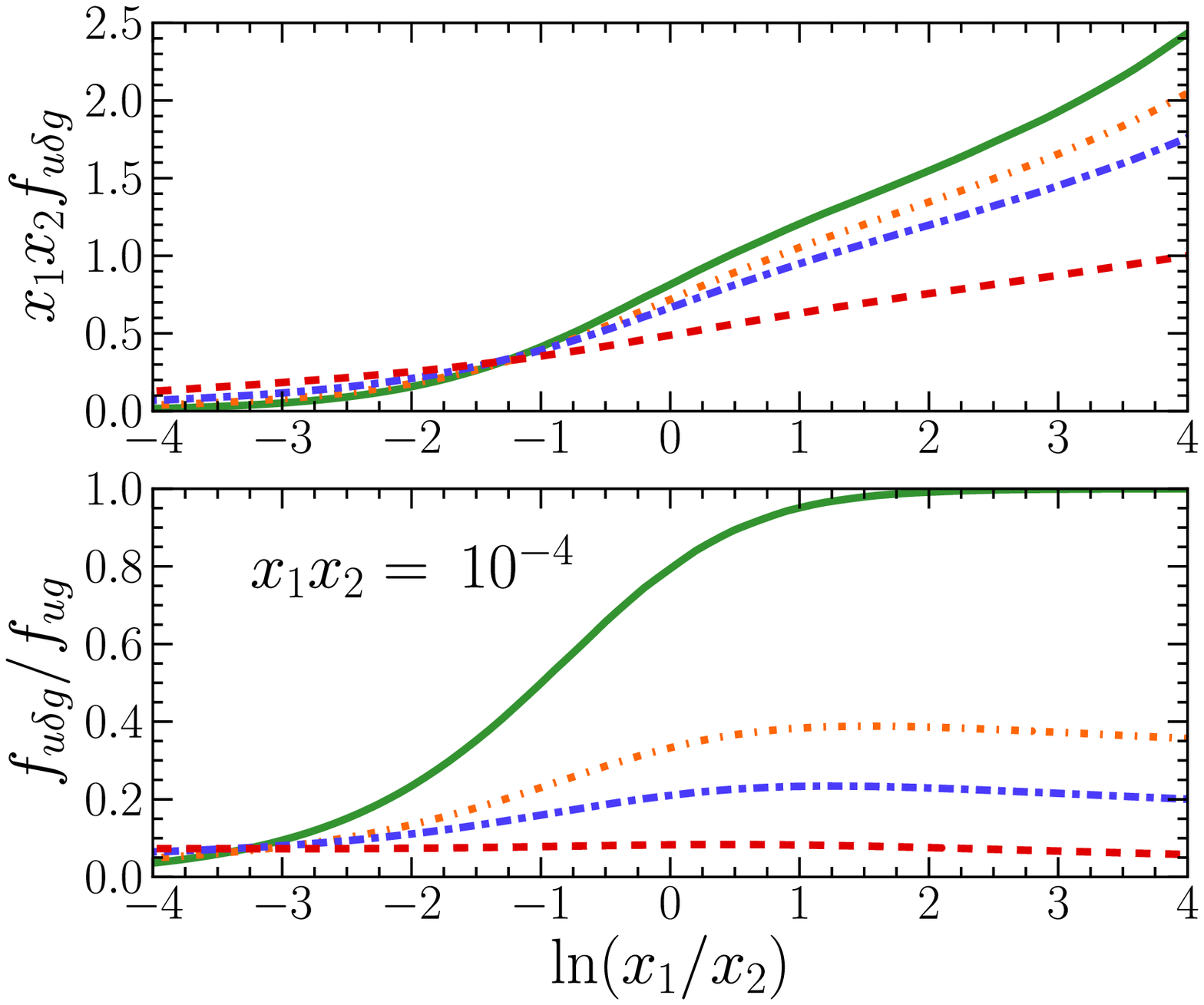}}
  \caption{\label{fig:u-ling-mdl} As figure~\protect\ref{fig:u-ling-max}
    but in the splitting scenario.  The factor $(y M)^2$ in the starting
    conditions \eqref{eq:split-coeff-single} has been set to $1$ for
    simplicity.}
\end{figure}

Let us finally discuss the distributions for an unpolarized quark and a
linearly polarized gluon.  Figure~\ref{fig:u-ling-max} shows that in the
max scenario we have a moderate degree of polarization after evolution, in
particular when $x_2 > x_1$.  In the splitting scenario, shown in
figure~\ref{fig:u-ling-mdl}, the starting conditions provide maximal
linear polarization in the limit where the gluon momentum is soft $(x_2
\ll x_1)$, in analogy to what we observed earlier for $f_{g\ms \delta g}$.
This trend is preserved by evolution up to moderately high scales.


\subsection{Very low starting scale}

The studies presented so far in this section have taken an initial scale
of $Q_0^2=1 \gev^2$ for evolution.  One may ask how our findings change if
we assume the starting conditions of the max or the splitting scenario to
hold at a much lower scale.  This question can be addressed if we
construct the initial conditions from the GJR PDFs, which are available
down to $Q^2=0.3 \gev^2$.  A very low scale is also typically associated
with quark models, which may be used to calculate unpolarized and
polarized DPDs for two quarks
\cite{Chang:2012nw,Rinaldi:2013vpa,Broniowski:2013xba}.  We note that very
large spin correlations were found in the bag model study
\cite{Chang:2012nw}.

\begin{figure}[tb]
  \centering
  \subfloat[]{\includegraphics[width=0.49\textwidth]{%
      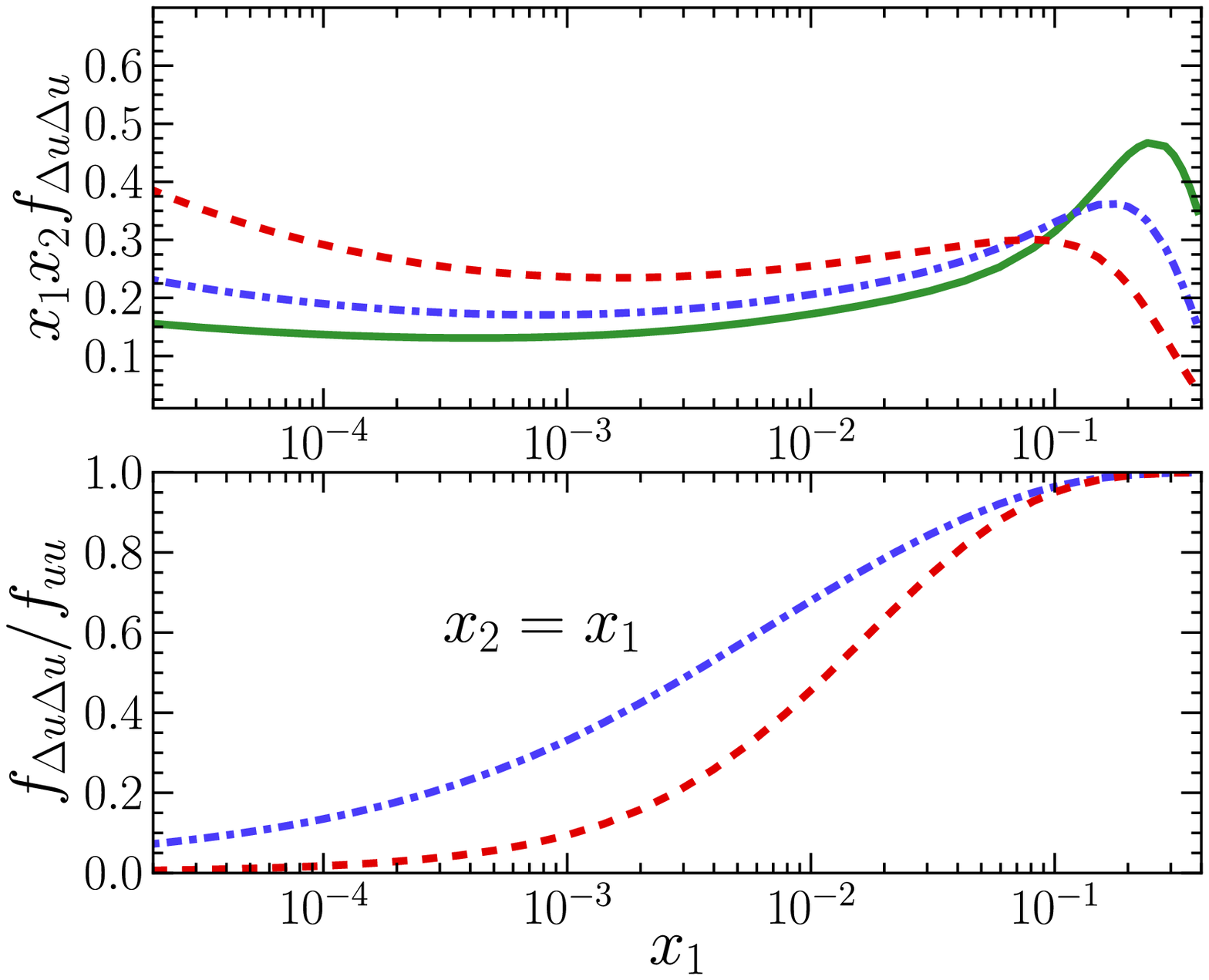}}
  \subfloat[]{\includegraphics[width=0.49\textwidth]{%
      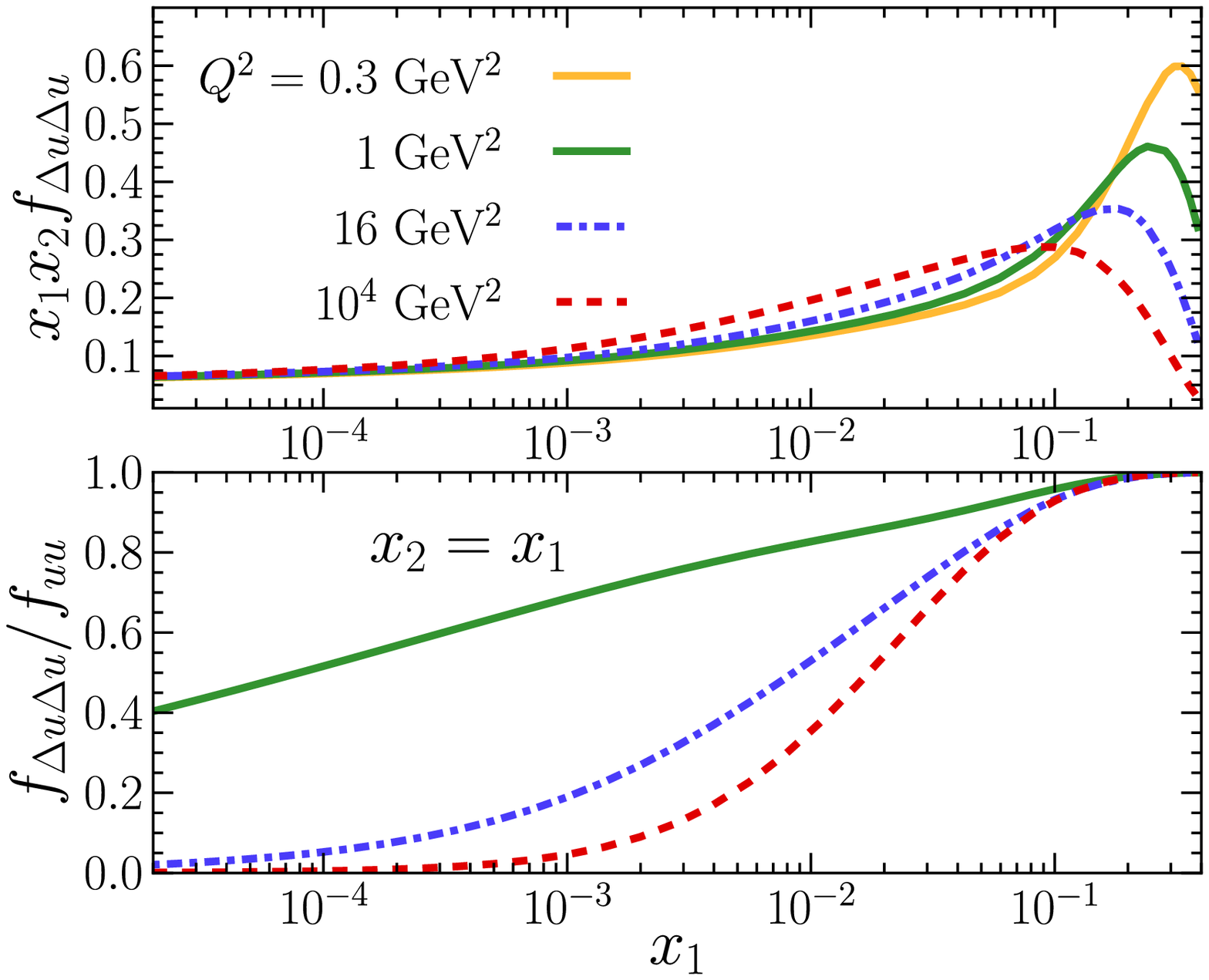}}
  \caption{\label{fig:lowQ-up} Distribution of two longitudinally
    polarized up quarks in the max scenario, with initial conditions using
    the GJR PDFs at $Q_0^2 = 1 \gev^2$ (a) or at $Q_0^2=0.3 \gev^2$ (b).}
\end{figure}

In figure~\ref{fig:lowQ-up} we compare the DPDs for two longitudinally
polarized up quarks obtained in the max scenario with the two different
starting scales just mentioned.  Worth noting is that the peak in the
valence-like distribution $f_{\Delta u \Delta u}(x,x)$ at $Q^2=0.3 \gev^2$
migrates only slowly to smaller $x$ under evolution.  In the case of two
transversely polarized quarks (not shown here) the position of that peak
stays at $x \geq 0.1$ for all scales, and the degree of transverse
polarization is smaller than the degree of longitudinal polarization after
evolution.  At $Q^2 = 10^4 \gev^2$ and $x_1=x_2=10^{-2}$ the degree of
polarization for transversely polarized up quarks is 15\%, compared with
35\% for longitudinal quark polarization.  Comparing
figures~\ref{fig:lowQ-up}(a) and (b) we note that with the higher starting
scale, $f_{\Delta u \Delta u}(x,x)$ shows a moderate growth with $Q^2$ at
small $x$, whereas with the low starting scale it barely evolves at all
for $x$ below $10^{-2}$.  The degree of polarization shows little
difference between the two cases for $Q^2 \ge 16 \gev^2$ and is largely
controlled by the rise of the unpolarized DPD with~$Q^2$.

\begin{figure}[t]
  \centering
  \subfloat[]{\includegraphics[width=0.496\textwidth]{%
      Plots3/LpGJRhigh_00_1_max.eps}}
  \subfloat[]{\includegraphics[width=0.496\textwidth]{%
      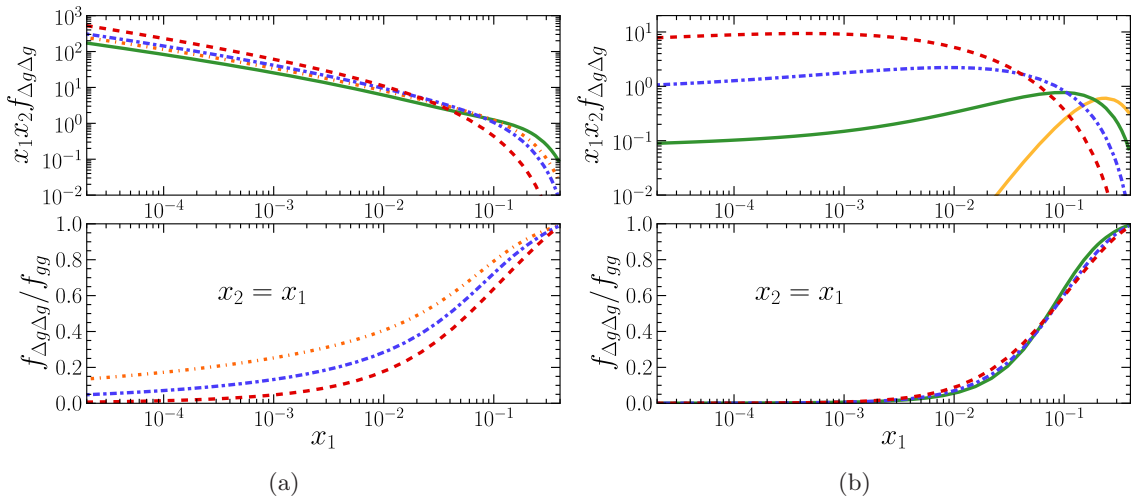}}
  \caption{\label{fig:lowQ-glu} As figure~\protect\ref{fig:lowQ-up} but
    for two longitudinally polarized gluons.}
\end{figure}

Figure~\ref{fig:lowQ-glu} shows the distribution for two longitudinally
polarized gluons obtained by starting evolution in the max scenario with
the lower or the higher starting scale.  With $Q_0^2=0.3 \gev^2$ the
degree of polarization becomes suppressed already at $Q^2 = 1 \gev^2$ and
then remains rather stable.  At low $x_i$ it becomes altogether
negligible, in contrast to the case where we start evolution at $Q_0^2 = 1
\gev^2$.  While $f_{gg}$ increases dramatically at low $x_i$ when evolved
from $Q_0^2=0.3 \gev^2$, its polarized analog $f_{\Delta g \Delta g}$
rises only slowly and never even reaches the size it has in the max
scenario at $Q_0^2 = 1\gev^2$.  Only for $x_i$ well above $10^{-2}$ do we
find a significant degree of gluon polarization in the scenario with a low
starting scale.
Comparing the evolution of $f_{\delta g \delta g}$ from $Q_0^2=0.3 \gev^2$
(not shown here) with the one of $f_{\Delta g \Delta g}$ in
figure~\ref{fig:lowQ-glu}(b), we find that $f_{\delta g \delta g}$ does
not increase with $Q^2$ at small $x_i$.  The resulting degree of linear
polarization is hence even smaller than the degree of longitudinal
polarization.

Taking the initial conditions of the splitting scenario at $Q_0^2 =
0.3\gev^2$ we find the same general pattern as in the max scenario: little
change in the degree of polarization for quarks at $Q^2 \ge 16 \gev^2$ and
a very small degree of polarization for gluons at low $x_i$.

%% file: Factorize.tex
\section{Approximation of independent partons}
\label{sec:factorize}

\subsection{Effect of the kinematic limit}
\label{sec:int-lim}

\begin{figure}
  \centering
  \subfloat[]{\includegraphics[width=0.49\textwidth]{%
      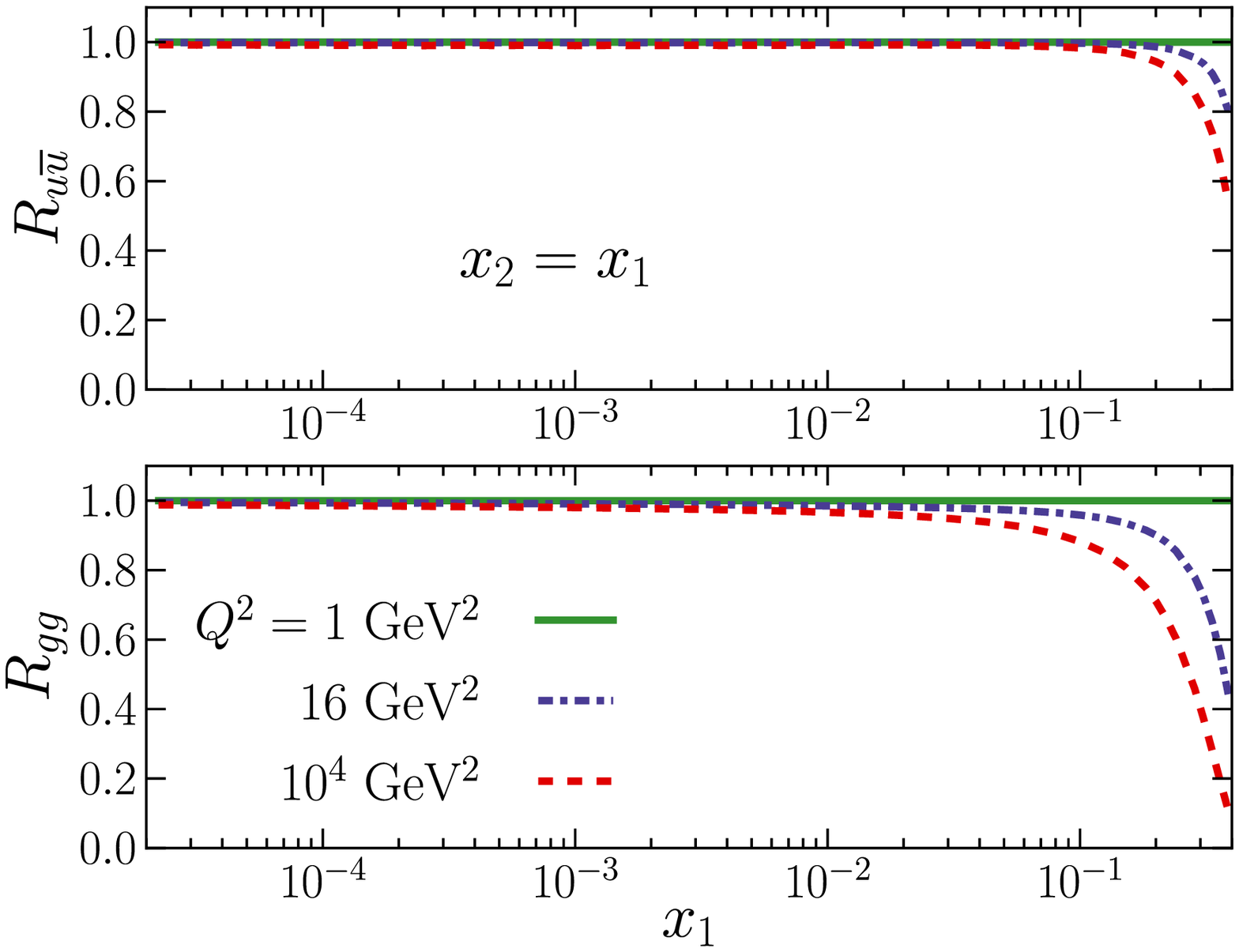}}
  \subfloat[]{\includegraphics[width=0.49\textwidth]{%
      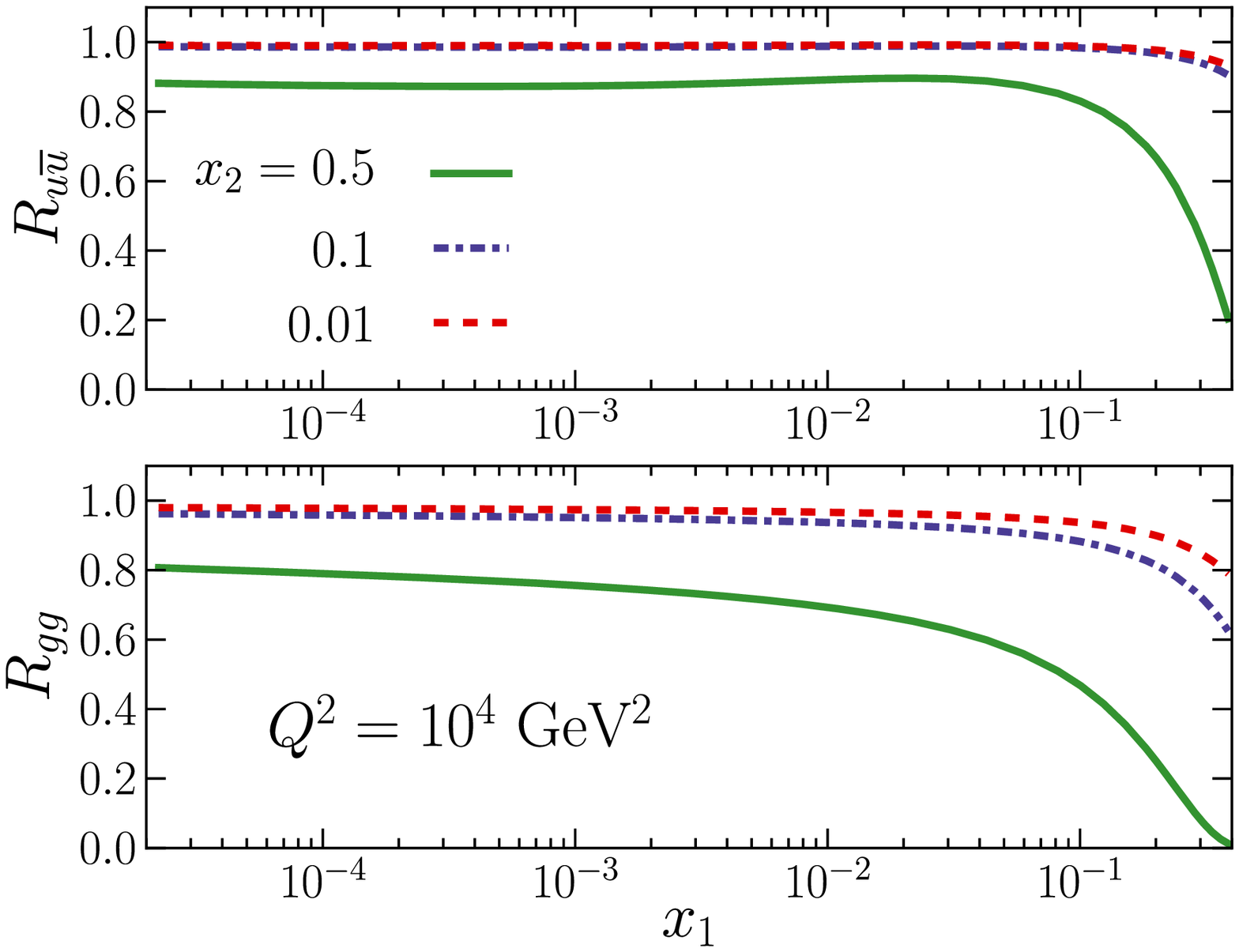}}
  \caption{\label{fig:inteffect} The ratio $R$ defined in
    \protect\eqref{eq:ratio-def}, which quantifies the effect of the
    kinematic limit on parton radiation in the double DGLAP equation.  The
    starting distribution at $Q^2 = 1\gev^2$ is the product of two PDFs
    from the MSTW set.}
\end{figure}

Perhaps the most immediate difference between single and double DGLAP
evolution is the maximal momentum fraction that can be carried by a
parton.  While the evolution equation for a single PDF involves an
integral of $f_b(z;Q)$ over momentum fractions $z$ all the way up to 1,
the corresponding integration in the double DGLAP equation
\eqref{eq:evol-parton-1} is limited by momentum conservation to
\begin{align}
  \int_{x_1}^{1-x_2} \frac{dz}{z} P_{ab}\left( \frac{x_1}{z} \right)
                     f_{bc}(z,x_2,y;Q)
\end{align}
for the evolution of the parton with momentum fraction $x_1$.  It is
obvious that the reduced integration limit has an impact for very large
$x_i$ values, but through evolution the effect can propagate down towards
smaller $x_i$.  We investigate this effect by evolving a product of MSTW
distributions, $f_{ab}(x_1,x_2; Q_0) = f_a(x_1; Q_0)\ms f_b(x_2; Q_0)$,
with the DPD evolution equations and comparing the result with the product
of evolved single parton distributions.  The ratio
\begin{align}
  \label{eq:ratio-def}
R_{ab}(x_1,x_2; Q) &=
 \frac{f_{ab}(x_1,x_2; Q)}{f_{a}(x_1; Q)\ms f_b(x_2; Q)}
\end{align}
would be equal to $1$ at all scales if the phase space effect just
described was absent.  Figure~\ref{fig:inteffect}(a) shows
$R_{u\bar{u}}$ and $R_{gg}$ at $x_2=x_1$ for three different values of
$Q^2$. We can see how the effect of the integration limit propagates down
towards lower $x_i$, especially for the gluon distribution.

In figure~\ref{fig:inteffect}(b) we show $R$ as a function of $x_1$ for
three different $x_2$ values at high~$Q^2$.  Overall, the effect of the
integration limit is not large, except at large $x_i$.  Indeed, at the
kinematic limit $x_1+x_2 = 1$ the DPD should vanish, which is ensured by
evolution even though it is not satisfied for our oversimplified starting
conditions.  Our exercise illustrates that even if one assumes that a DPD
factorizes into the product of single parton distributions at some scale,
this factorization cannot strictly hold at higher scales.


\subsection{Independent partons and $y$ dependence}
\label{sec:fact-y-dep}

The ansatz \eqref{eq:dpd-gpd} presented in \sect{sec:init-cond} assumes
that the two partons are distributed independently of each other, even if
for each of them we have correlations between longitudinal momentum and
transverse position.  It is natural to ask whether this form persists
under evolution, provided that one is sufficiently far from the kinematic
limit $x_1+x_2 = 1$ just discussed.

To answer this question is easy if one transforms both single and double
parton distributions from transverse position to the Fourier conjugate
transverse momentum.  As was pointed out in \cite{Blok:2010ge}, the
convolution in \eqref{eq:dpd-gpd} then turns into a simple product
\begin{align}
  \label{eq:dpd-gpd-r}
F_{ab}(x_1,x_2,\vek{r}) = f_a(x_1,\vek{r}) \, f_b(x_2,-\vek{r}) \,.
\end{align}
The distributions in $\vek{r}$ space do not have a probability
interpretation since $\vek{r}$ is a momentum difference between partons on
the left- and right-hand sides of the final-state cut.  This is discussed
for instance in section~2.1 of \cite{Diehl:2011yj} (where also the
normalization factors in the Fourier transforms are specified).  The
ansatz \eqref{eq:tran-gau} in impact parameter space turns into a Gaussian
\begin{align}
f_a(x,\vek{r}) = f_a(x)\, 
    \exp\bigl[ -h_{a}(x)\, \vek{r}^2 \bigr]
\end{align}
in momentum space with an $x$ dependent width.  We can now use the results
of the previous subsection to conclude that for each $\vek{r}$ the
factorized form \eqref{eq:dpd-gpd-r} will be preserved by evolution to
higher scales to good accuracy as long as $x_1 + x_2$ is not too close to
$1$.  The convolution form \eqref{eq:dpd-gpd} in transverse position space
remains of course valid to the same extent.

%% file: Conclusions.tex
\section{Conclusions}
\label{sec:concl}

Correlations between partons can have a large impact in double parton
scattering processes, both on the overall cross section and on the
distribution of particles in the final state.  We have shown that the
effect of double DGLAP evolution, where each of the two partons develops
its own parton cascade, generally suppresses such correlations at higher
scales.  The strength of this suppression varies widely, with a rapid
decrease of correlations in some cases and a slow decrease in others.

At a certain degree of accuracy, the dependence of DPDs on the transverse
distance $y$ between the two partons is expected to depend on the type of
the partons and on their momentum fractions.  We have studied the
evolution of a $y$ dependence motivated by the phenomenology of
generalized parton distributions at the initial scale.  We find that a
Gaussian $y$ dependence at the initial scale is approximately preserved
under evolution, with a noticeable but relatively slow change of the
effective Gaussian width.  Despite the mixing between gluons and quarks in
the singlet sector, the differences between their distributions persist up
to high scales.

Spin correlations between two partons in the proton are described by
polarized DPDs.  Positivity constrains these to be at most as large as the
unpolarized DPDs for the same parton types, a property that is preserved
under evolution to higher scales.  For the initial conditions of
evolution, we have either assumed maximum polarization or a degree of
polarization as it is obtained when the two partons originate from the
perturbative splitting of a single, unpolarized one.  In the latter
scenario, the degree of polarization strongly depends on the ratio
$x_2/x_1$ of momentum fractions for certain parton combinations.  We find
that the DPDs for two longitudinally or two transversely polarized quarks
decrease slowly with the evolution scale $Q^2$ or even remain
approximately constant.  Given the rise of the corresponding unpolarized
DPD, the degree of longitudinal or transverse quark polarization shows a
rather pronounced decrease with $Q^2$.  The DPD for two longitudinally
polarized gluons rises slowly with the scale, as does the DPD for a
longitudinally polarized gluon and a longitudinally polarized quark.
However, due to the very rapid increase of unpolarized gluon distributions
with $Q^2$, the degree of longitudinal polarization decreases with the
scale both for $gg$ and $qg$ distributions, in particular at small $x$.
The distribution for two linearly polarized gluons decreases under
evolution, and the associated degree of polarization quickly becomes
negligible for $x$ below $10^{-2}$.  In certain processes like double
charm production, the DPD for one unpolarized and one linearly polarized
gluon is relevant.  Except at high $x$, it increases with $Q^2$ and the
corresponding degree of polarization decreases rather gently.  Our
quantitative results in the gluon sector depend strongly on the initial
scale of evolution and on the gluon densities used in the initial
conditions, which entails a much stronger model dependence than for
quarks.  Broadly speaking, we find that almost all polarization effects
become small for $x \le 10^{-2}$ and $Q^2 \ge 10^{4} \gev^2$, whereas for
$x$ above a few $10^{-2}$ many polarization correlations remain sizeable
even at high scales.

The phase space available for parton radiation in DPDs is reduced compared
with the case of single parton densities.  As a result, evolution does not
conserve the factorization of DPDs into separate functions of the momentum
fractions $x_1$ and $x_2$ if one assumes this property at a certain scale.
Quantitatively, we find that this effect is important only if at least one
of the momentum fractions is of order $0.3$ or larger, otherwise one
retains a factorized form to a good approximation.  This result
generalizes to the convolution ansatz specified in \eqref{eq:dpd-gpd}
because this ansatz corresponds to a product in the momentum space
representation \eqref{eq:dpd-gpd-r}.

In summary, we find that the effect of scale evolution on parton
correlations is important and should be included in quantitative
estimates.  The assumption that parton radiation will quickly wash out
correlations is true in a few cases but cannot serve as a general
guideline.  How this affects double parton scattering processes remains to
be studied in future work.

%% file: Appendices.tex
\appendix


\section{Choice of single parton densities}
\label{ap:pdfs}

The models we use for the initial conditions of unpolarized DPDs in
\sects{sec:trans}, \ref{sec:pol} and \ref{sec:factorize} are constructed
from products of ordinary single parton distributions.  For this we use LO
PDFs, so as to match the leading-order evolution we perform for the DPDs.
We need these PDFs at low scales, down to $Q_0 = 1\gev$, where different
PDF sets significantly deviate from each other.

\begin{figure}[tb]
\centering
\subfloat[]{\includegraphics[width=0.49\textwidth]{%
    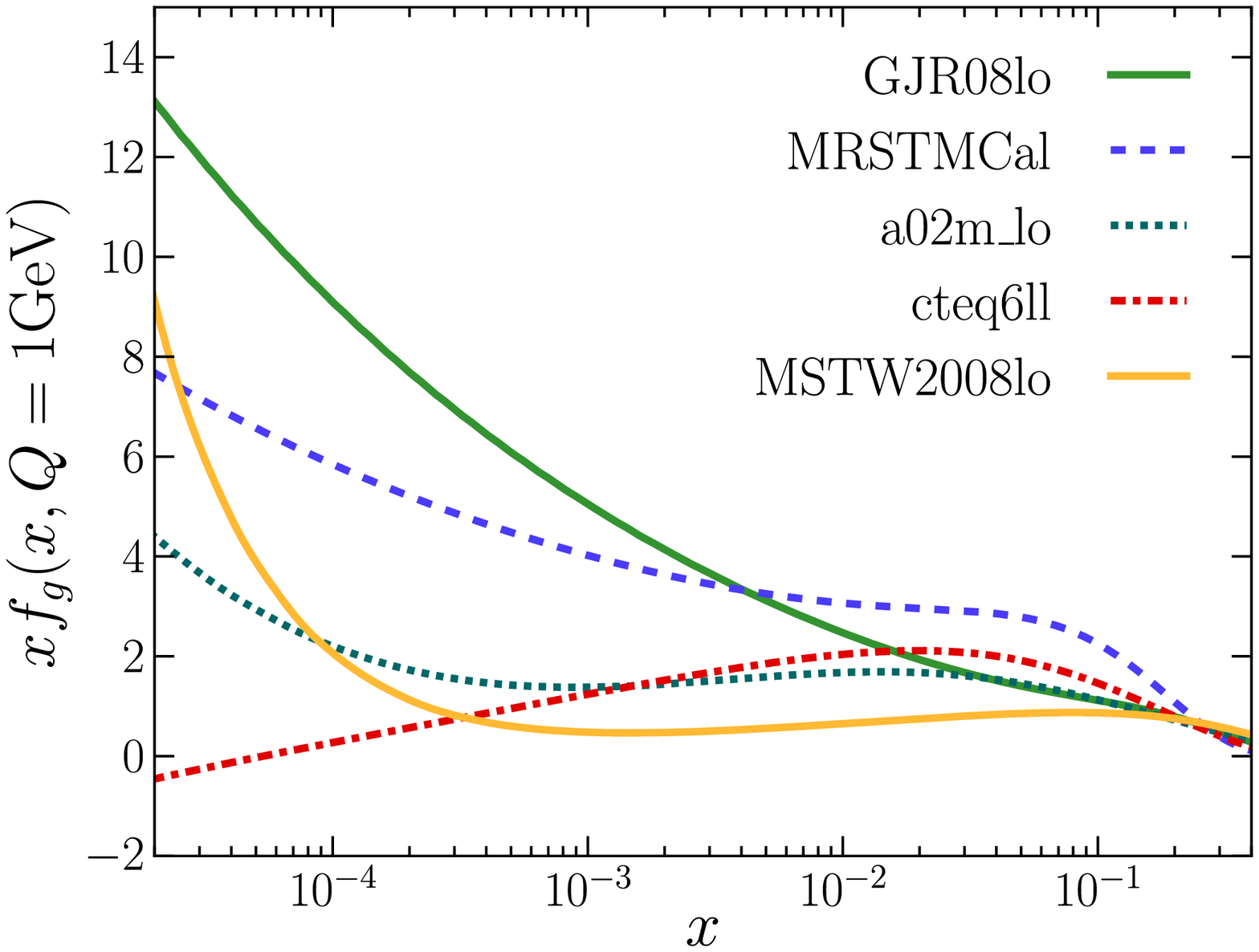}}
\subfloat[]{\includegraphics[width=0.49\textwidth]{%
    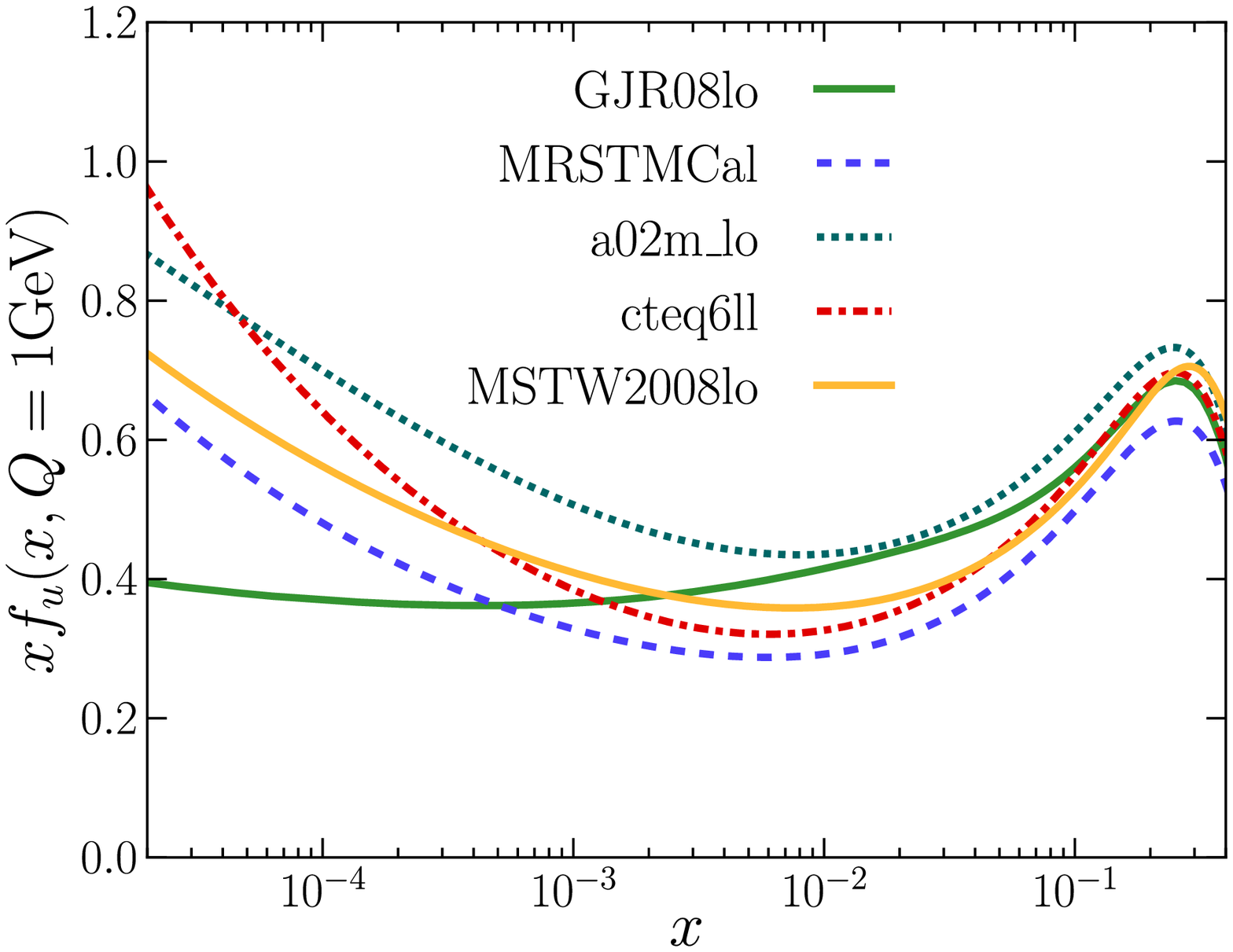}}
\caption{\label{fig:pdfs} Comparison of recent LO PDF sets at scale $Q =
  1\gev$ for gluons (a) and for $u$ quarks (b).  Note the different $y$
  ranges in the two panels.}
\end{figure}

This is clearly seen in figure~\ref{fig:pdfs}, where we show LO PDFs from
Alekhin (a02m\_lo) \cite{Alekhin:2002fv}, CTEQ6 (cteq6ll)
\cite{Pumplin:2002vw}, GJR (GJR08lo) \cite{Gluck:2007ck} and MSTW
(MSTW2008lo) \cite{Martin:2009iq}.  We also show the dedicated Monte Carlo
PDF set MRSTMCal \cite{Sherstnev:2008dm}; the other set of that study
(MRST2007lomod) looks similar.  All PDF values are generated using the
LHAPDF interface \cite{LHAPDF}.  Not included in the figure are the LO
PDFs of NNPDF2.1 \cite{Ball:2011uy} since they are not available at
$Q=1\gev$ via LHAPDF.  The CTEQ6 gluon distribution turns negative at low
$x$ and is hence not suited for our purpose (one of our two models for
polarized DPDs in \sect{sec:pol} builds on the positivity of parton
distributions).  For the same reason we discard the dedicated Monte Carlo
PDFs of CT09 \cite{Lai:2009ne} (not shown here), although they are set to
zero below some value of $x$ instead of going negative.  The LO gluon
distribution of HERAPDF1.5 \cite{HERAPDF:2013} is very close to the one of
CTEQ6 at $Q=1\gev$ (and hence not shown in the figure).  It also turns
negative at low $x$.

Among the positive gluon PDFs shown in the figure, the sets of GJR08 and
MSTW2008 represent extremes in the sense of having a very steep or a very
flat behavior over a wide $x$ range.  We chose these two sets for our
investigations of DPDs, expecting that results obtained with different
PDFs should approximately lie within the range covered by the two
representatives we have selected.
In the right panel of the figure we see that the spread of $u$ quark
distributions in the different PDF sets is notable, but not as large as
for the gluons.

For the evolution of DPDs, we adjusted the values of $\alpha_s$ and of the
quark masses to those used by the two PDF sets we have selected.  This is
\begin{align}
m_c & = 1.30 \gev \,, & m_b & = 4.2 \gev \,, &
\alpha_s(Q=1 \gev) & = 0.4482
\intertext{for GJR and}
m_c & = 1.40 \gev \,, & m_b & = 4.75 \gev \,, &
\alpha_s(Q=1 \gev) & = 0.6818
\end{align}
for MSTW.


\section{Double parton distributions from perturbative splitting}
\label{ap:split}

For small interparton distances $y$, or more precisely in the limit $y
\Lambda \ll 1$, where $\Lambda$ is a typical hadronic scale, the dominant
contribution to DPDs is given by the short-distance splitting of a single
parton into two \cite{Diehl:2011yj}.  To leading order in $\alpha_s$, one
can then express the DPD as the product of a usual PDF with an expression
for the perturbative splitting; at higher orders the product turns into a
convolution.  From the results of section 5.2 in \cite{Diehl:2011yj} we
can readily extract the corresponding expressions for the collinear
color-singlet DPDs we are studying in the present work.

For unpolarized or doubly polarized DPDs, we have at leading order in
$\alpha_s$
\begin{align}
  \label{splitting-1}
f_{p_1\ms p_2}(x_1,x_2,y) &=
  \frac{\alpha_s}{2\pi^2}\, \frac{1}{y^2}\,
    \frac{f_{p_0}(x_1+x_2)}{x_1+x_2}\,
       T_{p_0\to p_1 p_2}\Bigl( \frac{x_1}{x_1+x_2} \Bigr)
\end{align}
from the splitting process $p_0\to p_1\ms p_2$, with kernels
\begin{align}
T_{g\to q\bar{q}}(z) &= \frac{1}{2} \bigl( z^2 + \bar{z}^2 \bigr) \,,
&
T_{g\to \Delta q \Delta\bar{q}}(z) &= - T_{g\to q\bar{q}}(z) \,,
&
T_{g\to \delta q \delta\bar{q}}(z) &= - z\bar{z} \,,
\nonumber \\[0.2em]
T_{g\to g g}(z) &= 2 N_c
  \biggl[ \frac{\bar{z}}{z} + \frac{z}{\bar{z}} + z\bar{z} \biggr] \,,
&
T_{g\to \Delta g \Delta g}(z) &= 2 N_c\, (2 - z\bar{z}) \,,
&
T_{g\to \delta g \delta g}(z) &= 2 N_c\ms z\bar{z} \,,
\nonumber \\[0.2em]
T_{q\to q g}(z) &= C_F\, \frac{1+z^2}{\bar{z}} \,,
&
T_{q\to \Delta q \Delta g}(z) &= C_F\, (1+z) \,,
\nonumber \\
T_{\bar{q}\to \bar{q} g}(z) &= T_{q\to q g}(z) \,,
&
T_{\bar{q}\to \Delta\bar{q} \Delta g}(z) &= T_{q\to \Delta q \Delta g}(z) \,,
\end{align}
where $\bar{z} = 1-z$.  Corrections to \eqref{splitting-1} are suppressed
by further powers of $\alpha_s$ or by powers of $y\Lambda$.  In $T_{g\to
  q\bar{q}}$, $T_{g\to gg}$ and $T_{q\to qg}$ we recognize the familiar
DGLAP splitting functions for $z<1$.  Further DPDs are obtained by
permuting the parton labels and momentum fractions.  All other unpolarized
or doubly polarized distributions, including $f^t_{p_1\ms
  p_2}(x_1,x_2,{y})$, do not receive any contribution from perturbative
splitting at this accuracy.

DPDs with one polarized and one unpolarized parton arise by perturbative
splitting only for linearly polarized gluons,
\begin{align}
  \label{splitting-2}
f_{a\ms \delta g}(x_1,x_2,y) &=
  \frac{\alpha_s}{2\pi^2}\, \frac{1}{y^4 M^2}\,
    \frac{f_{a}(x_1+x_2)}{x_1+x_2}\,
       T_{a\to a\ms \delta g}\Bigl( \frac{x_1}{x_1+x_2} \Bigr)
\end{align}
with
\begin{align}
T_{g\to g\ms \delta g}(z) &= 2 N_c\ms \frac{z}{\bar{z}} \,,
&
T_{q\to q\ms \delta g}(z) &= 2 C_F\ms \frac{z}{\bar{z}} \,.
\end{align}
Because $f_{g\ms \delta g}$ and $f_{q\ms \delta g}$ are multiplied by two
vectors $\vek{y}$ in their definitions, the associated distributions
$F^{jj'}_{g\ms \delta g}$ and $F^{jj'}_{q\ms \delta g}$ diverge like
$1/y^2$ for $y\to 0$, just as the distributions in \eqref{splitting-1}.
The distributions $f_{g\ms \delta q}$ and $f_{\bar{q}\ms \delta q}$ do not
receive contributions from perturbative splitting due to the chiral
invariance of massless QCD.

%% file: Bibliography.tex
\providecommand{\href}[2]{#2}